Topical Review

# Nanoparticle-enhanced Multifunctional Nanocarbons as Metal-ion Battery and Capacitor Anodes and Supercapacitor Electrodes - Review

*Subrata Ghosh[1*], S. R. Polaki[2], Andrea Macrelli[1], Carlo S. Casari[1], Suelen Barg[3], Sang Mun Jeong[4], and Kostya (Ken) Ostrikov[5]*

[1] *Micro and Nanostructured Materials Laboratory – NanoLab, Department of Energy, Politecnico di Milano, via Ponzio 34/3, Milano I-20133, Italy*

[2] *Surface and Nanoscience Division, Materials Science Group, Indira Gandhi Centre for Atomic Research, Homi Bhabha National Institute, Kalpakkam, TN - 603102, India*

[3] *Institute for Materials Resource Management, Augsburg University, 2, 86159 Augsburg, Germany*

[4] *Department of Chemical Engineering, Chungbuk National University, Cheongju, Chungbuk, 28644 Republic of Korea*

[5] *School of Chemistry and Physics and QUT Centre for Materials Science, Queensland University of Technology (QUT), Brisbane QLD 4000, Australia*

ORCID ID:
Subrata Ghosh https://orcid.org/0000-0002-5189-7853
S R Polaki https://orcid.org/0000-0002-6344-4472
Andrea Marcelli https://orcid.org/0000-0002-5307-5124
Carlo S Casari https://orcid.org/0000-0001-9144-6822
Suelen Barg https://orcid.org/0000-0002-0723-7081
Sang Mun Jeong https://orcid.org/0000-0002-3694-3110
Kostya (Ken) Ostrikov https://orcid.org/0000-0001-8672-9297

Corresponding author email: subrataghosh.phys@gmail.com, subrata.ghosh@polimi.it

Topical Review

***Abstract***

As renewable energy is becoming a critical energy source to meet the global demand, electrochemical energy storage devices become indispensable for the efficient energy storage and reliable supply. The electrode material is the key factor determining the energy storage capacity and the power delivery of the devices. Carbon-based materials, specifically graphite, activated carbons etc., are extensively used as for electrodes, yet their low energy densities impede the development of advanced energy storage materials. Nanoparticle decoration of the carbon structures is one of the most promising and easy-to-implement strategy to enhance the charge-storage performance of carbon-based electrodes. Decoration by nanoparticles of metals, metal oxides, nitrides, carbides, phosphides, chalcogenides, and bimetallic components lead to significant enhancements in the structural and electronic properties, pore refinement, charge-storage, and charge-transfer kinetics of both pristine and doped carbon structures, thereby making their performance promising for next-generation energy storage devices. Structuring the materials at nanoscale is another probable route for better rate performance and charge-transfer kinetics. This review covers the state-of-art nanoparticle decorated nanocarbons as materials for battery anode, metal-ion capacitor anode, and supercapacitor electrode. A critical analysis of the elemental composition, structure, associated physico-chemical properties and performance relationships of nanoparticle-decorated nanocarbon electrodes is provided as well to inform the future development of the next generation of advanced energy storage materials and devices.

## 1. Introduction

Electrochemical energy storage technology is one of the promising solutions for sustainable and green energy in the period of global energy crisis.[1] Batteries, supercapacitors, and metal-ion capacitors are the three major types of devices that have drawn a significant attention to the industrial and academic community.[2–5] However, these devices suffer from several limitations as elucidated by the Ragone plot (Figure 1A). Briefly, the battery is capable to deliver a high energy density but a poor power density, whereas the supercapacitors is well-known for high-power applications yet it possesses low energy density and the metal-ion capacitors performance lies in between the former two (Figure 1A). Unfortunately, none of these devices can be used independently to store/deliver energy effectively. The common constituents of these devices are electrodes (anode and cathode) and electrolyte, and each of them has a distinct role to play. As the performance of the energy storage device greatly depends on the properties of the electrode materials, engineering advanced materials is the foremost challenge and also highly desirable in pursuing the energy storage technology.

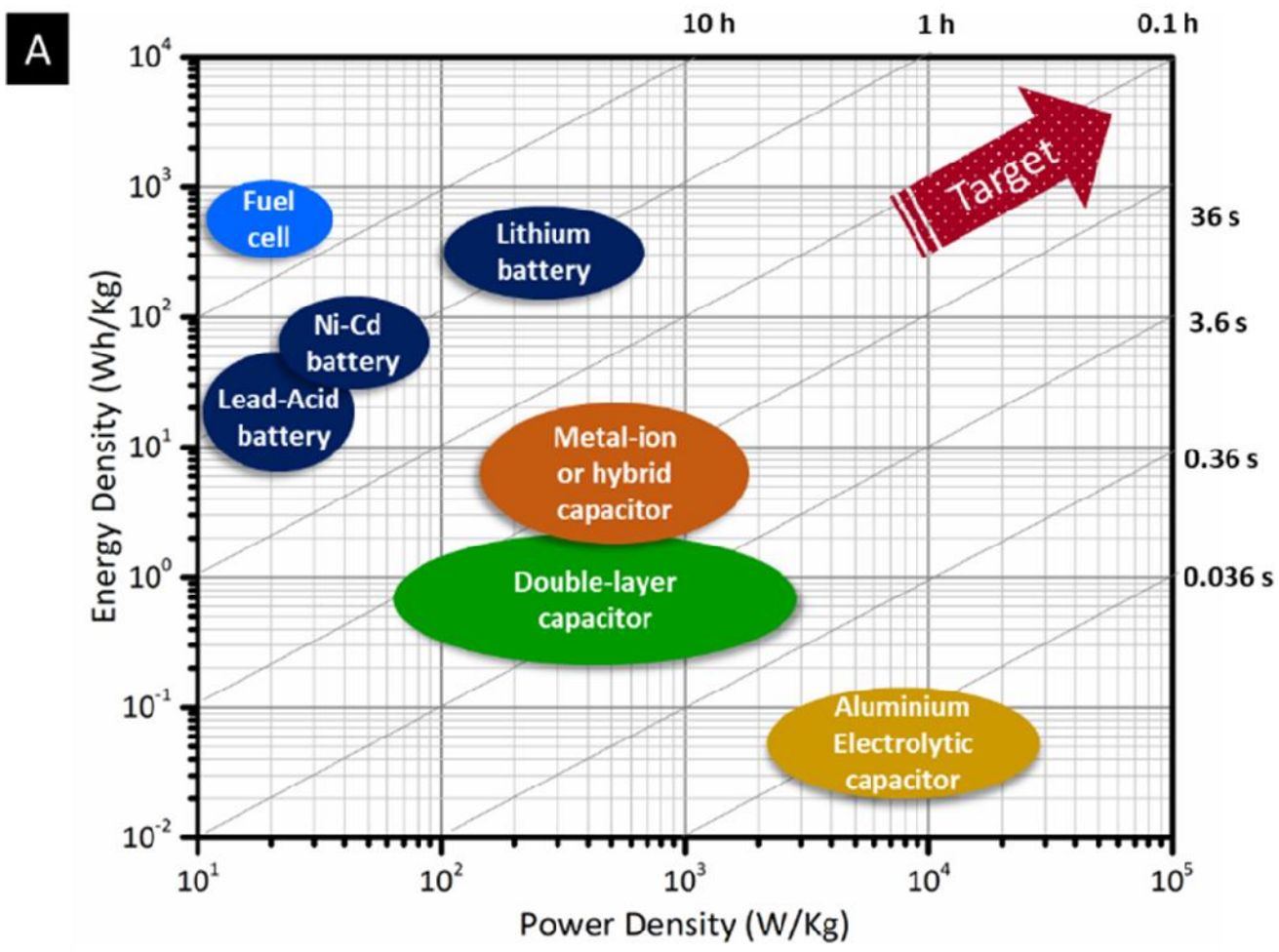

B

| | Li-ion battery | Supercapacitor | Li-ion capacitor |
|---|---|---|---|
| Storage mechanism | Chemical | Physical | Chemical & Physical |
| Internal resistance | High | Low | Medium |
| Minimum operating voltage (V) | 2.5 | 0 | 2.2 |
| Maximum rated voltage (V) | 4.1-4.3 | 2.3-3.0 | 3.8 |
| Energy density (Wh/Kg) | 75-250 | 2.5-15 | 10-100 |
| Power density (W/Kg) | 150-315 | 500-10000 | 300-5000 |
| Cycle life | 1000-10000 | 100000-10000000 | 10000-100000 |
| Operating temperature | -20 – 60 °C | -40 – 70 °C | -25 – 85 °C |
| Self-discharge | Very low | Very high | Low |
| Safety | Need improvement | Good | Good |
| Voltage monitoring | Needed | Not required | Needed |

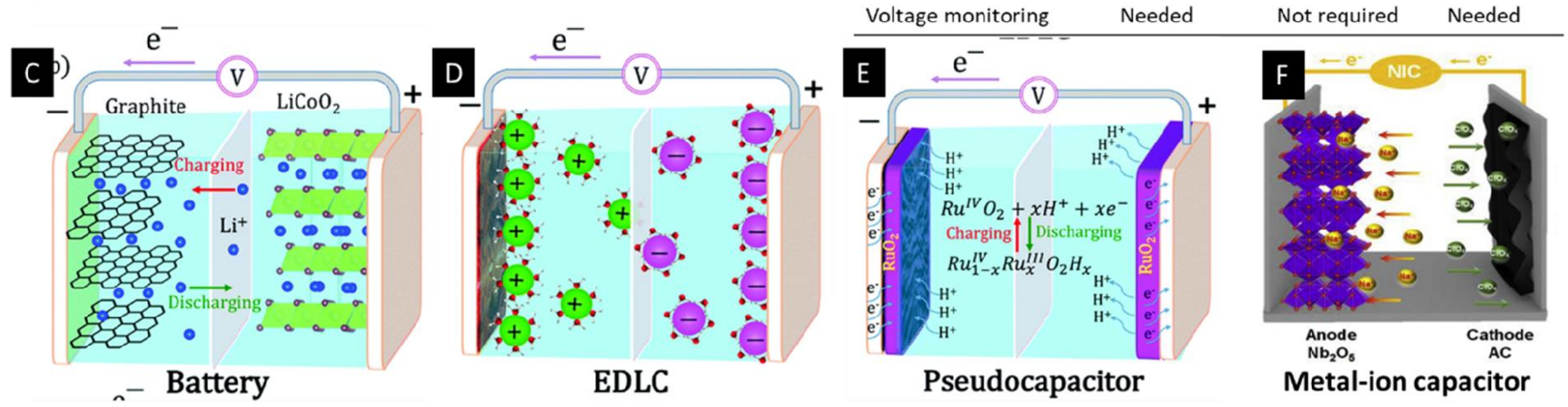

*Figure 1: (A) Ragone plot and (B) comparison table of electrochemical performances of battery, supercapacitor and metal-ion capacitor. Data of the comparison table is taken from Ref. [2]. Schematic of charge-storage for (a) battery, (D) electric double-layer capacitor or EDLC, (E) Pseudocapacitor and (F) Metal-ion capacitor. Figure (C-E) is reproduced from Ref.[6], © 2014 with permission from Royal Society of Chemistry. Figure 1(F) is reproduced from Ref.[7], © 2018, with permission from Elsevier)*

Nanocarbons (NC) are defined as carbon materials with at least one dimension in nanoscale such as carbon quantum dots, graphene, carbon nanotubes, carbon fibers, porous carbon, activated carbons. NC are well-appreciated as an energy storage electrode due to their huge surface area, excellent electrical/thermal conductivity, thermal/electrochemical stability, abundancy, easily scalable synthesis process and ability to grow heterostructure. However, limited charge-storage capacity/capacitance, chemical inertness, hydrophobicity, etc. are the major shortcomings of the NCs[8] On the other hand, metal-based structures (metal oxides, metal nitrides, metal chalcogenides etc.) are attractive to provide the higher storage capacity, however, poor electrical conductivity and poor electrochemical stability are their drawbacks. Therefore, fusing metal-based materials into the carbon matrix is considered as a promising solution to obtain the synergistic effect and to improve the electrochemical performances.[9][10] Despite higher reversible capacity than the commercially used graphite electrode, the severe structural damages of transition metal oxides-based electrodes during the $Li^+$-ion storage (for example due to the $Li_2O$ formation and volume expansion) are quite common. Hence, structuring the materials at the nanoscale could be a promising solution to enhance the reversible capacity and rate performances by shortening the pathways of electrolyte ions. Thus, among the possible strategies, there are significantly increasing trends in NPs incorporation into the NC or vice-versa, termed as 'NPs/NCs composite' throughout this review, to design an electrode for electrochemical energy storage applications.[2][11]

***Scope of the review***

NPs/NC composite were extensively explored as promising materials for battery anode,[12–15] battery cathode,[16–19] metal-ion capacitors anode,[20–23] and supercapacitors electrodes [24–27]. The present review concentrates on the NPs/NC-based anode materials for battery and metal-ion capacitors. In the case of supercapacitor, the electrode materials can be used in both symmetric and asymmetric configurations which is discussed in details. While NPs/NC composite are emerges as potential electrode materials for the electrochemical energy storage devices,[3-5] there are limited number of reviews available with this specific focus.[11] Till now, the available reviews in the literature are on metal/metal oxide decorated graphene for supercapacitors[11] metal/metal oxide NPs composited with porous carbon for supercapacitors[28] and partially discussions on metal oxide/graphene composite anode for $Na^+$-ion battery[29], graphene-NPs for supercapacitors and $Li^+$-ion battery[30] etc. Several missing areas where the present review provides a reasonably ample coverage and *in*-depth discussions are:

- The role of metal, metal oxides and other metal-based (metal nitride, carbide, chalcogenide, phosphide, etc.) NPs on the energy storage performance of NPs/NC composites;
- The role of NPs with different shape, size and morphology on the multifunctional NC. The morphology of NPs includes solid, hollow, porous, core-shell, yolk-shell etc.

- The need for multicomponent composite, metal *vs* metal-based NPs, dual NCs coating, etc.;
- Other important and NPs-specific points on the 'choice of NC' for the NPs, 'reason of increased capacity with cycle life' of the composites, 'synthesis process dependency', 'Origin of the higher-than-theoretical capacitance', 'clarification on Ni- and Co-like composites' for supercapacitors, etc.;
- A critical analysis on challenges and future opportunities for the NPs decorated NC as a viable energy storage electrode material.

***Organization of the review***

Based on the aim of this review defined above, the review is constituted of four major sections. Section 2 will brief the basic principles of three main electrochemical energy devices. Section 3 will introduce the methodology of NPs incorporation into the NC matrix with specific examples. The synthesis process of the preparation of composite has been listed in table 1 to table 5. Readers are advised to follow the cited references and/or existing reviews for the detailed synthesis methodology of NPs[31][32] and NC.[33]The section 4, 5 and 6 is categorized into the battery, supercapacitors and metal-ion capacitors, respectively. Each part discusses the effect of NPs size and its distribution, surface area, mass loading of NPs, hollow *vs* solid NPs, NPs on doped NC, multicomponent composite, and the choice of NC. This section uses a few examples to demonstrate the impact of the composite. Wherever appropriate, a correlation or link between the physico-chemical changes and the electrochemical performance of electrode materials is established. Furthermore, the challenges, guidelines, and future directions for the NPs/NCs composite as anode materials are highlighted in the final section.

While summarizing the electrochemical performances for comparison sake, half-cell test results are taken from the literature (Table 2-5). The experimentally reported testing conditions and half-cell test results in the research laboratory makes a huge gap with the real device used in commercial practice, which need a significant attention.[34]. Moreover, in order to show the electrodes suitable for desired energy storage device, presenting volumetric and areal matrices for specific capacity (or capacitance), energy density and power density are also preferrable along with their gravimetric counterpart.[35] Mass loading of the active materials is another important parameters for the electrochemical energy storage performances. To show the potentiality of the electrode materials in commercial application, the mass loading of active materials should be at least 10 mg/cm$^2$ and the specific capacity (or capacitance) measurement has to be carried out at the minimum current density of 1 A/g and scan rate of 10 mV/s.

The present review aims to serve as a one-stop reference on metal-based NPs/multi-functional NC composite for next-generation clean energy applications and will be of interest to the scientific and industry community of applied materials and electrochemical energy storage.

## 2. Basic principle of Electrochemical energy storage devices:

The electrochemical energy storage devices discussed here mainly metal-ion battery, supercapacitor, and metal-ion capacitor. The naming of those devices is based on the way they store the charge. A tabulated summary of the basic difference between these three energy storage devices is highlighted in Figure 1B. We encourage readers to follow the popular articles for *in*-depth knowledge over the subject.[6,36–38] Here, we just outline the principles of each energy storage devices.

Secondary rechargeable batteries are most used energy storage device in our day to day life usage like mobile, laptop. They stores the charge either in the form of chemical energy inside the electrode materials via chemical bonds or converting the chemical energy to electrical through Faradaic redox reaction or intercalation (Figure 1C).[6] There are four types of reactions such as (i) intercalation/deintercalations, (ii) alloying/dealloying, (iii) conversion and (iv) alloying and conversion type.[37] Presently, research focuses on the development of novel and more efficient electrode materials for $Li^+$-ion battery, however alternative battery chemistries like Li-sulfur, $Na^+$-ion, $K^+$-ion, $Al^{3+}$-ion, $Mg^{2+}$-ion, $Zn^{2+}$-ion are also progressing rapidly as potential contenders to conventional $Li^+$-based technology. In addition, investigation into other battery technologies like metal-air batteries[39], redox-flow batteries[40], aqueous batteries[41], and all-solid-state batteries[42] is proceeding.

Supercapacitor is another class of electrochemical energy storage device which can store much higher charge than the conventional capacitor and deliver the charge rapidly for prolonged cycle-life.[6] Based on the charge-storage mechanism, which is purely physical, there are two types of supercapacitors: electric double-layer capacitor (Figure 1D) and pseudocapacitor (Figure 1E). The electric double layer charge storage relies on the electric double layer formation at the electrode/electrolyte interface. Carbon-based materials exemplify this behaviour. It should also be noted that graphite and hard carbons undergo intercalation process. For the pseudocapacitors, the charge-storage mechanism is based on the adsorbed monolayer formation of metal ions or protons on the electrode surface, rapid surface/near-surface redox reactions, and fast ionic intercalation into the near-surface atomic layers without phase transformation in the electrode materials.[43] Metal-based materials and conducting polymers are falling into this category.

To bridge the gap between batteries and supercapacitor, metal-ion capacitors (also known as metal-ion hybrid supercapacitor) are becoming competitive (Figure 1F).[38,44,45] The metal-ion capacitors is assembled by sandwiching a battery anode and a carbon-based structure as a cathode, separated by a metal salt-containing organic electrolyte. The charge-storage mechanism in metal-ion capacitors, e.g. for CNF@$CoNi_2S_4$//AC $Li^+$-ion capacitors [20], is as follows: $Li^+$ ions from the electrolyte (1 M $LiPF_6$ in the mixture of ethylene carbonate, EC, and diethyl carbonate, DEC) are intercalated into the CNF@$CoNi_2S_4$ nanocomposite during the charging, whereas $PF^{6-}$ ions are adsorbed on activated carbons.

## 3. Strategies of NP decoration into nanocarbons

The decoration of NPs on the NC can be done in two most generic ways as follows (Figure 2A): *ex-situ* growth and *in-situ* growth. Figures 2B and 2C illustrate typical examples of *in-situ* and *ex-situ* processes for NPs/NC composite preparation, respectively.

### 3.1. *in-situ* process

Herein, the NPs nucleate and grow directly on NC employing thermal, physical or chemical reactions (Figure 2B). Prior to the growth, the precursor of carbons and NPs are mixed together and the final composite are obtained. The *in-situ* growth techniques result in uniform distribution of NPs on NC, due to a controlled growth process. A variety of the growth methods, such as chemical reduction, sol-gel, hydrothermal, thermal, physical or chemical vapour deposition methods, electrochemical reduction or deposition methods etc. are categorized under the *in-situ* crystallization process.[46–49] In addition, the *in-situ* crystallization or growth of NPs on NC and their derivatives is further categorized as a chemical route and a physical route based on the process involved and the precursors used.

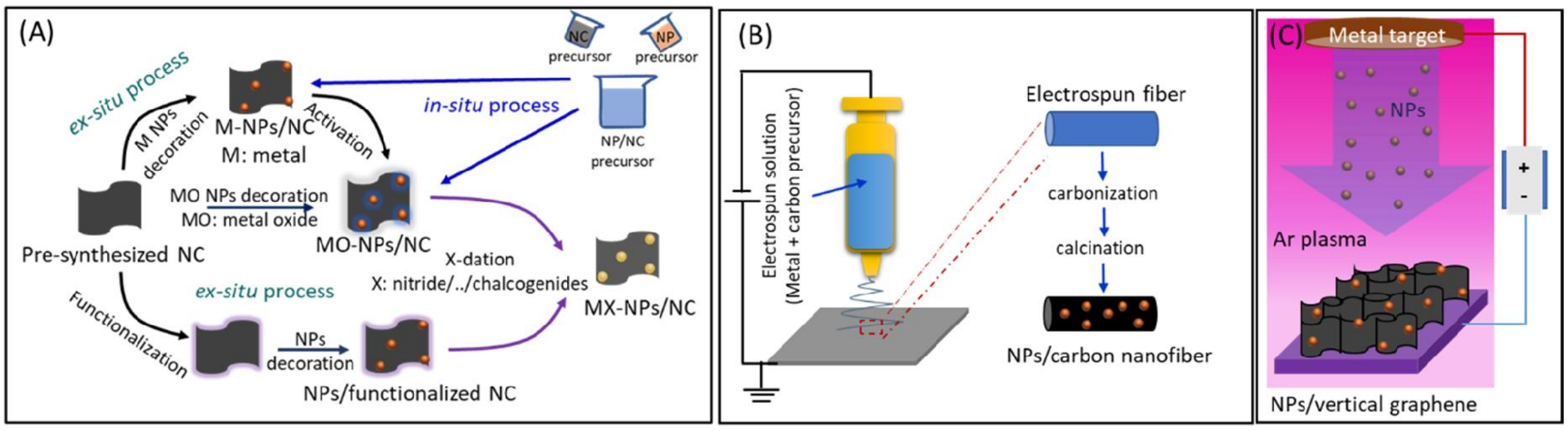

*Figure 2: (A) Schematic of the strategy of NPs/NC composite preparation methodologies; (B) typical example of in-situ process of NPs/carbon nanofiber composite preparation and (C) typical example of ex-situ process of NPs decoration by a low-temperature DC plasma on vertical graphene.*

### 3.2. *ex-situ* process

In the *ex-situ* process, NPs are decorated on pre-synthesized or commercially available NC to grow the hybrid structures or composites. NC serves as a mechanical platform here on which NPs are decorated.[50–53] Often, before mixing, the surface functionalization of NPs and/or NC was adopted to bind them. This process involves π-π stacking to bind the NPs to NC or vice-versa, through chemical bonding or in a non-covalent manner. For example, the Au, CdS NPs functionalized with 2-mercaptopyridine and benzyl mercaptan, respectively, were successfully attached to graphene oxides (GO) or reduced graphene oxide (RGO).[54] Alternatively, RGO/$TiO_2$ composites were prepared by binding $TiO_2$ NPs to Nafion-coated RGO.[55] Furthermore, the RGO coated with bovine serum albumin protein, serves as a universal adhesive layer for Au, Ag, Pt and Pd NPs.[56] In the *ex-situ* growth, the geometry and structure of NCs almost remain the same even after the NPs decorations.

### 3.3. Synthesis process dependent electrode materials

There are several strategies to decorate NPs on the NC platform or to synthesize the NPs/NC composite. Figure 3A summarizes the progress of the electrode materials based on the NPs/NC composite. In Table 1, the various morphology of NPs/NC composite prepared by different methodology is highlighted. Table 1 also consists of synthesis process of NPs and NC separately. Furthermore, the synthesis techniques for various NPs/NC composite are also listed in Table 2, 3, 4 and 5 along with their electrochemical performances. It is important to choose the proper synthesis process to obtain high-performance electrode materials. As the process parameters of the synthesis, such as the precursor ratio, synthesis temperature, and type of precursor, can be varied, the final structures with tunable morphology and structural properties can be obtained, which directly effects the electrochemical performance (Figure 3B-G). The synthesis process does decide the bonding between the NPs and NC too which is essential for the structural stability, electrochemical performances, and electrochemical stability of the electrode. In this section, we are highlighting few case studies to give the glimpse of such facts.

*Table 1: Methodology of NPs, NCs and composite preparation with process category and final morphology.*

| **Synthesis techniques** | | | **Process category** | **Final morphology and remarks** |
|---|---|---|---|---|
| **Nanocarbons (NC)** | **Nanoparticles (NPs)** | **Nanocomposites (Ns/NC)** | | |
| - | Bare NiO microspheres: Spray pyrolysis of Ni nitrate hexahydrate | NiO/C microspheres: One-pot spray pyrolysis using Ni nitrate hexahydrate, polyvinylpyrrolidone, polystyrene nanobeads + heat treatment | *in-situ* | Coral-like yolk-shell structures [12] |
| GO: Modified Hummers method; B-doped RGO: thermal treatment of a mixture of GO and boric acid | - | Ag-NPs embedded B-RGO: Reduction of $AgNO_3$ in the presence of Tollens' reagent | *ex-situ* | Spherical Ag NPs (10-15 nm) decorated Few-layer stacked sheets of rGO [13] |
| - | $Co_3O_4$ : Thermal oxidation of chemically bonded Co NPs with porous carbon nanosheets (Co@PCNS) | chemically bonded Co NPs with porous carbon nanosheets (Co@PCNS): NaCl template method starting from glucose and Co nitrate precursors + carbothermal | *in-situ* | PCNS with a 2D architecture, very low degree of agglomeration, and uniformly decorated with Co NPs (25±5 nm) [15] |
| - | | Physically bonded Co-PCNS: Removal of Co NPs from Co@PCNS by HCl treatment followed by mixing with commercial Co NPs (20-50 nm) | *ex-situ* | Nanoporosity and 2D microstructure but lack of Co-C bonds [15] |
| Carbon nanofibers (CNF): Electrospinning of polyacrylonitrile solution + carbonization | - | CNF@$CoNi_2S_4$ NPs: Electrodeposition of CoNi-sulfide on CNF surface | *ex-situ* | NPs with the diameter of 10-15 nm on the surface of CNFs [20] |
| GO: Modified Hummers method; RGO aerogel : Freeze-drying of GO by Hummers method + thermal treatment under Ar | - | $SnO_2$-RGO : Freeze/freeze-drying of GO suspension in the presence of $SnSO_4$ at different ratios | *in-situ* | Nanosized NPs (< 10 nm) homogeneously distributed on RGO sheets. Macroporous structure [21] |
| GO: Modified Hummers method | $Fe_3O_4$ NPs: Hydro-thermal method | $Fe_3O_4$/RGO: Hydrothermal method with $Fe^{3+}$:$Fe^{2+}$= 2:1 molar solution in the presence of pre-synthesized GO solution | *in-situ* | NPs (≈15 nm) attached to the RGO nanosheets [24] |
| GO: Modified Hummers method; RGO: Reduction of GO by $NaBH_4$ | $SnS_2$ nano-sheets: Hydro-thermal method | $SnS_2$/RGO: Hydro-thermal method | *in-situ* | Hexagonal $SnS_2$ NPs (145-155 nm) embedded in RGO sheets [25] |
| N-doped carbon: Hydrothermal treatment + post-annealing | - | Birnessite $MnO_2$ embedded with N-doped carbon: Hydrothermal treatment of egg albumin and $KMnO_4$ + post-annealing | *in-situ* | Spherical NPs (15-20 nm) uniformly embedded into hierarchically porous structure [49] |
| Vertical graphene: plasma enhanced chemical vapor deposition | - | Au NPs - VG: Drop-casting of commercial Au NPs colloidal solution over VG + infra-red heating | *ex-situ* | Uniform dispersion of Au NPs on vertical graphenes [52] |
| Graphene nanosheets (GNS): Reduction from GO by $NaBH_4$ | Benzene-anchored CdS quantum dots: Colloidal synthesis | CdS functionalized GNS: Mixture of GNS in dimethylformamide with quantum dots suspension | *ex-situ* | Homogeneous distribution of quantum dots (≈ 3 nm) on GNS. Non-covalent strong interactions between CdS and GNS [54] |
| GO: Modified Hummers method RGO: Reduction of GO by bovine serum albumin (BSA) | Au NPs, Pt NPs, Pd NPs, Ag NPs: Colloidal synthesis | Assemblies of NPs-BSA-RGO: Mixing of BSA-rGO and excess of NPs | *ex-situ* | Uniform distribution of NPs on GO/RGO [56] |
| - | $SiO_2$ nano-spheres: Stöber method | Double carbon shells coated Si NPs: Chemical vapor deposition + magnesio-thermic reduction of $SiO_2$ NPs to Si | *in-situ* | Si NPs (≈20 nm) are well-dispersed in the inner carbon shell [57] |
| | | Carbon-coated Si NPs: Polymerization of dopamine-hydrochloride on commercial Si particles | *ex-situ* | Single carbon layer on Si NPs [57] |
| GO: Modified Hummers method | - | ZnO NPs embedded holey RGO (ZnO/H-rGO): Freeze-drying of GO in $Zn(NO_3)_2$ solution + calcination under Ar at 700°C (etching process) | *in-situ* | ZnO NPs (20 nm) uniformly dispersed on rGO surface, with individual ZnO NP located at one etched hole [58] |
| | | ZnO NPs embedded reduced graphene oxide (ZnO/rGO): Freeze-drying in $Zn(NO_3)_2$ solution + calcination under Ar at 600°C | *in-situ* | ZnO NPs uniformly dispersed on rGO surface but absence of well-defined holes [58] |
| Functionalized CNTs: Dispersion of homemade CNTs in $HNO_3$ | Pure $Co_3O_4$ NPs: Chemical co-precipitation $Co^{2+/3+}$ ions in alkaline solution | $Co_3O_4$/CNT: Chemical co-precipitation of $Co^{2+/3+}$ ions in alkaline solution in the presence of CNTs | *ex-situ* | NPs (15-30 nm) homogeneously coated on the surface of CNTs [59] |
| - | T-$Nb_2O_5$: Solvothermal synthesis using ethanol solution of $NbCl_5$ + calcination | T-$Nb_2O_5$/ graphene: Solvothermal synthesis using an ethanol solution of GO and $NbCl_5$ + calcination | *in-situ* | Agglomerated particles (129 nm) [60] |
| | | T-$Nb_2O_5$/N-doped graphene: Solvothermal method using ethanol solution of GO, urea, $NbCl_5$ + calcination | *in-situ* | NPs (average 17 nm) uniformly anchored on N-doped RGO surface. No agglomeration. [60] |

### *3.3.1. Comparison of synthesis process*

Comparing the synthesis methods is important to choose the most effective one. For instance, chemical and γ-radiation methods are used to prepare RGO-Au NPs based nanocomposite. In both the cases, GO (25 mg/ml) and chloroauric acid were taken as main precursors with different additional solvent in each technique. The charge-storage performance of RGO-Au NPs based supercapacitor electrode was found to rely on the specific technique used. For example, the gravimetric capacitance of RGO-Au NPs nanocomposite prepared by the chemical and γ-radiation methods were 100 and 500 F/g, respectively, in 6M KOH electrolyte, whereas it was only 50 F/g for bare RGO. The variation in the charge-storage capacitance was attributed to the presence of defects and the number of layers of the as-prepared graphene structures.[61] It should be noted that the intrinsic properties and surface properties of nanocomposites may also depends on the type of solvents, chloroauric acid concentrations, reduction time etc. used during the synthesis.

### *3.3.2. Physically-bonded vs chemically-bonded*

Two types of Co NPs/porous carbon nanosheets have been prepared. In one case, NaCl template was mixed in a homogeneous solution of glucose and $Co(NO_3)_2$ followed by carbothermal reactions, which resulted in chemically bonded 25 nm Co NPs on 100 nm amorphous porous carbon nanosheets. In the second method, Co NPs was removed from the chemically bonded composite by acid treatment and the resulting porous carbon nanosheets were physically mixied with commercially available Co NPs (Table 1). The $Li^+$-ion storage performance of chemically bonded composites in terms of specific capacity, rate capability and cyclic stability was found to be much better than that of physically bonded composite (see Figure 9B-D).[15] The better performance of Co NPs bonded amorphous porous carbon nanosheets was attributed to the morphology, limited $Li_2O$ formation during the electrochemical process, and Co-C bonds which facilitate the effective charge transfer between the NPs and NC.

### *3.3.3. Precursor dependency*

The type of precursors used during the synthesis also plays a key role in the final structure of the nanocomposite, while the morphology remains similar. The variations in the final nanocomposites can be observed in terms of surface area, size of NPs, conducting properties etc which has significant impact on the electrochemical performances. For example, the lower $Li^+$-ion storage performance of the Co@ porous carbon nanosheets, obtained using Co-acetate as a precursor in contrast to the Co-nitrate, was attributed to the reduced porosity of the composite, which limits the contact with electrolyte solution and the $Li^+$-ion diffusion (Figure 3E).[15] In similar manner, other synthesis parameters like synthesis temperature, and deposition time are also the determining factors for the final structure, morphology and electrochemical performances (Figure 3F-G). The storge mechanism for this composite is a combination of capacitive and intercalation mechanism of NC and conversion mechanism of NPs. Thus, a balance between the contribution from each of them is crucial to obtain high Coulombic efficiency, rate performance, and capacity retention after prolonged cycles from an energy storage device which can be tailored by tuning the synthesis parameters.

### *3.3.4. Ratio of NPs and NCs content*

Another factor that controls the charge-storage performance of the electrode is the ratio of NPs and NC content. The ratio can be tailored by controlling the weight ratio of precursors. For example, different concentration of $KMnO_4$ (0.1 wt%, 0.2 wt% and 0.3 wt%) was taken with 100 mg of commercial CNTs to synthesize CNTs@$Mn_3O_4$ hybrid materials.[62] As can be seen from the Figure 3(H-J), there are changes in NPs size, density of NPs, and weight ratio of NPs and NCs. These changes are responsible for the change in the surface area, pore distribution, electrical and thermal conductivity, degree of graphitization etc. which play a decisive role in electrochemical stability and rate performance of the final electrode material. For instance, poor rate performance and poor electrochemical stability have been seen from the CNTs@$Mn_3O_4$ prepared with 0.3 wt% $KMnO_4$.[62] Thus, it is crucial to choose a proper ratio to obtain high-storage performances from electrode materials.

#### 3.3.5. *Layering vs mixing*

However, all innovative strategies may not be useful during the operation of the electrode materials. For example, $Fe_3O_4$ NPs were decorated on N-doped RGO surface by two individual approaches, namely, mixing and layering, but maintaining the ratio of N-doped RGO and $Fe_3O_4$ NPs constant. With the same mass loading (2 mg/cm$^2$), the composite that was synthesized by the layering method (201 mF/cm$^2$ at 2 mV/s in 0.5 M $Na_2SO_4$ *vs.* Ag/AgCl) was found to outperform as supercapacitor electrode compared to the composite prepared by mixing (166 mF/cm$^2$ at 2 mV/s).[63] In the layering approach, the $Fe_3O_4$ NPs were directly exposed to the electrolyte, whereas not all the NPs were available for the surface redox reaction in the case of electrode prepared by the mixing method. This result indicates that exposing the NPs to the electrolyte is much preferable to improve the storage capacity rather than simply making a composite where some of the NPs are not exposed to the electrolyte. This is not true for battery electrodes, since a supercapacitor relies on surface phenomena different from the diffusion/intercalation-based mechanisms of a battery.

#### 3.3.6. *Planar vs 3D NCs structure for NPs decoration*

Three dimensional structures are always preferential for the energy-storage applications due to the high surface-to-volume ratio. For the *ex-situ* method, the vertically oriented NC are preferential to decorate the NPs in well-controlled manner in contrast to their planar counterpart.[50–52] For instance, planar nanographitic structures and vertical graphenes were prepared by the plasma enhanced chemical vapour deposition. Thereafter, Au-precursor solutions of same concentration were drop-casted.[52] One can easily visualize better NPs distributions with higher loading on the vertical graphenes compared to the planar nanographitic structures (Figure 3K-L). It is important to note that, both surfaces of graphene in vertical graphene can be used for NPs decoration. As a result, Au NPs/vertical graphene exhibited infra-red emission (Figure 3M). This result reveals that one can get exciting properties from the composite where NPs are decorated onto three-dimensional NC.

It is well-known that each synthesis technique results in a nanocomposite with unique morphology and structure. The materials prepared at different conditions by various methodologies are characterized by a variation in porosity, size and shape of NPs, loading of NPs, ratio between the content of NPs and NC, electrical conductivity, thermal conductivity, wettability. These changes in structural properties also have significant impact on the final electrochemical performance of the electrode materials, which is discussed in the following sections.

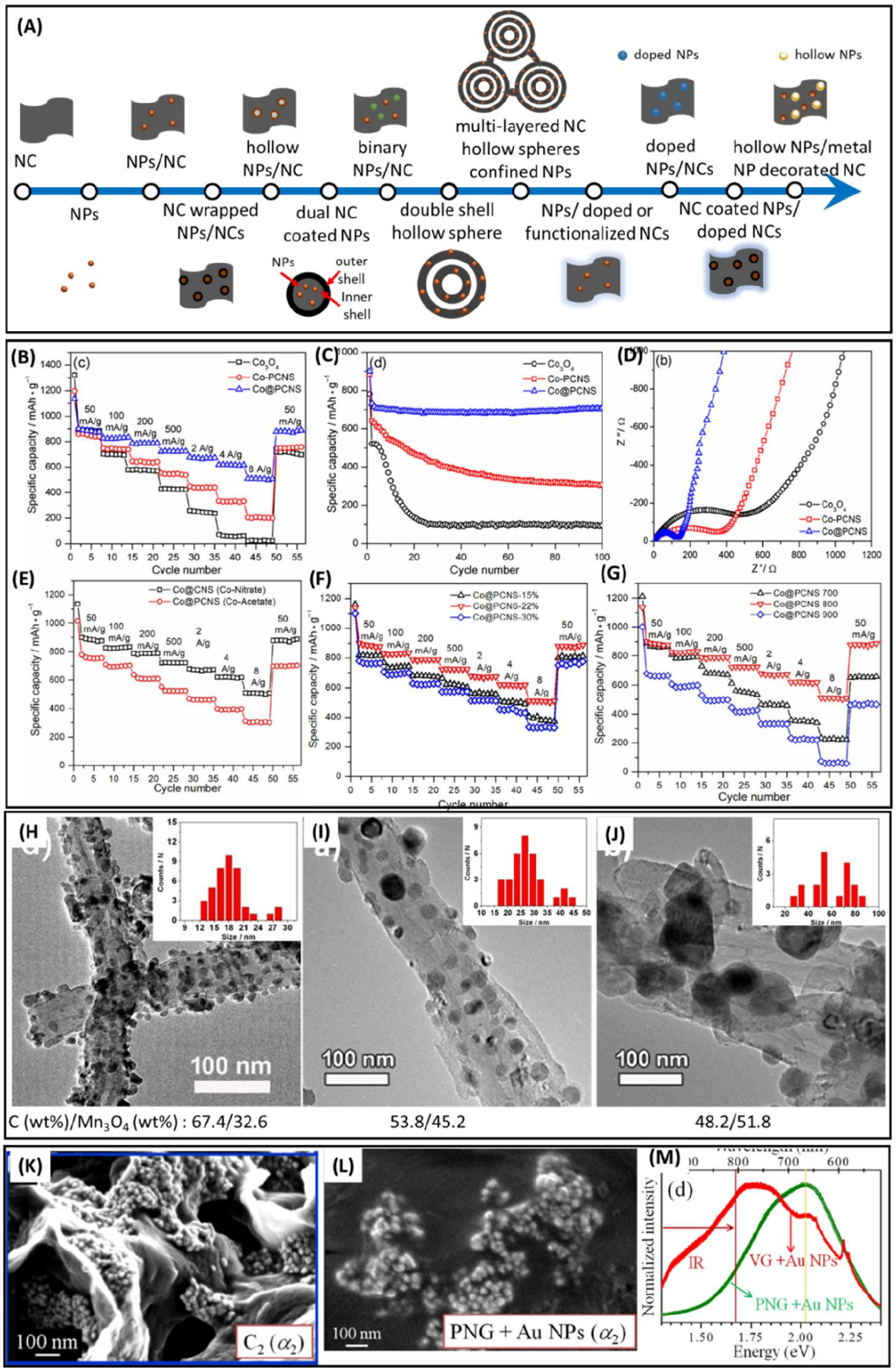

*Figure 3: (A) Evolution of NPs/NC composite. (B-G) Half-cell $Li^{+}$-storage performance of $Co_3O_4$ NPs, physically mixed Co NPs-PCNS and chemically bonded Co NPs@PCNS prepared with different synthesis parameters; PCNS: Porous Carbon Nanosheets. Reproduced from Ref. [15], © 2018, with permission from American Chemical Society. Transmission electron micrographs of CNTs/$Mn_3O_4$ nanocomposite prepared with different $KMnO_4$ concentration of (H) 0.1 wt%, (I) 0.2 wt% and (J) 0.3 wt%. Reproduced from Ref. [62] © 2018, with permission from Elsevier. Au NPs decorated on (K) vertical graphene and (L) planar nanographitic structures and (M) their photoluminescence spectra. Reproduced from Ref. [52] © 2017, with permission from IOP publishing Ltd.*

### 4. Battery anode

The anode materials for batteries should (i) sustain the fast Faradaic reactions for the higher capacity, better power density while a good balance with cathode materials should be ensured, (ii) have the low working potential for high voltage and higher energy density, and (iii) have the least possible volume expansion for better cycle life.

### 4.1. Li-ion Batteries (LIBs)

#### 4.1.1. *Size of NPs*

The size of NPs greatly matters in the charge-storage performance. It has been reported that $SnO_2$ NPs with size < 10 nm in $SnO_2$/RGO have shorter Li-ion diffusion pathways which lead to the higher rate performance.[21] Smaller NPs can also be decorated in a NC matrix with higher mass loading, which can provide plenty of electrochemically active sites to accommodate the stress produced by the volume expansion of anode materials during the charging/discharging cycles.[21] In other words, larger NPs have a strong tendency to agglomerate and NC lose their ability to prevent the agglomeration. As a result, poor rate performance and cycle stability of the device are obvious. Briefly, the Li-ion storage performance of $Fe_2O_3$ NPs decorated nanomesh graphene with different NPs sizes indicates that the smaller size of NPs decoration has impressive charge-storage performance (Figure 4A-C).[64] Moreover, smaller NPs also reduce the chances of NPs crack and fracture.[57] The size of NPs above which crack and fracture can happen is known as critical size. The critical size may vary with the composition and other parameters of the NPs. For example, the critical size for the Si NPs is 150 nm.[65]

#### *4.1.2. Porosity and surface area*

Other factors affecting the charge-storage performances of anode materials are the porosity and surface area. It has been observed that Sn NPs decorated porous carbon nanofiber (PCNF) outperforms as anode material for Li-storage compared to Sn NPs/CNF due to the higher surface area and pore volume in Sn/PCNF.[46] Moreover, no agglomeration of Si NPs on PCNF has been evidenced. Indeed, NPs were found to be dispersed quite homogeneously on PCNF in contrast to the CNF.[66][67] In turn, Si/PCNF exhibited higher capacity retention after 100 cycles (58.5%*) than Si/CNF (11.1%*) (asterisk represents the data estimated either from the plot or available data from the cited reference).[66] The porous conducting NC was also credited for the higher Coulombic efficiency of Si/PCNF in the first charge/discharge cycle for $Li^+$-ion storage (71.7%) compared to that of Si NP (57.2%).[66] It has also been reported that the structures with micropores are highly undesirable here due to the large mass-transfer resistance.[68] Thus, designing carbon-coated mesoporous Si NPs was found to be the promising solution to deliver a very high capacity of 2482 mAh/g at 0.2 A/g and to keep the material stable after the electrochemical process.[68] Importantly, carbon-coated NPs are doped with nitrogen, which also has a significant impact on the charge-storage performance as discussed in the next section.

Basically, mesoporous structures, high surface area and smaller NPs are beneficial for fast ionic diffusion.[69] Since the NiO nanocrystal was surrounded by carbon structures, the core of yolk-shell structures contains interconnected mesopores and plenty of hollow space between the yolk and the shell is available, the coral-like yolk-shell NiO/C microspheres exhibited excellent cycle stability and rate performance (Figure 4D-F).[12] Moreover, to provide the space for volume change for metal-based NPs, a large number of holes on the graphene surface was created.[58] In this composite (ZnO NPs embedded holey RGO), ZnO NPs not only contributed to charge-storage but were also used to etch the graphene surfaces to create the holes. On the contrary, a large irreversible capacity loss of Ge/RGO/C was attributed to the high surface of graphene.[70] This result suggests that NC with a very high surface area is not preferable but the space to accommodate the volume change of NPs is essential.

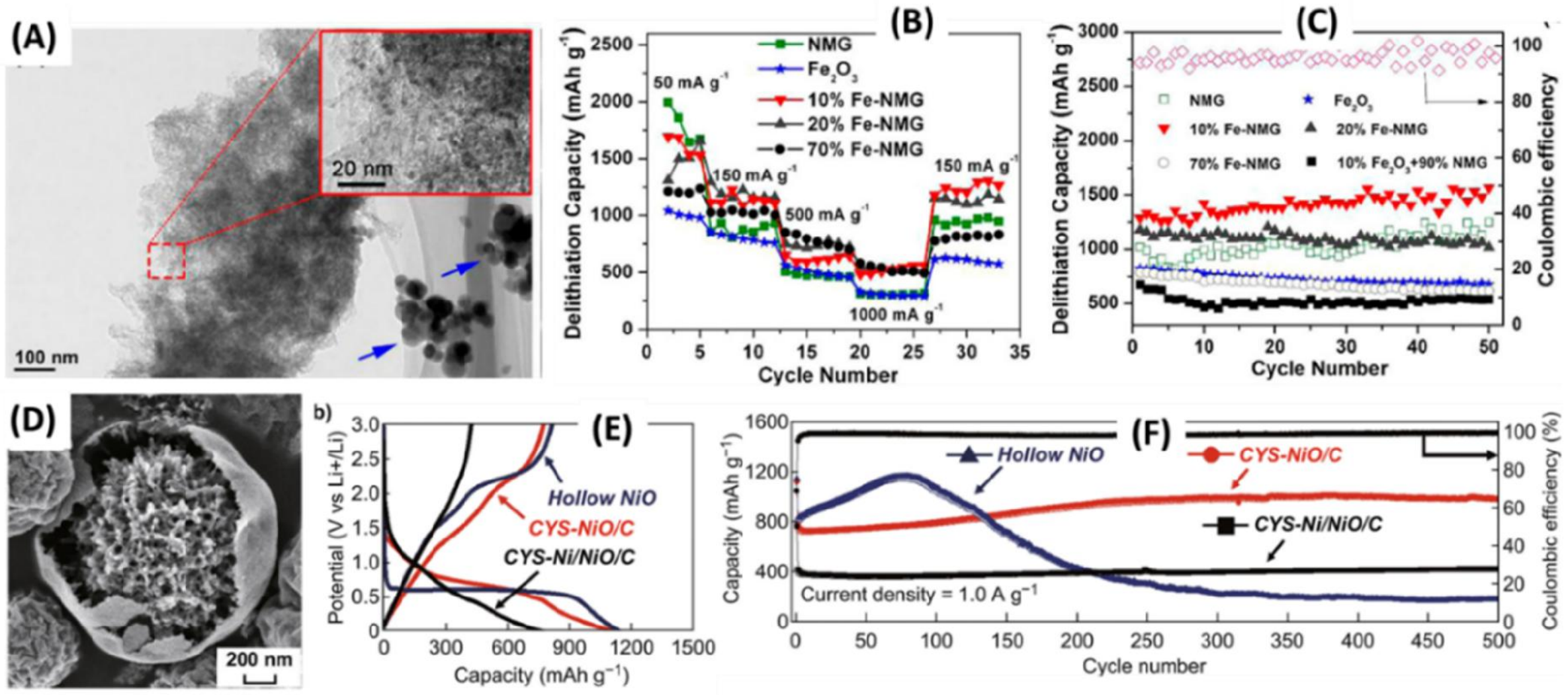

*Figure 4: (A) Transmission electron micrographs of 70% Fe-NMG (20-100 nm), (B-C) $Li^+$-storage performance of $Fe_2O_3$ NPs, NMG, $Fe_2O_3$ NPs/NMG. NMG: nanomesh graphene. (A-C) is reproduced from Ref. [64], © 2014 with permission from American Chemical Society. (D) Scanning electron micrograph, (E) charge-discharge profile and (G) capacity and Coulombic efficiency of coral-like yolk-shell NiO-C composite microsphere $Li^+$-storage. (D-G) is adopted from Ref. [12], © The author(s) 2019, Springer Publishers.*

#### *4.1.3. Loading of NPs on NCs*

Another important factor is the loading of NPs which depends on the surface area of NC and the size of NPs. The loading of NPs on NC should be performed in such a way that the high reversible capacity of the anode with the excellent Coulombic efficiency, rate capability and capacity retention are maintained.[71][72] It has been reported that 22% loading of Co NPs onto porous carbon nanosheets gave the best $Li^+$-storage performance.[15] Moreover, the $Li^+$-ion storage capacity was increased and better capacity retention was obtained for the composite of $Fe_2O_3$ NPs/carbon aerogel when the loading of Fe-content was increased from 14.2 to 32 wt%.[73] In other words, an increase in graphene content in the $Fe_3O_4$ NPs/RGO aerogel composite resulted in reduced Coulombic efficiency.[74] It is noteworthy that there should be an upper limit for NP loading beyond which the utilization efficiency of NPs in the composite is lower. A high reversible capacity of 731.5 mAh/g at 200 mA/g after 50 cycles was obtained from the $SnO_2/NiFe_2O_4$/RGO composite with 50 wt% $SnO_2$ content, beyond which (70 wt% $SnO_2$), an agglomerated structure and electrode pulverization during the charge/discharge process were observed. The agglomeration of NPs after several charge/discharge and higher mass loading of NPs (52 wt%)[66] could be responsible for the drastic discharge capacity loss of Si NPs/CNF from 1880 mAh/g to 172 mAh/g after 100 cycles. It has also been reported that the higher loading of $Mn_3O_4$ NPs in the composite lowered the electrical conductivity and the void spaces of the composite resulting in a major structural deterioration and poor electrochemical performance.[75] Thus, one has to balance the NPs loading and conducting carbon content to ensure the best charge-storage performance.

#### *4.1.4. Hollow NPs on NCs*

In addition to the design of porous NC, there is an upscaling interest in hollow NPs, [76–78] mesoporous structures,[59] yolk-shell structures[12] to be encapsulated into the NC matrix. It has been shown experimentally that the hollow NiO microsphere alone failed to hold the charge-storage capacity for prolonged cycle life, which confirms the necessity of NC incorporation (Figure 4F).[12] It can be seen that hollow $Co_3O_4$/P-RGO delivered much higher gravimetric capacity, rate performance, and better cycle life than $Co_3O_4$/P-RGO. Among the intrinsic properties, hollow $Co_3O_4$/P-RGO possesses a higher surface area and higher $I_D/I_G$ ratio, where $I_D/I_G$ is the intensity ratio of D and G peak.[79] D-band represents the presence of defects and disorder and G-band is the signature of $sp^2$ content in NC.[80][81] Moreover, structural defects in the composite also serve as micro-voids to accommodate excessive $Li^+$-ion.[68] More importantly, not only the void space of/within graphitic nanotubes but also the empty spaces of hollow metal oxide NPs are promising to provide sufficient space to deal with the volume expansion of NPs during charging/discharging.[77]

#### *4.1.5. NPs with dual NCs*

Although various strategies of NPs decorations onto NC have been implemented, a few major concerns exist in the preparation of the composite materials: (i) NPs on the NC surface are exposed directly to the electrolyte which leads to the NPs dissolution into the electrolyte, thick solid-electrolyte-interfacial layer formation, and hence low Coulombic efficiency.[83]; (ii) due to the different volume expansion coefficient of NC and NPs, there is a huge chance of peeling off the NPs from the graphene source after several charge/discharge cycles; (iii) possibility of crack formation during lithiation/delithiation as shown in Figure 5A.[57][82][84] Thus, the concept of dual NC incorporations is becoming more and more popular (Figure 5A-D).

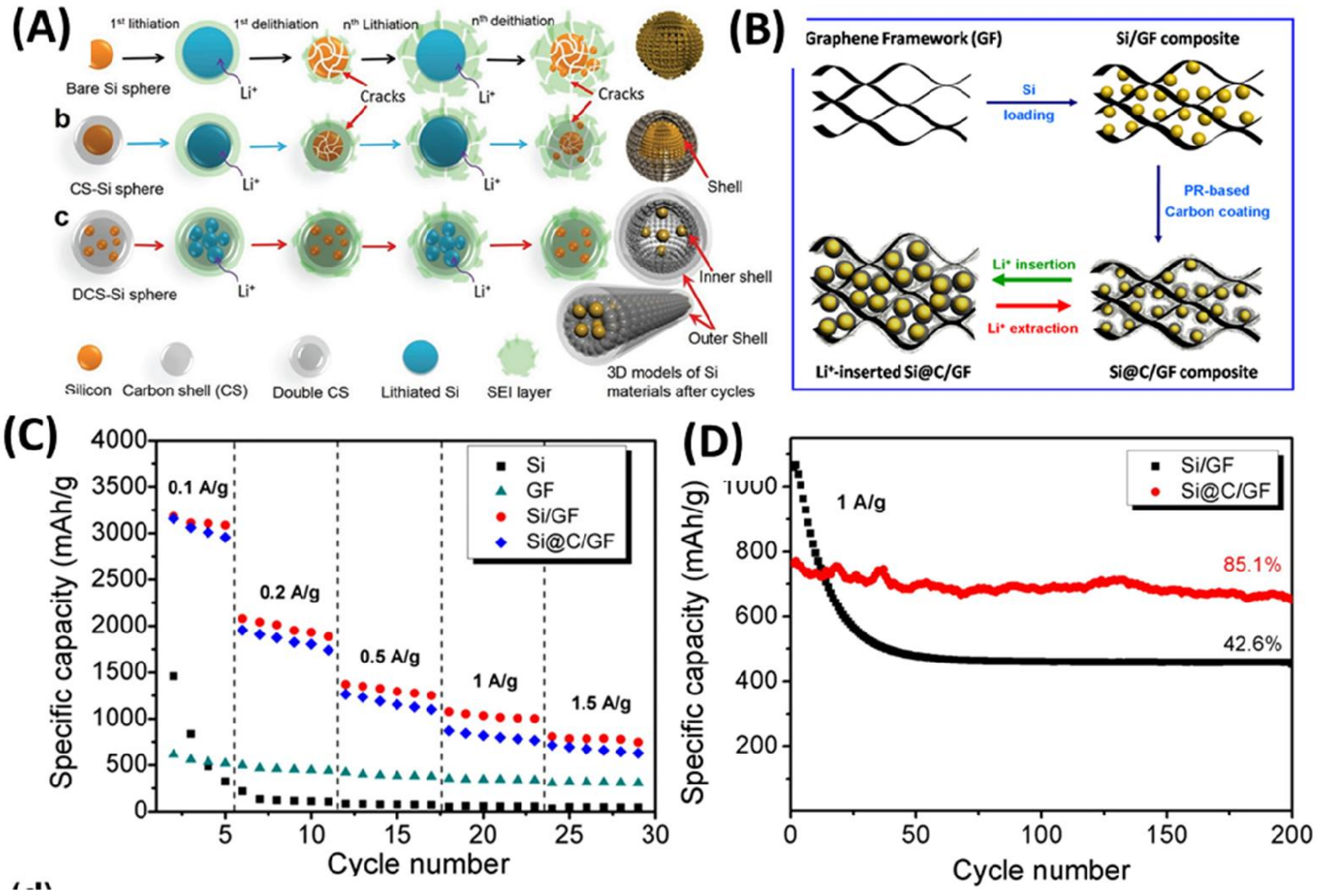

*Figure 5: (A) Schematic of morphological evolution of Si NPs and Si NPs/carbon shell after repeated lithiation/delithiation cycles. Reproduced from Ref. [57] © 2017 with permission from WILEY-VCH Verlag GmbH & Co. KGaA, Weinheim. (B) Schematic strategy of Si@C/graphene foam preparation with $Li^+$ insertion, (C) Charge capacity of Si, graphene foam (GF), Si/GF and Si@C/GF at different current density based on Si-weight; (D) cycle test of Si/GF and Si@C/GF. Reproduced from Ref. [82] © 2014, with permission from Elsevier Ltd.*

Importantly, each NC plays a distinct role. In the case of N-doped carbon-coated MnO NPs anchored on the interconnected graphene ribbons, N-doped carbon protects MnO NPs from the direct interaction with the electrolyte and interconnected graphene ribbons serves as a conducting and mechanical platform. In turn, NPs/NCs composite showed much better performance than interconnected graphene ribbons-MnO and MnO NPs, when they are employed in LIB applications.[86] For Si@C/RGO, where Si NPs were coated by thin amorphous carbon (*a-C*) layer and embedded in the 3D RGO network, the 3D network of RGO provides effective space for the volume expansion, whereas *a-C* reduces the volume changes, offers high reversible capacity and maintains good contact with Si and RGO framework.[82] As a result, the Si@C/RGO framework outperformed compared to Si, RGO, Si/RGO framework as a $Li^+$-storage anode material (Figure 6B-D). Moreover, both the carbon structures improved the cycle stability and preserved the structural integrity. In the dual-carbon shell coating of Si NPs, the inner carbon shell affords the space for the volume change of NPs whereas the outer shell facilitates the stable solid-electrolyte-interfacial layer formation and inter-shell spaces ease the mechanical stress from inner carbon shells and volume changes.[57] In the Si-CNT/RGO composite, carbon nanotubes (CNTs) connect the graphene layers to prevent the Si NPs agglomeration. As a result, this 3D composite outperformed compared to Si-RGO, Si-CNT and mechanically mixed RGO-CNT-Si.[84] It is also noteworthy that the surface area of the composite (M-$Nb_2O_5$@C/RGO) has been increased by two times after the dual carbon coating than M-$Nb_2O_5$@C[87], which may reflect on the higher charge-storage capacity. An important point to be addressed here is the thickness of carbon coating on NPs, which is the subject of further research.

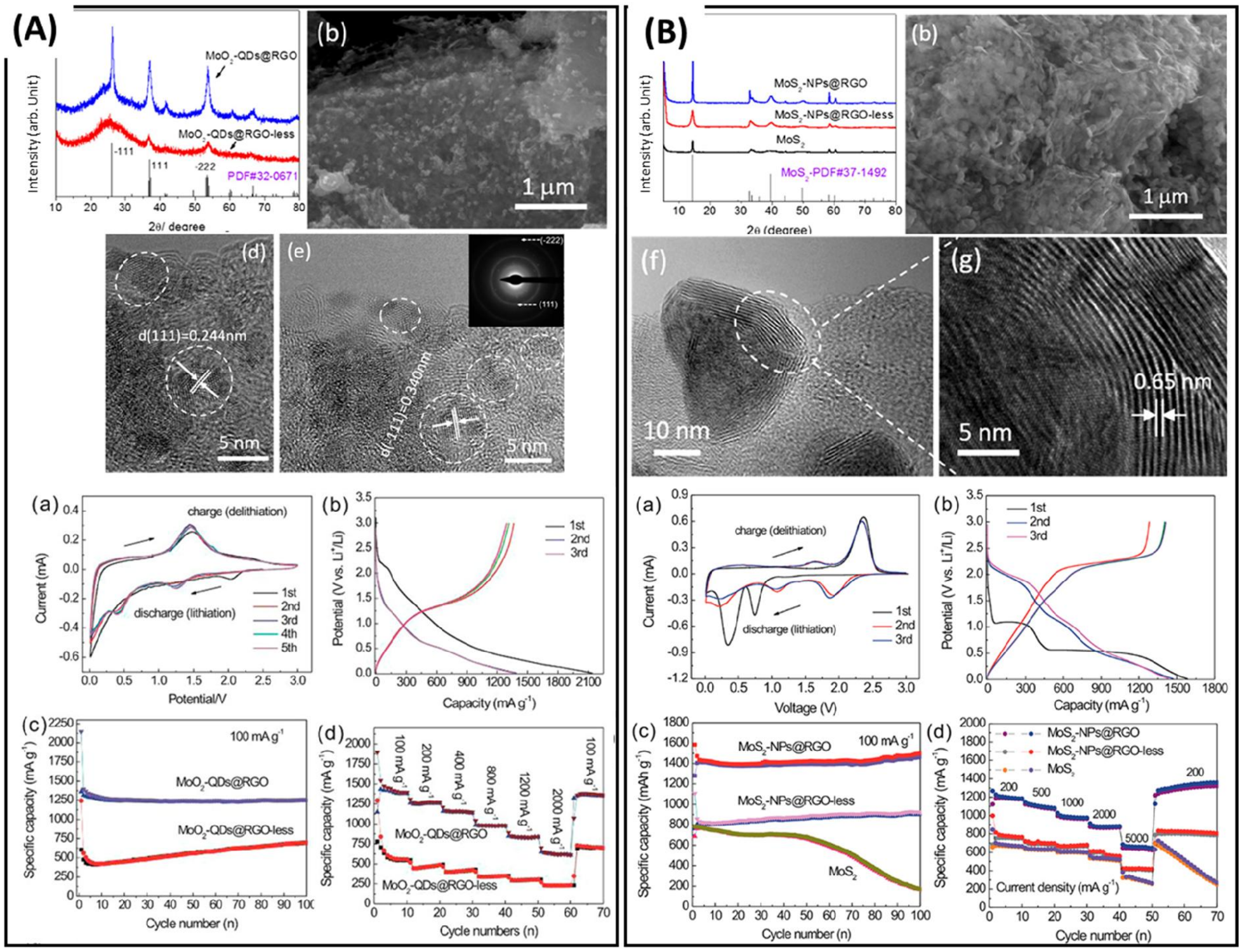

*Figure 6: X-ray diffraction pattern, high-resolution transmission electron micrographs and half-cell $Li^+$-storage performance of (A) $MoO_2$-based QD/RGO and (B) $MoS_2$-based NPs/RGO. Reproduced from Ref. [85]. © 2017, with permission from American Chemical Society. According to the convention, a. u. is actually stands for astronomical units. Hence, the unit of intensity is expressed as arb. unit by replacing a. u. from original cited references.*

### *4.1.6. Non-oxide NPs decorated NCs*

Attention on the carbide/sulphides/nitrides/phosphides/chalcogenides-based NPs decorated NC has also been paid.[20,85,88,89] As an example, a much higher $Li^+$-storage capacity of 497 mAh/g at 0.05 A/g is reported for CNF@$CoNi_2S_4$ than that of graphite and $Li_4Ti_5O_{12}$ .[20] The strong covalent and ionic bonds of metal oxides are responsible for the huge energy consumption during the $Li^+$-ion intercalation/deintercalation and limit the availability of free electrons and hence reduce the electrical conductivity. On the other hand, alloy-type materials (metal carbides, nitrides, and sulphides) with metallic bonds possess numerous free electrons and hence fast charge transfer kinetics in contrast to their oxide counterparts. As a result, $MoS_2$ NP/RGO (1497 mAh/g at 0.1 A/g after 100 cycles) has been found to deliver higher $Li^+$-storage capacity than the $MoO_2$ quantum dots/RGO (1269 mAh/g at 0.1 A/g after 100 cycles) (Figure 6).[85] It is important to note that $MoS_2$ NPs size was 20 - 60 nm which is larger than $MoO_2$ quantum dots(< 5 nm). While transforming oxide-based NPs/NC into non-oxide-based NPs/NC, the morphological, surface and intrinsic properties also change, which makes the $Li^+$-storage performance of the composite different from the parent composite.

#### *4.1.7. Choice of NCs*

The charge-storage performance does rely on the intrinsic properties of NC, like electrical conductivity, graphitic quality, structural stability, along with their morphology. Thus, the choice of NC is also important. The graphitic carbon shell was found to be beneficial for the better conductivity, enhanced Li-ion transportation to the Si-core, and better capacity retention over amorphous carbon shell for the Si-core-carbon-shell NPs.[90] For the Si NPs/RGO, in addition to the sufficient space for volume accommodation for NPs and the improved conductivity, RGO also traps and stabilizes the SiNPs inside the electrode by forming strong chemical bonds via Si-OH functional groups and by avoiding unstable solid-electrolyte-interfacial formation on the Si surface.[72] Even, the flake size of graphene's (edge sites and sheet disorder) is also found to be decisive for the cycle performance of anode materials.[91] More specifically, the RGO with 238 nm graphene flakes showed the optimized electrochemical performance for $Li^+$-ion storage compared to the other flake sizes (113 and 160 nm).[91] This study suggests that one need to pay attention on the flake size of graphene to be composited with NPs. The choice of NC was also found to be significant for the $NiCo_2S_4$-NiS/NC composite.[92] Compared to the flower-like network of carbon (888 mAh/g), bowknot-like carbon-based composite exhibited a higher Li-ion storage capacity of 994 mAh/g at 0.2 A/g after 200 cycles and a better rate performance.[92] Although there was no discussion on the impact of NCs, we believe that the geometry of NC and hence the corresponding surface and structural features are responsible for the different $Li^+$-ion storing abilities.

*Table 2: Half-cell $Li^+$-ion storage performances of NPs/NCs composite. (asterisk symbol represents the data estimated either from plot using WebPlotDigitizer copyright 2010-2020 Ankit Rohatgi or available data from the reference)*

| **Ref.** | **NPs/NC composite** | **Synthesis method of active materials** | **Particle size (nm), Specific surface area ($m^2/g$)** | **Mass loading ($mg/cm^2$)** | **Type of battery** | **Reversible discharge capacity (mAh/g)** | **Rate performance (%)** | **Cycle life (Retention, current density, cycle number)** | **Charge transfer resistance (Ω)** |
|---|---|---|---|---|---|---|---|---|---|
| [46] | Sn-carbon nanofiber | Electrospinning + carbonization | -, 708.22 | | $Li^+$-ion | 834.2 at 0.8 A/g | 0.06%* at 4 A/g | 42.6%* at 0.8 A/g, 200 | |
| | Sn/porous carbon nanofibres | | 100-300, 872.16 | - | | 792.7 at 0.8 A/g | 44%* at 4 A/g | 81.5%*, 0.8 A/g, 200 | - |
| [100] | Ge/SWCNT | Sonication and filtration | 60, - | - | $Li^+$-ion | 760* at 0.1 A/g | 72.4%* at 1 A/g | - | - |
| [82] | Si-Carbon@graphene foam | Solvothermal reaction of GO and Si NPs + coating by amorphous carbon layers | 100, -, | 0.5 | $Li^+$-ion | 2956 at 0.1 A/g 1176 $mAh/cm^3$ at 0.2 A/g | 24%* at 1.5 A/g | 85.1%, 1 A/g, 200 | - |
| [21] | RGO aerogel (RGOA) | Hummers method+ freeze-dry | - | | $Li^+$-ion | 258 at 0.1 A/g | 13.6%* at 3 A/g | - | - |
| | $SnO_2$-RGOA-50 | Freeze-drying the mixed solution and calcination | 7 – 8, 46 | 2 | | 450 at 0.1 A/g | 25.5%* at 3 A/g | - | - |
| | $SnO_2$-RGOA-75 | | 7 – 8, 101 | | | 600 at 0.1 A/g | 18.3%* at 3 A/g | - | - |
| [64] | Nanomesh graphene (NMG) | Chemical vapour deposition | - | - | $Li^+$-ion | 1675* at 0.05 A/g | 295 at 1 A/g | 34.6%* at 1 A/g, 100 | |
| | $Fe_2O_3$ NPs | Adsorption-precipitation on NMG | 30 – 50, - | | | 980* at 0.05 A/g | 30.1%* at 1 A/g | 32.7%*, 1 A/g, 100 | - |
| | 10%Fe-NMG | | 1.5 – 5, - | | | 1530* at 0.05 A/g | 36.3%* at 1 A/g | 108.5%*, 1 A/g, 100 | 61 |
| | 20%Fe-NMG | | 20 – 30, - | | | 1650* at 0.05 A/g | 30%* at 1 A/g | 78.7%*, 1 A/g, 100 | - |
| | 70%Fe-NMG | | 20 – 100, - | | | 1240* at 0.05 A/g | 39.9%* at 1 A/g | - | - |
| [79] | $Co_3O_4$/P-RGO | Freeze-drying the mixed solution + annealing at 800 °C in $N_2$ + annealing at 280 °C in air | -, 184 | - | $Li^+$-ion | 580 at 0.1 A/g | 15.5%* at 10 A/g | 117.7%* at 0.2 A/g, 200 | - |
| | Hollow-$Co_3O_4$@P-RGO | | -, 230 | | | 855.7 at 0.2 A/g | 60.7%* at 10 A/g | 121.9%*, 0.2 A/g, 200 | - |
| [76] | Hollow-$Co_3O_4$@N,S-RGO | Co-Metal organic framework/GO precursor by Solvothermal + annealing at 800 °C for 2 hrs in Ar | 20 – 25, 168 | - | $Li^+$-ion | 1105 1t 0.2 A/g | 62.2%* at 10 A/g | - | - |
| | Co-$Co_3O_4$/N,S-RGO | Above steps + annealing at 250 °C for 2 hrs in air | 10 – 18, - | | | 808* at 0.2 A/g | 64.1%* at 10 A/g | 112.1%*, 1 A/g, 700 | - |
| [59] | CNT | - | - | - | $Li^+$-ion | 316 at 0.1 A/g | - | 68.7%*, 0.1 A/g, 50 | - |
| | Mesoporous $Co_3O_4$ | co-precipitation + calcination | 200-300, 146 | - | | 713.3* at 0.1 A/g | 18.6% at 0.5 A/g | 89.1%*, 0.1 A/g, 50 | Larger |
| | Mesoporous $Co_3O_4$/ functionalized CNT | | 15-30, 188 | - | | 986.7* at 0.1 A/g | 77.5%* at 0.5 A/g | 89%*, 0.1 A/g, 50 | smaller |
| [87] | M-$Nb_2O_5$@C | Reaction at room temperature with or without GO to obtain Nb-based Metal organic framework + heat treatment | 200, 80.5 | - | $Li^+$-ion | 165* at 0.05 A/g | 40.6%* at 5 A/g | 84.8%, 2 A/g, 1000 | - |
| | M-$Nb_2O_5$@C/RGO | | 10, 168.3 | | | 190* at 0.05 A/g | 51%* at 5 A/g | 90.5%, 2 A/g, 1000 | - |
| [94] | $TiO_2$ | Commercial | 5 – 10, - | 0.6 | $Li^+$-ion | 136* at 0.3C | 41.9%* at 50C | - | 85.9 |
| | S-doped $TiO_2$ | Thermal conversion of $TiS_2$ nanosheets with organics adsorbed | 5 – 10, - | | | 288* at 0.3C | 20.8%* at 50C | - | 53.4 |
| | S-$TiO_2$/N-carbon nanosheets | | 5 – 10, 70.5 | | | 550 at 0.3C | 18.5%* at 50C | 89.1%, 0.6C, 500 | 49 |
| [86] | MnO NPs | Microwave reaction of IGR and Mn-based precursors (without IGR for bare NPs) + polymerization + calcination at 800 °C for 2 hrs in $Ar/H_2$ | 200 – 500, 19 | 1 | $Li^+$-ion | 739 at 0.1 A/g | 29.9%* at 2 A/g | 38%, 0.5 A/g, 200 | 177.4 |
| | interconnected graphene ribbons (IGR) -MnO | | -, 56 | | | 774 at 0.1 A/g | 56.6%* at 2 A/g | 73%, 0.5 A/g, 550 | 77.8 |
| | N-carbon-coated IGR-MnO | | -, 115 | | | 1055 at 0.1 A/g | 51.8%* at 2 A/g | 113%, 0.5 A/g, 550 | 64.9 |
| [77] | B,N-graphitic nanotube (BNG) | Two-step pyrolysis and oxidation | -, 602 | 0.54 – 0.65 | $Li^+$-ion | 412 at 0.1 A/g | - | - | 87.83 |
| | CoO@BNG | | 15, 146 | | | 1451 at 0.1 A/g | 17%* at 3 A/g | 96%, 1.75 A/g, 480 | 95.08 |
| | $Ni_2O_3$@BNG | | 25, 97 | | | 823 at 0.1 A/g | - | - | 153.4 |
| | $Mn_3o_4$@BNG | | 30, 84 | | | 710 at 0.1 A/g | - | - | 171.7 |
| [88] | $Ni_2P$ @ P-porous carbon | One-pot preparation through high temperature annealing process | -, 337.08 | - | $Li^+$-ion | 768* at 0.1 A/g | 31.3%* at 5 A/g | - | 100.8 |
| | $Ni_2P$ @ N/P-porous carbon | | 20 – 200, 454.68 | | | 1137.2 at 0.1 A/g | 38.7%* at 5 A/g | 88.9%*, 2 A/g, 1700 | 29.07 |
| [99] | $CoS_2$/N-porous Carbon Shell | Nano MOF-derived synthesis | 15, - | - | $Li^+$-ion | 701 at 0.1 A/g | 58.5%* at 2.5 A/g | 79.9%*, 0.1 A/g, 50 | - |
| [85] | $MoO_2$ QDs/RGO | Solvothermal method | < 5, - | 1.8 - 2 | $Li^+$-ion | 1257 at 0.1 A/g | 43.7%* at 2 A/g | 98%, 0.1 A/g, 100 | - |
| | $MoS_2$/RGO | Solvothermal + sulfidation | 20 – 60, - | | | 1497 at 0.1 A/g | 73.3%* at 2 A/g | - | - |
| [92] | Ni-Co-S@carbon bowknots | Surfactant-assisted co-precipitation method | -, 26 | 1.6 – 2 | $Li^+$-ion | 914 at 0.2 A/g | 22.6%* at 16 A/g | 113.7%*, 0.2 A/g, 200 | - |
| | Ni-C-S@carbon flowers | | -, 10 | | | 912 at 0.2 A/g | 18.8%* at 16 A/g | 95.9%*, 0.2 A/g, 200 | - |
| [47] | Co@ mesoporous carbon | Hydrothermal method + calcination | -, 18.5 | - | Li-S | 1291* at 0.5C | - | 38.1%*, 0.5C, 200 | 184.7 |
| | Ni@ mesoporous carbon | | -, 135 | - | | 1241* at 0.5C | - | 45.2%*, 0.5C, 200 | 176.5 |
| | CoNi@mesoporous carbon | | 5 – 20, 190.6 | - | | 1383.9* at 0.5C | 61.6%* at 4C | 71.5%, 0.5C, 200 | 18.9 |
| [98] | $TiO_2$@C CNs | Electrospinning + heat treatment | 20 – 200, 256 | - | Li-S | 1094 at 2C | 66.5%* at 6C | 44%*, 4C, 2000 | 40.7 |

#### 4.1.8. Doping

The research has been extended towards designing NPs/doped-NC composite, since NC has the potential to further improve the structural properties.[76][77] Dopants are mostly boron (B), nitrogen (N), phosphorous (P), sulfur (S), etc. For instance, Ag/B-RGO showed the initial reversible capacity of 1484 mAh/g at 50 mA/g.[13] Even, $Li_4Ti_5O_{12}$ NPs with a size of 50 nm have been prepared and coated by N-doped carbon to enhance the $Li^+$-storage capability.[93] The choice of NPs on the doped NC are crucial to obtain high-performance energy storage. It has been reported that encapsulating hollow CoO NPs in B/N co-doped graphitic nanotubes was better rather than encapsulating hollow $Ni_2O_3$ NPs and $Mn_3O_4$ NPs. This is due to the fast charge-transfer kinetics, high surface area and adequate NP size, nanotube diameter and wall thickness, presence of defects due to the co-doping and stoichiometry of each element of hollow CoO NPs encapsulated B/N co-doped graphitic nanotubes.[77] We would like to emphasize that, although the experimental procedure was the same and the precursor ratio was maintained, the NPs-to-NC ratio was different which makes significant differences on the intrinsic properties as well as on the charge-storage performance.

In addition to the NC doping, significant attention has been paid to the doping of metal oxide NPs. It has been reported that S-doped $TiO_2$ NPs embedded carbon nanosheets exhibited much higher $Li^+$-storage capacity than the commercial $TiO_2$ and S-$TiO_2$ NPs.[94] It is important to note that the $Li^+$-ion diffusion coefficient was found to be the highest for the S-$TiO_2$/N-NC ($3.19 \times 10^{-9}$ cm$^2$ S$^{-1}$) than the S-doped $TiO_2$ ($1.25 \times 10^{-9}$ cm$^2$ S$^{-1}$) and commercial $TiO_2$ ($1.63 \times 10^{-10}$ cm$^2$ S$^{-1}$). The enhanced $Li^+$-storage capacity and the rate performance of the composite were attributed to the electronic and ionic conductivity, and hence charge-storage kinetics, induced by S-doping. Although not highlighted in the manuscript, we emphasize the role of N-doping on the NC for the improved performance.

#### 4.1.9. Multi-component composites

To obtain the synergy of the physico-chemical properties and hence the high electrochemical performance, there is also an increasing trend in designing multi-component electrode materials. The multiple components may consist in both metal NPs, metal-based NPs, a combination of a metal and metal-based NPs, dopants of NC, etc. Here, metal-based NPs are the metal oxide, nitride, phosphides, chalcogenide NPs. In the case of binary metal NPs/NC composite, Co NP assist in the size buffering effect in CoSn NPs supported on commercial carbon black (CoSn@C)[95], which resulted in around four times higher $Li^+$-ion diffusion coefficient, higher Li-on storage capacity, and the improved cycle stability compared to the CoSn and Sn@C. The performance was even found to be better than the NiSn@C which is attributed to the higher pseudocapacitance from CoSn@C and higher oxidation rate of smaller NiSn@C. In the composite of $SnO_2$/$NiFe_2O_4$/RGO, the observed high capacity was attributed to the metal oxide parts, $SnO_2$ serves as a buffer layer to reduce the pulverization, and graphene plays multiple roles. Graphene not only improved the structural stability, but also reduced the agglomeration of metal oxides and pulverization during the charge-discharge process, and hence resulted in better cycle stability and rate capability.[96] It is important to note that, although it is not reflected in the name of the composite, RGO was doped with nitrogen which improved the electronic conductivity and charge-transfer kinetics. In the Co-CoO/MnO hetero-structured nanocrystals anchored on N/P-doped 3D porous RGO,[97] each component plays a unique role: the dopants provide plenty of electrochemically active sites for $Li^+$ adsorption, the RGO serves as a conducting platform for electron transport and mechanical platform the for heterostructures, the MnO offers pseudocapacitance (which is essential to improve the rate capability), while the metallic Co enhances the electrical conductivity and the rate performance of the composite.

Thus, one needs to be extremely careful in balancing the stoichiometry between the metal NPs, metal-based NPs, carbon content and doping concentration to achieve the best impact on total capacity, rate capability and cycle-life. Moreover, the advantages of metallic NPs in the composite on the charge-storage capacity have to be clarified.

### 4.2. Li-S batteries

As an alternative to LIBs, the Li-S battery is emerging as a promising solution. In spite of the high theoretical capacity of 2567 mAh/g, poor sulfur utilization due to the insulating $Li_2S/Li_2S_2$, the shuttle effect with the ~80% volume expansion during the discharge are the biggest challenges of the Li-S battery. Several NPs/NC composites are explored as anode materials for Li-S batteries. Owing to the strong affinity towards polysulfides, ability to adsorb lithium polysulfide species, and low cost, $TiO_2$/carbon composite nanofiber emerges as one of the promising anode materials for Li-S battery. The nanofiber composite delivered a specific capacity of 978 mAh/g at a very high rate of 6C.[98] It is important to note that the distribution of sulfur was quite uniform on the nanofiber surface without any agglomeration, as confirmed from the EDX mapping. Bimetallic CoNi NPs anchored on petal-like mesoporous carbon were found to be more efficient in terms of higher capacity, alleviated shuttle effect, enhanced Li-ion diffusion, reduced polarization of battery and low interfacial impedance in contrast to the individual NPs anchored NC.[47] Doping is another adoptable strategy, since dopants such as nitrogen can adsorb the polysulfide intermediates and hence improve the cycle stability of the electrode.[99]

### 4.3. $Na^+$-ion batteries

Despite advances in LIBs, the crucial factors such as limited sources, abundancy, price and environment-friendliness drive to some the alternatives, and hence the sodium-ion battery (SIB) received significant attention.[101][14] The main challenges here are a limited choice of anode materials for the effective diffusion of $Na^+$-ions, fast charge-transfer kinetics and accommodability for large volume expansion due to the larger radius and molecular weight of $Na^+$-ion compared to $Li^+$-ion.

With a high theoretical capacity of 847 mAh/g for $Na^+$-ion storage, Sn has shown a promising electrochemical performance. Importantly, de-sodiation potentials of $Na_xSn$ are lower than the de-lithiation potentials of $Li_xSn$ which suggests that Sn as an excellent candidate for the $Na^+$-ion storage compared to the $Li^+$-ion storage.[102] The probable solution to the volume expansion from Sn to $Na_{15}Sn_4$ (around 420 %) is to prepare a composite with carbon while reducing the size of Sn. Scanning electron micrograph and energy-dispersive X-ray spectroscopy mapping suggested that the composite with 8 nm Sn NPs shows a very little volume expansion, while uniform distribution of NPs remains in the carbon matrix after 200 charge/discharge cycles. On the other hand, composite with 50 nm Sn NPs showed a rapid capacity decay after 200 cycles.[102] Other approaches are embedding Sn NPs in nitrogen-doped carbon microcages [103], preparing metal-metal oxide NPs/NC composite[104], bimetallic oxide/NC composite[105], etc. A ternary $SnO_2$@Sn core-shell decorated N-RGO aerogel was also found to outperform as anode material for SIBs compared to $SnO_2$/RGO aerogel and $SnO_2$/N-RGO aerogel. The improved $Na^+$ ion storage was attributed to the thin $Na_2O$ layer formation which prevents NPs from agglomeration, improves the $Na^+$-ion diffusion pathways, and confirmed the reversible Sn-$SnO_2$ conversion.[104] In the composite of bimetal oxides with 9.6% graphene content,[105] cubic $SnO_2$-CoO@graphene composite exhibited better rate capability and excellent cycle life with higher $Na^+$-ion storage capacity compared to other combinations of bimetallic oxide and graphene.[105]

The problems of the slow diffusion and electron transfer kinetics were resolved by considering hetero-interfaces between the oxides rather than their single oxide counterpart (Figure 7).[106] In particular, the diffusion coefficient of $Na^+$-ion in $Fe_2O_3/Fe_3O_4$ nano-aggregates/N-doped RGO was found to be higher ($1.34\times10^{-11}$ $cm^2/s$) than that of $Fe_2O_3$/N-doped RGO ($1.65\times10^{-12}$ $cm^2/s$). The defined role of each component of $Fe_2O_3/Fe_3O_4$ nano-aggregates anchored on nitrogen-doped RGO in improved electrochemical performances are (i) faster electron transport through $Fe_3O_4$, (ii) improved $Na^+$-ion transport through phase-boundaries and voids present in nano-aggregates, (iii) promoted electron transport and electrochemical stabilization by Fe-O-C bonds and (iv) increased reversible capacity by N-doping.[106] Importantly, unlike $Fe_2O_3$/N-RGO, $Fe_2O_3/Fe_3O_4$/N-RGO did not show any cracks and retained the same particle size even after 300 charge/discharge cycles. This result suggests that the binary metal oxide NPs with metal-oxygen-carbon bonds can be a material of choice for energy storage electrodes over the single metal oxide NPs/NC composite.

The research on $Na^+$-ion battery has also been directed towards non-oxide-based anode materials.[107][108] It has been shown that the carbon coating on the NPs surface is not only necessary to improve the electrochemical performances, but the carbide formation at the interface of oxide and carbon is also beneficial.[109] Metal sulphide NPs decorated graphene exhibited more reversible sulphide formations and smaller volume pulverization compared to their oxide counterpart.[107] For instance, $Sb_6O_{13}$NPs@*a-C* delivered the highest $Na^+$-ion storage capacity of 239 mAh/g at 1A/g with a capacity retention of 89.64 % after 170 cycles and the maximum obtained Coulombic efficiency was 98.6% after 17 cycles.[48] Unfortunately, agglomeration and structural degradation of $Sb_6O_{13}$NPs@*a-C* have been evidenced after the electrochemical process. In contrast, with the theoretical capacity of $Sb_2S_3$ for $Na^+$-ion storage of 946 mAh/g, $Sb_2S_3$/RGO exhibited the first discharge capacity of 1050 mAh/g at 0.05 A/g with 98.7% efficiency and >95% capacity retention after 50 cycles. It is important to note that, the lower limit of energy density of RGO/$Sb_2S_3$–$Na_2$/$3Ni_1$/$3Mn_2$/$3O_2$ full cell is found to be 80Wh/kg.[107] In the absence of RGO, the crystalline $Sb_2S_3$ was found to be transformed into an amorphous phase after the discharge cycles. Moreover, graphene coatings are essential in the composite for effective dealloying and metal-sulfide recombination during $Na^+$-ion removal.

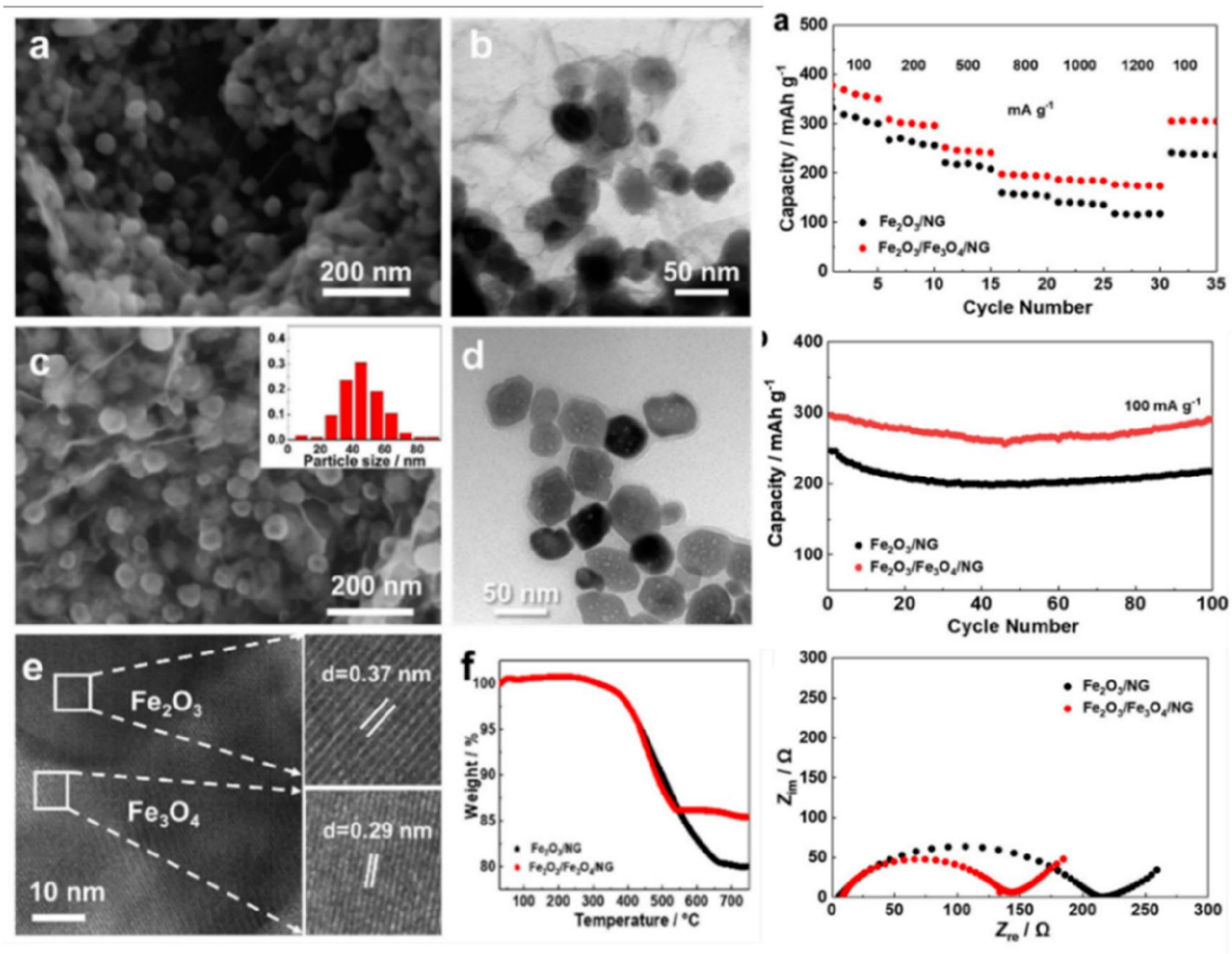

*Figure 7: Morphology, structure, thermal stability profile and $Na^+$-storage performance of $Fe_2O_3$/NG and $Fe_2O_3$/$Fe_3O_4$/NG. Adopted from Ref. [106], © 2020 by the authors. License MDPI, Basel, Switzerland.*

### 4.4. Other metal-ion batteries

As possible substitutes for $Na^{+}$-ion batteries, potassium-ion batteries, magnesium-ion batteries or aluminum-ion batteries are also getting attention due to their beneficial characteristics. The challenging part here is constructing suitable anode materials to host the electrolyte ions with higher ionic size (except the $Al^{3+}$).[112][113]

In this context, anode materials explored so far are metal nitride NPs/NC, sulfides and selenides-based composites such as ZnSe NPs embedded N-doped porous carbon matrix[114], nanosized $MoSe_2$@Carbon Matrix[111], three-dimensional carbon network confined antimony NPs [115], etc. However, $K^{+}$-ion diffusion into the bulk material was found to be difficult for the initial 5-20 cycles of charge/discharge.[114] Impressively, VN NP-assembled hollow microsphere/N-CNF showed excellent $K^{+}$-ion storage capacity of 834.2 mAh/g at 0.1 A/g for the second cycle (Figure 8A-E). In the composite, VN NP-assembled hollow microspheres were in series and connected within the nitrogen-doped CNF (Figure 8A-C), where N-doped CNFs prevent the agglomeration of NPs, provide conducting pathways and promote the structural stability. The N-doping provides electrochemical active sites for $K^{+}$-ion and the hollow structure increases the contact areas of NPs and electrolyte, accommodate the volume expansion and allows the electrolyte to penetrate effectively.[110] In addition to the electrode material design, other influencing factors for $K^{+}$-ion storage are the selection of potassium salt, solvent, electrolyte concentration, and additives (Figure 8E). [110] To accommodate electrolytes effectively and overcome the problem of bare NPs, two NCs were implemented to construct a hybrid structure. For instance, amorphous carbon coating (~ 3 nm thick) on the mesoporous $Co_9S_8$ (avg. diameter of 23.5 nm) not only protected from the electrolyte, but it also reduced the side reactions. On the other hand, RGO provided a 3D network for efficient charge transportation and restricted the movement of NPs during the electrochemical process.[116] This structure also exhibited its capability to host sulfur for Li-S batteries. As can be seen from table 3, the rate performance of NPs/NC composite for $K^+$-ion storage is not that great. It is important to note that the nano-sized $MoSe_2$@Carbon matrix served as an excellent $K^+$-ion and $Al^+$-ion storage (Figure 8F-I) even in the temperature range of -10° to 50 °C.[111]

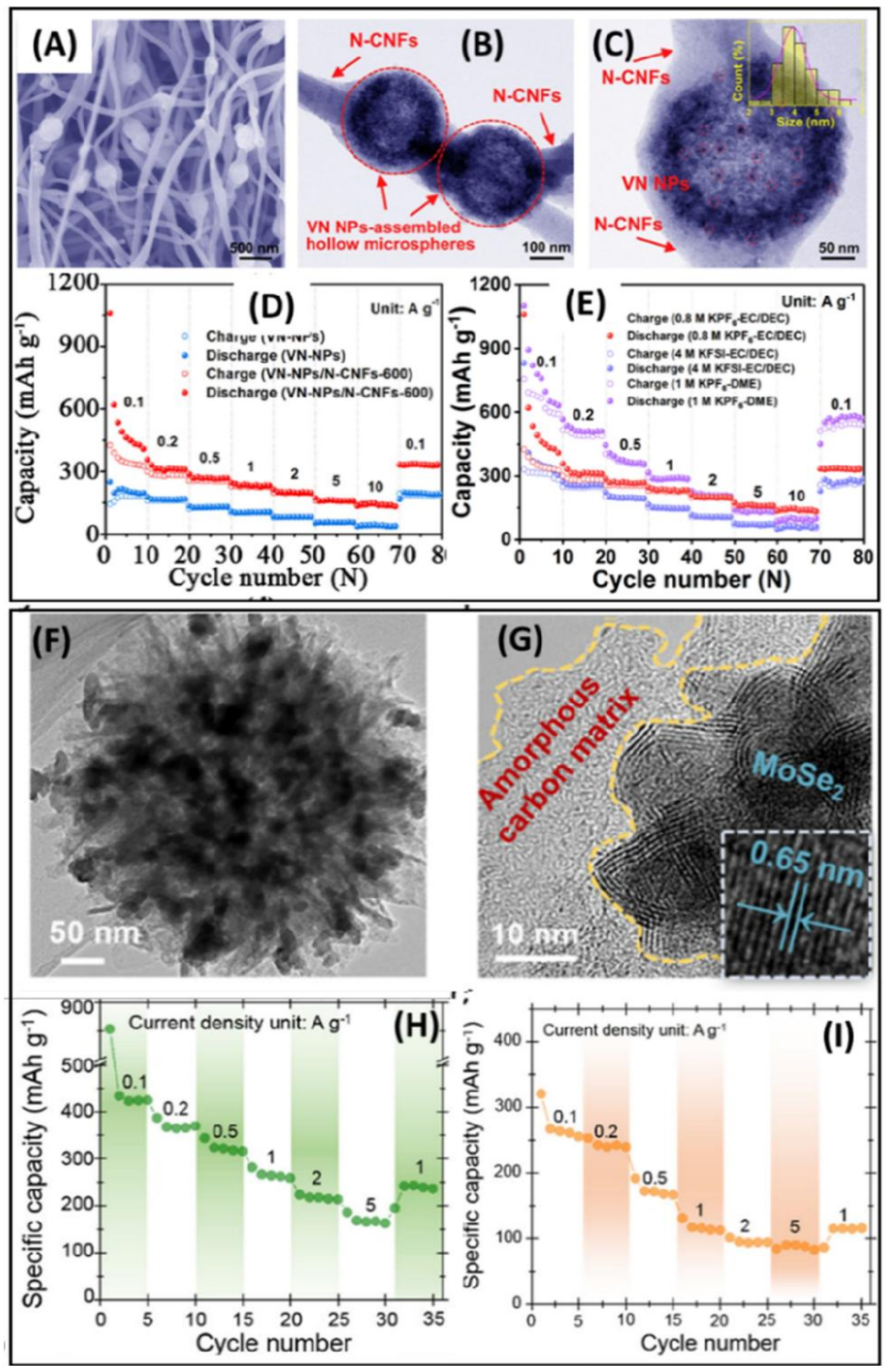

*Figure 8: (A) Scanning electron micrographs and (B-C) transmission electron micrograph of VN-NPs/N-doped CNFs-600 after 2000 cycles of charge/discharge, (D) Rate property of VN-NPs/N-CNFs-600 and VN-NPs for $K^{+}$-ion storage, and (E) Rate performance of the VN-NPs/N-CNFs-600 electrode in the 0.8 M $KPF_6$-EC/DEC, 4 M KFSI-EC/DEC and 1 M $KPF_6$-DME electrolytes. CNF: carbon nanofiber. Adopted from Ref. [110], © 2021 Chongqing University. Publishing services by Elsevier B.V. on behalf of KeAi Communications Co. Ltd. (F-G) transmission electron micrographs, (H) $K^{+}$-ion storage performance and (I) $Al^{3+}$-ion storage performance of N-$MoSe_2$@amorphous carbons electrode. Reproduced from Ref. [111], © 2019, With permission from American Chemical Society.*

For the $Mg^+$-ion and $Zn^+$-ion storage, the NP/NC composites are mainly used as cathode materials.[117][118] There are limited studies on the NP/NC anode materials such as Bi/RGO[119], graphene supported SnSb NPs[120]. It has shown that the effective $Mg^+$-ion storage can be obtained for the composite with Sn NPs of <40 nm size. Although $Mg^+$-ions are removed from $Mg_{1.2}Sn$ particles but trapped in Sb-rich alloy domains during the demagnesiation. Moreover, $Mg^{2+}$ and $Al^{3+}$ diffusions are sluggish for the graphene supported $Co_3O_4$ nanocube, whereas an obvious diffusion-controlled heterogeneous surface reactions during the lithiation has been evidenced from in-situ transmission electron miscropic study.[121] During the charging, no conversion reaction at room temperature has observed for both these multivalent ions and there is a formation of thin film of Mg NPs on the graphene surface. The challenges for $Mg^{2+}$-ion storage are finding suitable electrode materials and limited electrolytes.

### 4.5. LIBs *versus* other MIBs

The performances of the composite for various metal-ion storage are summarized in Table 3 and compared with $Li^+$-ion storage. It can be seen from Table 3 that the NPs/NC composite showed higher storage capacity, better rate performance and lower charge-transfer resistance for the $Li^+$-ion than the other ions. This fact can be correlated with the morphology and structure of the composite with the lower ionic radius which promotes the better $Li^+$-ion diffusion into the composite. Remarkably, the cycle life of the composites is found to better in the case of $Na^+$-ion storage (Table 3). In spite of significant advances on other metal-ion batteries, the improvement in the ion storage performance is still a subject of further research and developments.

*Table 3: Half-cell post $Li^+$-ion storage performances of NPs/NC composite and their comparison with $Li^+$-ion storage performances. (symbol with an asterisk represents the data estimated either from the plot using WebPlotDigitizer copyright 2010-2020 Ankit Rohatgi or available data from the cited reference)*

| Ref. | NPs/NC composite | Synthesis method for the composite | Size of NPs (nm), BET surface area ($m^2/g$) | Mass loading ($mg/cm^2$) | Type of battery | Reversible discharge capacity (mAh/g) | Rate performance (%) | Cycle life, current density, cycle number | Charge transfer resistance (Ω) |
|---|---|---|---|---|---|---|---|---|---|
| [102] | 8-Sn@carbon | Aerosol spray pyrolysis | 8, 150.43 | 1.5 – 2 | $Na^+$-ion | 493.6 at 0.2 A/g | 70.7%* at 4 A/g | 97.6%*, 1 A/g, 500 | 47.8 |
| | 50-Sn@arbon | | 50, 60.17 | | | 506 at 0.2 A/g | 12.6%* at 4 A/g | - | 138 |
| [48] | Carbon encapsulated $Sb_6O_{13}$ NPs | One-pot hydrothermal | 60, - | 0.25 | $Na^+$-ion | 338 at 0.2 A/g | 67.7%* at 1 A/g | 89.64%, 0.2 A/g, 170 | - |
| [122] | $GeO_2$ | Commercial | 100-200, - | - | $Na^+$-ion | 9.2 at 0.1 A/g | - | - | 317.7 |
| | Freeze-drying treated $GeO_2$ | Freeze-drying treatment | - | - | | 17 at 0.1 A/g | - | - | - |
| | $GeO_2$/RGO (25% RGO) | Freeze-drying method | - , 9.09 | - | | 335.8 at 0.1 A/g | 57%* at 1 A/g | 93.15%*, 1 A/g, 650 | 186.6 |
| [107] | rGO@$Sb_2S_3$ | sulphurisation of peroxoantimonate-coated GO in alcohol + vacuum annealing at 300 °C | 15 – 30, - | - | $Na^+$-ion | 730 at 0.05 A/g | 71.2%* at 3 A/g | > 95%, 0.05 A/g, 50 | - |
| [104] | $SnO_2$@N-RGO aerogel | Hydrothermal method | 30 – 200, - | - | $Na^+$-ion | 340* at 0.1 A/g | ≈25% at 2 A/g | 70.7%*, 2 A/g, 200 | 39.0 |
| | $SnO_2$@Sn core-shell/N-RGO aerogel | Hydrothermal method + microwave plasma | -, - | 1.5 – 2 | | 448.5 at 0.1 A/g | 35.7% at 2 A/g | 73%*, 2 A/g, 200 | 21.8 |
| [60] | T-$Nb_2O_5$ | Solvothermal method + calcination | 129, - | - | $Na^+$-ion | 48* at 0.05 A/g | 10.4%* at 2 A/g | 28%*, 0.2 A/g, 200 | - |
| | T-$Nb_2O_5$@G | | 30, - | - | | 118* at 0.05 A/g | 32.2%* at 2 A/g | 91.8%*, 0.2 A/g, 200 | - |
| | T-$Nb_2O_5$@NG | | 17, - | - | | 137 at 0.05 A/g | 46.7%* at 2 A/g | 85.4%*, 0.2 A/g, 200 | - |
| [106] | $Fe_2O_3$ nanoaggregates/NG | Solvothermal method | 50, - | 0.75 | $Na^+$-ion | 300 at 0.1 A/g | 38.7%* at 1 A/g | - | 210.6 |
| | $Fe_2O_3$/$Fe_3O_4$ nanoaggregates/NG [106] | Solvothermal method + microwave plasma | -,- | 1.13 | | 305 at 0.1 A/g | 49.6%* at 1 A/g | 84%, 1 A/g, 100 | 134.7 |
| [109] | Anatase $TiO_2$ | Fluidized-bed plasma enhanced chemical vapour deposition | 27 ± 6, - | 1.6 | $Na^+$-ion | 198.1 at 0.05 A/g | 37.4%* at 4 A/g | 90%, 4 A/g, 300 | - |
| | Anatase $TiO_2$@carbon | | 35 ± 8, - | 1.2 | | 290.2 at 0.05 A/g | 40.3%* at 4 A/g | 101.2%, 4 A/g, 300 | - |
| [103] | Sn nanodots@ N- carbon microcages | One-pot wet chemical method in GO solution | 10-20, 598 | 0.8 | $Na^+$-ion | 439 at 0.05 A/g | 33.9%* at 5 A/g | 75.6%*, 0.05 A/g, 300 | larger |
| | | | | | $Li^+$-ion | 835 at 0.05 A/g | 47.4%* at 5 A/g | 60.5%*, 0.2 A/g, 500 | smaller |
| [123] | $CO_3ZnC$@CNT-insertedN-carbon concave-Polyhedrons | Pyrolysis of zeolitic imidazolate frameworks | 2-26, 87 | 0.8 – 1.5 | $Na^+$-ion | 448 at 0.5 A/g | 56.7%* at 2 A/g | >98%, 0.2 A/g, 200 | - |
| | | | | | $Li^+$-ion | 805 at 0.5 A/g | 72.2%* at 2 A/g | 93.6%*, 0.5 A/g, 200 | - |
| [89] | MnS@Carbon fiber | Electrospinning + carbonization | 5, 254 | 1.25 | $Na^+$-ion | 132 at 0.1 A/g | 65%* at 1 A/g | 104%*, 0.02 A/g, 200 | - |
| | | | | | $Li^+$-ion | 723 at 0.1 A/g | 61.5%* at 1 A/g | 57.7%*, 1 A/g, 1000 | - |
| [108] | $Ni_3S_2$ @ GO | One-pot wet chemical method | 30, - | - | $Na^+$-ion | 249.2 at 0.1 A/g | 42%* at 5 A/g | 90.4%, 0.1 A/g, 50 | - |
| | | | | | $Li^+$-ion | 829.3 at 0.1 A/g | 30.7%* at 5 A/g | 108%, 0.1 A/g, 50 | - |
| [110] | VN | Hydrothermal + annealing | -, 19.8 | - | $K^+$-ion | 312* at 0.1 A/g | 16.5%* at 10 A/g | 39.6%*, 0.1 A/g, 100 | 4.3, 1250 |
| | N-CNF | Electrospinning hydrothermal | -, 5.7 | - | | 303* at 0.1 A/g | 24.3%* at 10 A/g | 65%*, 0.1 A/g, 100 | - |
| | VN assembled hollow microspheres @ N-CNFs | VOOH and Polyacrylonitrile (PAN)+carbonization @600 °C | 4.2, 33.4 | - | | 834.2 at 0.1 A/g | 25.8%* at 10 A/g | 44.6%, 0.1 A/g, 100 | 2.4, 583 |
| | Solid VN/N-CNFs | Electrospinning of VOOH and PAN + carbonization @600 °C | - | - | | 521.4* at 0.1 A/g | - | 34%*, 0.1 A/g, 100 | - |
| [115] | 3D carbon network confined Sb NPs | Freeze-drying method + carbothermic reduction | 60 – 100, 38 | - | $K^+$-ion | 478 at 0.2 A/g | 60.3%* at 1 A/g | 80%*, 1 A/g, 50 | - |
| [124] | Co@graphitic nanotube | Pyrolysis | 50, 34.18 | 0.9 | $K^+$-ion | 200 at 0.05 A/g | 3.5% at 1 A/g | 49%, 0.1 A/g, 300 | 689.45, - |
| [114] | ZnSe@N-porous carbon-600 | Pyrolysis + selenization. 600, 700 and 800 represents the calcination temperature in °C | -, 196.37 | - | $K^+$-ion | 261.9* at 0.1 A/g | 19.5%* at 5 A/g | 80.7%*, 0.1 A/g, 60 | - |
| | ZnSe@NDPC-700 | | 100 – 200, 210.39 | | | 271.3* at 0.1 A/g | | 99.6%*, 0.1 A/g, 60 | - |
| | ZnSe@NDPC-800 | | -, 187.09 | | | 202.8* at 0.1 A/g | | 87.3%*, 0.1 A/g, 60 | - |
| [95] | CoSn@carbon black | Mixed solution was subjected to Sonication + centrifuge + vacuum dry | 6 ± 0.8, 56.2 | - | $K^+$-ion | 286.9 at 0.2 A/g | 26.7%* at 5 A/g | 29.2%*, 0.2 A/g, 300 | 4549 |
| | | | | | $Li^+$-ion | 846.2 at 0.2 A/g | 35.2%* at 5 A/g | 116.9%*, 0.2 A/g, 100 | 203 |
| | NiSn@carbon black [95] | | 4.2 ± 0.7, 51.5 | - | $K^+$-ion | 259.8 at 0.2 A/g | 17.2%* at 5 A/g | 64.3%*, 0.2 A/g, 400 | 1495 |
| | | | | | $Li^+$-ion | 668 at 0.2 A/g | 24.3%* at 5 A/g | 66.6%*, 0.2 A/g, 400 | 113 |
| [111] | Nanosized $MoSe_2$ @Carbon Matrix | Ion complexation-induced method + selenization | -, 50.4 | - | $K^+$-ion | 368 at 0.2 A/g | 45.9%* at 5 A/g | 85%*, 1 A/g, 5000 | - |
| | | | | | $Al^{3+}$-ion | 242 at 0.2 A/g | 37%* at 5 A/g | ≈100%, 1 A/g, 5000 | - |
| [120] | graphene supported SnSb | Mixed solvothermal and reduction process of a graphene sheets, $SnCl_2$ and $SbCl_3$ dispersion | 70 – 1000, - | | $Mg^{2+}$-ion | 420 at 0.05 A/g | 83.3%* at 1 A/g | 83.1%*, 0.5 A/g, 200 | - |
| [119] | Bi/RGO | Solvothermal method | 500, - | | $Mg^{2+}$-ion | 326 at 0.02 A/g | ≈0%* at 0.7 A/g | 87.4%*, 0.02 A/g, 50 | - |

**4.6. Reason for the increased capacity with cycle life**

In most cases, the deterioration in charge-storage capacity with repeated cycles or higher current density has been observed (Figure 4F and 5D). This fact is obvious since there are morphological and structural changes after the electrochemical process. These changes can be seen clearly from cyclic voltammogram, charge-discharge profile, or impedance spectra before and after charge/discharge cycles (Figure 9). It is also clear from Figure 9 that the hollow NiO failed to hold its morphological and structural stability after repeated charge/discharge cycles (Figure 4F) whereas coral yolk-shell NiO/carbon microspheres showed excellent electrochemical stability. Moreover, the charge transfer was found to be better for coral yolk-shell NiO/carbon microspheres after 200 cycles of charge/discharge. This result also reflects the importance of NCs incorporation in the hollow NiO structures. One may assume the volume expansion of the yolk-shell structure after charging/discharging from transmission electron micrograph in Figure 9(E-F), but actually they do not belong to the same microspheres. Surprisingly, in many cases, there are also an increasing trend in the specific capacity of anode materials after the repeated charge/discharge cycles (Figure 4F). A morphological test using Scanning electron micrograph and corresponding energy-dispersive X-ray spectroscopy was used to confirm the unchanged morphology of composite with very ignorable changes.[12] For deeper insight, transmission electron micrograph has been conducted on the $Mn_3O_4$ NPs anchored on CNTs which exhibited the pronounced reconstruction of NPs into small nanodots and spread out on the CNT surface without any agglomeration.[62] The increased cycle stability of the composite is mainly due to the reconstruction of electrode materials such as the transformation of crystalline structure to amorphous, which enhanced the electrolyte ion insertion kinetics, provided additional pathways for electrolyte ion diffusion, enhanced the pseudocapacitance, and improved the rate of the lithiation/delithiation during the electrochemical processes.[12][62] Surprisingly, the increasing/decreasing trend of $Li^+$-ion storage capacity with number of cycles was found to be dependent on applied current density.[108] The $Li^+$-ion storage capacity of $Ni_3S_2$ NPs/GO was increased with cycle number at 1 A/g and decreased at 5 A/g. On the contrary, the $Ni_3S_2$ NPs/GO showed decreasing trend for $Na^+$-ion storage with cycle number at both 1 A/g and 5 A/g. This result [108] certainly provoke for further investigation.

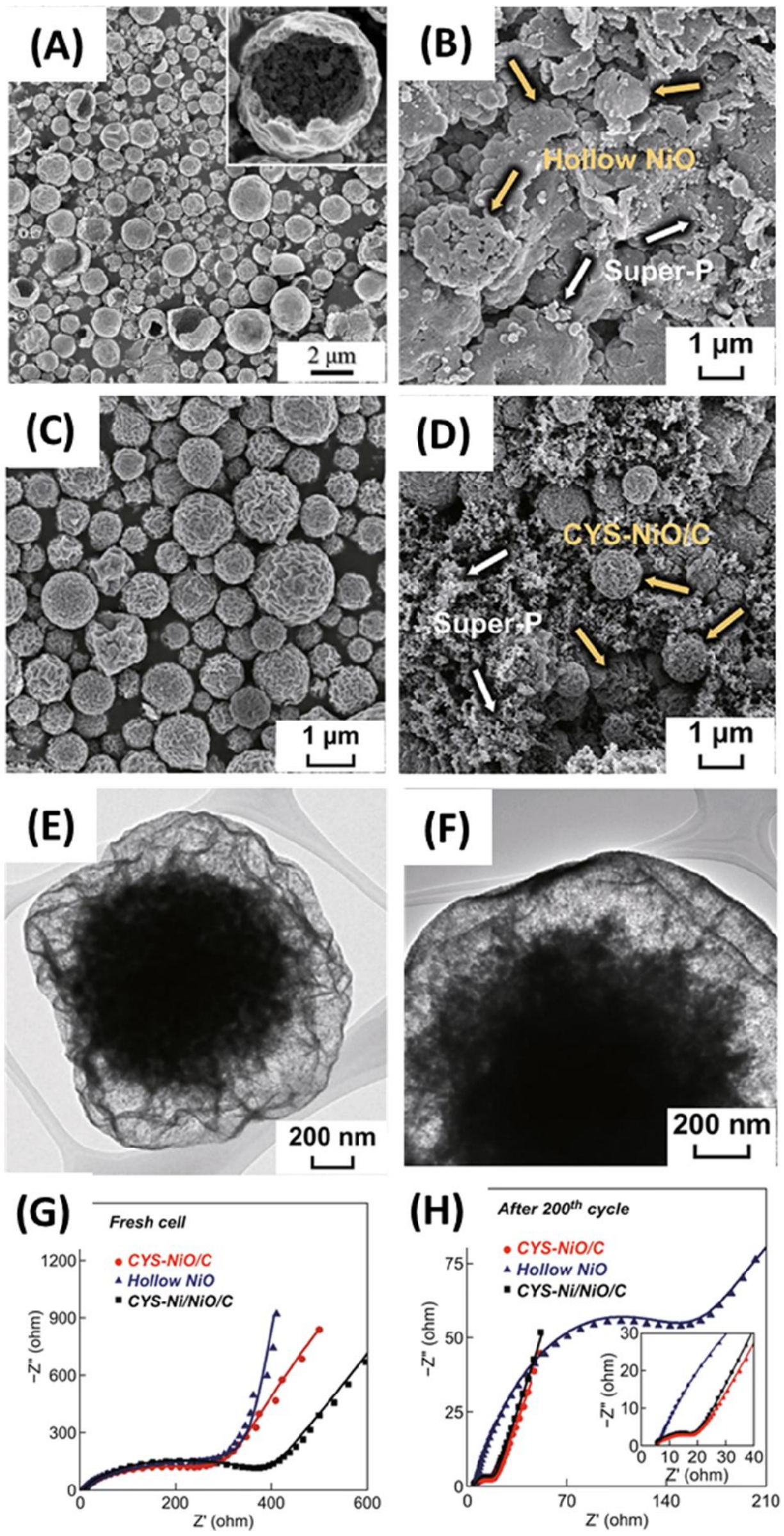

*Figure 9: Scanning electron micrograph of hollow NiO, scanning electron micrograph, transmission electron micrograph and Nyquist plot of coral-yolk shell NiO/carbon microspheres before (left panel) and after (right panel) half-cell electrochemical test. Adopted from Ref. [12] © The author(s) 2019, Springer Publishers.*

**4.7. Origin of the higher-than-theoretical capacity**

The NPs decoration onto NC or NPs/NC composite does not always increase the capacity than the capacity of NPs. Noteworthy, the theoretically estimated $Li^+$-ion storage capacity of NiO/C composite (588 mAh/g) is lower than the theoretical capacity of NiO (718 mAh/g) and higher than that of graphite (372 mAh/g).[12] Instead, the composite electrodes exhibited a better rate performance, cycle life and electrochemical stability. On the other hand, the introduction of pseudocapacitive materials such as MnO in the Co-CoO/MnO heterostructured nanocrystals anchored on N/P-doped 3D Porous RGO composite is anticipated for the higher initial charge/discharge capacity of 755/1125.8 mAh/g compared to the estimated theoretical capacity of the composite (642.8 mAh/g considering the weight contents of 48.4%, 22.9%, 2.0%, 26.7% for MnO, CoO, Co and C, respectively).[97] Nevertheless, many reports in the literature showed that the composite electrodes exhibited a specific capacity higher than the pure metal oxide NPs.[64][59][85][94] In such scenario, the changes in the surface area, NPs' size, graphitic qualities and other physico-chemical properties need to be considered. The higher storage capacity of the composite at low potential can be due to the electrolyte decomposition, excess storage of ions at the interface and/or defective sites of the composite. The contribution from the metallic NPs/lithiated matrix interface at low potential in conversion reaction materials is very small and negligible in comparison to the capacity of the composite.[125] The presence of metallic Ni in the coral yolk-shell NiO/carbon microsphere is very unlikely since metallic Ni is inactive for Li-ion storage and it showed the lowest Coulombic efficiency because of higher carbon-content and higher irreversible capacity loss.[12] Instead, bare metal NPs in the composites (for example, $Fe^0$ in a hierarchical $Fe_3O_4$@C core-shell composites) served as electrochemical catalysts for the reversible conversion of some solid-electrolyte-interfacial components ($Li_2CO_3$ among $Li_2O$, LiF, $Li_2CO_3$, $Li_2C_2O_4$, LiOH, and organic compounds), which adds to the excess capacity.[126]

## 5. Supercapacitor Electrodes

### 5.1. Size of NPs

The size of NPs greatly matters (Figure 10A) and there is a critical size. Briefly, smaller NPs (2 and 5 nm) were found to be unable to separate the graphene sheets whereas larger NPs (> 10 nm) containing more Au atoms decrease the number of NPs.[127] Meanwhile, Au NPs were used as nano-spacers to prevent the graphene flakes from stacking in RGO and outperform compared to the use of ZnO and $SnO_2$ nanospacers.[127] It has been reported that NPs of $Mn_3O_4$ smaller than 10 nm allow the aqueous electrolyte to access the electrode effectively and hence specific capacitance was found to be increased significantly.[128] On the other hand, increasing the Ag NPs size from 1 to 13 nm in CNTs leads to lowered gravimetric capacitance from 106 to 23 F/g at 1 mA/cm$^2$ in polyvinyl-alcohal/$H_3PO_4$.[129] A higher the NPs size results in increased blockage of pore in the composite when it is explored under the electrolyte.

### 5.2. Loading of NPs

It was found that the gravimetric capacitance increased for the cellular three-dimensional RGO/Ag-composite with increasing the NPs loading [130]. The highest gravimetric capacitance of 876 F/g at 1 A/g in 1M KOH with respect to the reference electrode was obtained for 40% loading of Ag NPs onto cellular three-dimensional RGO.[130] The enhanced gravimetric capacitance of CNTs after Ag NPs decoration is due to the additional pseudocapacitive redox process via the relation [129]: $Ag^0 \leftrightarrow Ag^+ + e^-; 2Ag + 2OH^- \rightarrow Ag_2O + H_2O$. Generally, the signature of redox reactions can be confirmed from the cyclic voltammogram at lower scan rates as shown in Figure 10B. It is noteworthy to say that, one can increase the specific capacitance of the composite by increasing the NPs on the NCs up to a certain limit. However, higher loading of $MnO_2$ NPs in the CNT fiber resulted in poor rate performance.[53] This is caused by the decreased electrical conductivity and electron flow from NPs to carbon matrix after crowded NPs on NC surface.[53] It has also been reported that a higher loading of Au NPs on NC blocks the electro-active surface area due to agglomeration, increases the conducting pathways for electron transfer, lowers the charge accumulation and hence lowers the capacitance value.[131][132] On the other hand, a higher amount of NC in the composite structure may improve the rate performance and the electrochemical stability, while the total specific capacitance will be less. Thus, there should be a balance between the NC and NPs content in the hybrid electrode material.

### 5.3. Metallic and metalloid NPs

There exists a series of metal and metalloid NPs such as Au, Ag, Cu, Ni, Co, Pt, Si, Sn etc. that can be incorporated in the NC matrix.[61,127,133,134] The physico-chemical features of these NPs are quite different from each other as well as the supercapacitors performance. Size of NPs also matters as shown in Figure 10A. The drastic enhancement in the specific capacitance of NCs after

NPs incorporation is observed(Figure 10B-C). Besides that, NPs decoration in RGO hydrogel reduced the voltage drop during the discharge (Figure 10D). The lower the voltage drop is, the better power density is exhibited by the device. Moreover, the metallic NPs decoration also reduces the charge-transfer resistance of the decorated NC, as it can be evidenced from the impedance spectra. Apart from those influences, it has been shown that the Si nanocluster decoration onto vertical graphenes reduces the relaxation time constant (i.e., the time required to deliver the stored charge) from 9.1 to 0.56 milliseconds when the symmetric device was tested under ionic liquid.[135]

*Do all metallic NPs decorations suit better to obtain high-performance electrode materials?* Since the change in mechanical or chemical nature of environment around the Au NPs after several scans has been evidenced, the ion intercalation occurs, which was irreversibly metastable.[136] Among the Ag, Pd and Pt NPs decoration onto CNF, CNF/Pt-0.5 h delivered the highest gravimetric capacitance but poor rate capability.[137] The enhanced gravimetric capacitance of Pt NPs/carbon nano-onions is due to the increased density of states of carbon nano-onions near the Fermi energy after Pt NPs decoration.[138] On the other hand, excellent cycle life has been seen from the CNF/Ag-1h and excellent rate capability due to the lowest equivalent series resistance has been evidenced from the CNF/Pd-0.5 h. Importantly, the enhanced capacitance of CNF after NPs decoration was attributed to the porous structure formed by the valley between the NPs.[137]

Looking at the advances in charge-storage performances of the composite after single-metallic NP decoration onto NCs, a bimetallic alloy of CoNi[139] or CoFe[140] decoration has been done on CNF. CoNi alloy decoration not only enhanced the charge-storage capacitance and reduced the charge-transfer resistance but it also reduced the nanofiber diameter and increased the porosity of the composite which is favourable for the electrolyte ion transportation into the electrode material.[139] On the other hand, after nucleating within the empty space of carbon nanofibers, CoFe NPs increased the diameter of nanofibers, the surface area and the total pore volume, leading to the enhanced gravimetric capacitance of the composite.[140] Apart from those positive influences, unfortunately, a higher leakage current has been observed in Cu-deposited carbon fiber, suggesting the possibility of electrolyte decomposition or the catalysis of the electrochemical oxidation/reduction of carbon.[141]

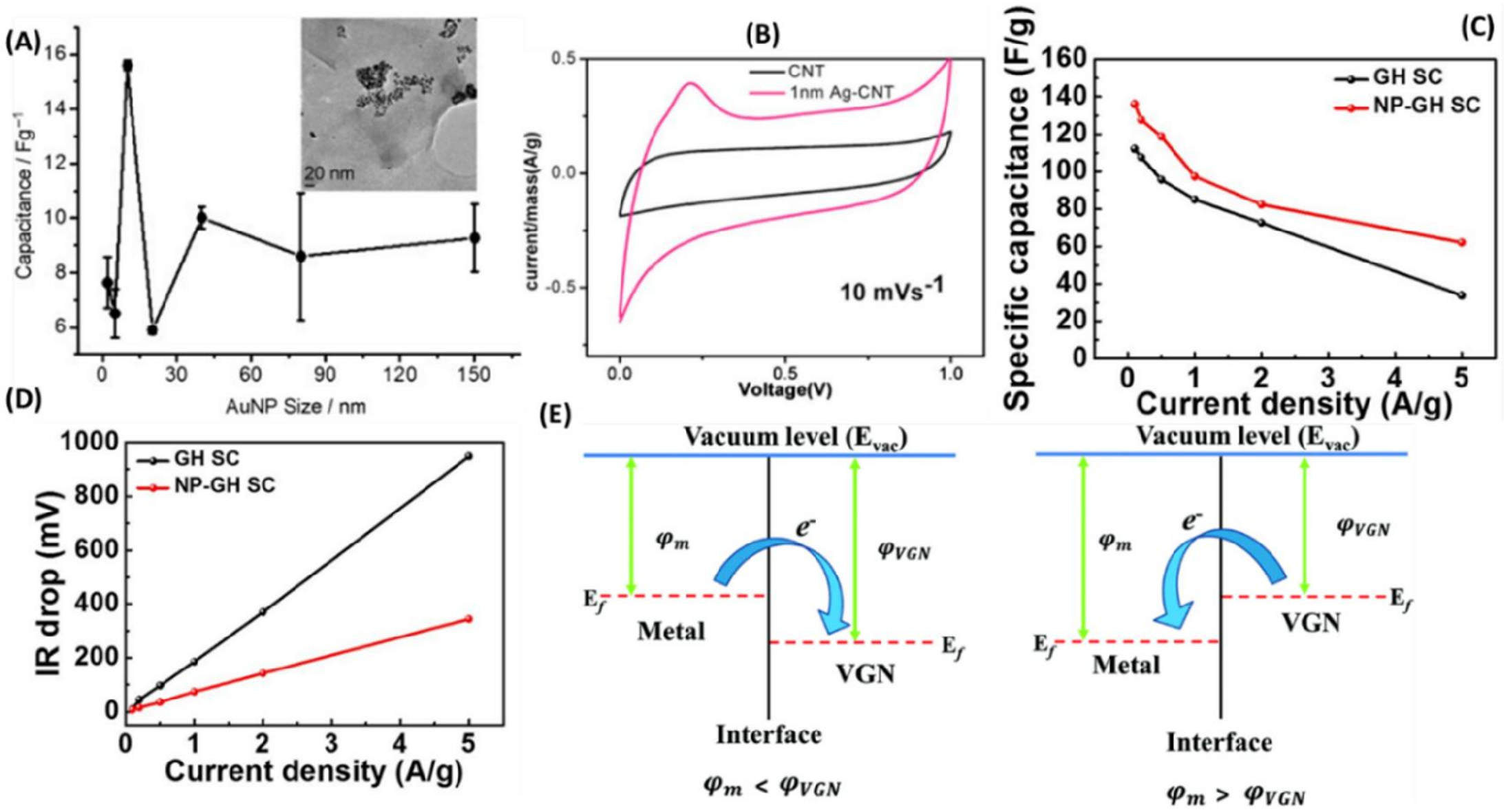

*Figure 10: (A) Gravimetric capacitance of Au NP/RGO with respect to the NP size. Reproduced from Ref. [127], ©2012, with permission from Wiley-VCH Verlag GmbH& Co. KGaA, Weinheim. (B) cyclic voltammogram of CNT and Ag/CNT at a scan rate of 10 mV/s. Reproduced from Ref. [129] © 2009 with permission from The Electrochemical Society. Plot of (C) specific capacitance and (D) potential drop of graphene hydrogel and Au NP/graphene hydrogel as a function of current density. Adopted from Ref. [142]. © 2017, The Author(s), Springer Nature publishers. (E) Schematic of charge-transfer between metal and vertical graphene surface. Reproduced from Ref. [133], ©the Owner Societies 2019.*

The enhancement in charge-storage performances of decorated NC is attributed to the lowered Schottky barrier and work-function of metals, enhanced electrical conductivity, increased pseudocapacitance due to the native oxide on NPs's surface and improved charge-transfer kinetics.[131][133] The charge-transfer between the NC and NPs occurs due to the mismatch in their work functions (Figure 10 E). Either electrons can flow from metal NPs to NC or vice-versa as shown in Figure 10(E). As a result, there is enhancement in the charge density of the NPs/NC composite. Basically, charge redistribution across the interface of metal NPs and NCs and charge accumulation take place at a lower mass loading of metal NPs on NC surface, which lead to the improved charge-storage capacity, electron-transfer kinetics and electrochemical stability.[131] The localized depletion region formed due to Au NPs decoration served as a charge-scattering center which reduces the charge mobility through the CNT network.[131] Moreover, the presence of metallic NPs on the surface of graphite sheets makes the heterostructure stable at higher total surface energy with respect to the pristine one and retains the metallic nature of the heterostructure.[143] It has been seen theoretically that the presence of metallic NPs on the graphitic-$C_3N_4$ enhances the localized electrons and results in overlapping state formation at the Fermi energy within the conduction band, resulting in enhanced electrical conductivity.[143] The charge-discharge profile of $FeNi_3$-NPs/g-$C_3N_4$ with no straight voltage curve and height changing curve reveals that the charge-storage mechanism is based on both intercalation and adsorption.

**5.4. Oxide NPs/NCs**

There are libraries of metal oxides that are considered as promising NPs to decorate the NC matrix. However, they have their pros and cons. Among the oxides, $RuO_2$ is the material of choice due to the highest theoretical capacitance (2000 F/g), electronic conductivity of 1 S/cm and high chemical stability.[51][144] With a coupled electron-proton transfer , reversible surface reactions of $RuO_2$ in an aqueous electrolyte can take place over a 1.2 V potential window. It is important to note that the hydrous and amorphous forms of $RuO_2$ have better capacitive behaviour than their anhydrous counterpart. Moreover, the hydrous component facilitates the permeation of protons into the bulk whereas interconnected $RuO_2$ is responsible for electronic conduction.[145] It has been reported that $RuO_2$ NPs/vertical graphene composite delivered higher gravimetric capacitance (650 F/g) than the pure vertical graphene (6 F/g) and $RuO_2$ NPs (320 F/g) in 1 M $H_2SO_4$ *vs* Ag/AgCl.[51] It is noteworthy to mention that the composite with an optimized content of NPs and NC, e.g. $RuO_2$/CNT = 6:7 wt%, is very crucial to obtain higher gravimetric capacity (953 F/g at 1 mV/s for $RuO_2$/CNT).[146] $RuO_2$ (1-2 nm in size)-RGO composite-based device in $H_2SO_4$ delivered a higher energy density of 16.7 Wh/kg at 1 KW/kg compared to the device with $Na_2SO_4$ electrolyte (15 Wh/kg).[147] However, the device operating with $Na_2SO_4$ showed better electrochemical stability (94.5% after 2000 cycles) and a higher voltage of 1.5 V than the device with $H_2SO_4$ (86.8% and 1.0 V).[147] It can be seen in many reports that the specific capacitance of the composites decreases initially in 0.5M KOH due to the low wettability, which then started to increase.[148] In order to avoid this issue, one can soak the electrode overnight or more for better wettability by electrolyte[149] and bubble $N_2$ or $O_2$ through the electrolyte for a certain amount of time before the test and then continue the flow during the test. In summary, the limited potential window in aqueous electrolyte, cost and toxicity are the major shortcomings of the use of $RuO_2$ in a small-footprint electronic device.

Mn-based oxides are another popular choice to decorate the NC surface for the improved supercapacitors performance due to high surface area, high theoretical specific capacitance (1100-1300 F/g), variety of oxidation states (MnO[150], $MnO_2$[53][151], $Mn_2O_3$[152], and $Mn_3O_4$[153]), existence of different crystallographic $MnO_2$ polymorphs (β, γ, δ, ε), abundancy, low cost, and easiness of synthesis. It should be noted that $MnO_2$ possesses higher specific capacitance than $Mn_3O_4$, and $Mn_3O_4$ is relatively insulating (~$10^{-7}$ – $10^{-8}$ S/cm). However, the enhanced specific capacitance and power characteristics of the $Mn_3O_4$ NPs/CNT arrays were attributed to the hydrophilicity of the composite and the size of NPs.[128] It is also important to note that the oxidation of $Mn_3O_4$ to $MnO_2$ took place during the charge/discharge cycles, which led to the increase in the specific capacitance after 200-300 cycles.[128] $MnO_2$-based electrode was found to operate well within the potential window of 0 to 1 V *vs* Ag/AgCl (saturated KCl) electrode, whereas RGO and N-doped RGO were operated within -0.2 to 0.8 V. As a result of synergistic effect from those two constituents ($MnO_2$ and RGO), it was found that $MnO_2$ NPs on N-doped RGO can operate within -0.2 to 1 V in 0.5 M $Na_2SO_4$.[154] Although 1M $H_2SO_4$ was found to be more effective for $Mn_2O_3$-$Mn_3O_4$-AC composite than the KOH electrolyte in terms of capacitance, rate capability and cycle life[155], dissolution in acidic electrolyte is one of the major challenges with Mn-based nanostructures.[151] To avoid the dissolution of the electrode material, an electrolyte with a lower concentration was recommended.[148] MnOx-carbon dot/graphene composite exhibited more ideal capacitive behaviour in 1M KOH and 1M $Na_2SO_4$ than that in an acidic electrolyte (1M $H_3PO_4$) as reflected from cyclic voltammogram result.[151] In spite of a higher ionic radius for $Na^+$ than $H^+$ and $K^+$, a much higher area under the cyclic voltammogram was observed in $Na_2SO_4$ than that in the other two aqueous electrolytes. The acidic or base electrolyte medium activates the carbon surface which helps in electrolyte permeation but neutral electrolyte does not do. Moreover, the neutral aqueous electrolyte has a high overpotential for hydrogen and oxygen evolution and resulted in higher voltage of the device. Other way to improve the energy density of a supercapacitor device is the use of organic or ionic electrolytes. Unlike the NC, porous akhtenskite ε-$MnO_2$ NPs/CNTf showed the comparable specific capacitance of 84 F/g at 5 mV/s vs Ag/AgCl in $PYR_{14}$TFSI ionic electrolyte compared to that in $Na_2SO_4$ (87 F/g at 5 mV/s) due to the open porosity of the hybrid composite.[53] In spite of the enhancement of total capacitance and energy density, the poor capacitance retention and rate performance of α-$MnO_2$/MWCNT compared to MWCNT[156] suggests that attention needs to be paid for improving it. Another problem related to the Mn-based materials is the use of toxic and hazardous solvents and reducing agents during the synthesis process.

$TiO_2$ could be an alternative promising metal oxide to be decorated onto the NC surface. However, $TiO_2$ suffers from a wide bandgap, poor electrical conductivity and poor electrochemical activity. Those problems were addressed by hydrogenating it or by introducing hydroxyl groups into the $TiO_2$ nanostructures. Thus the hydrogenated $TiO_2$-based, carbon composite has gained attention.[157] It is important to note that the hydrogenation temperature plays a crucial role in the final composite and hence supercapacitor properties. For example, hydrogenated -$TiO_2$ NPs/RGO hydrogenated at 400 ℃ demonstrated higher specific capacitance, better rate performance and good electrochemical stability compared to the samples hydrogenated at 300 ℃ and 500 ℃.[157]

There are also reports about the decoration of other metal oxide NPs such as ZnO,[158] $La_2O_3$[159], $Bi_2O_3$[160], CuO,[161], $Cu_2O$[162] on NC. However, they have some limitations. Zn-based composite mostly exhibited low specific capacitance[163] and limited potential window. Although an enhancement in the gravimetric capacitance (611 F/g at 1 A/g in 1M $H_2SO_4$ *vs* Ag/AgCl), charge-transfer kinetics and other parameters have been observed, non-uniform distribution of Cu-oxides on NC matrix,[162] and limited potential window of 0.8 V are the major challenges with composite electrodes based on Cu-oxides.

### 5.5. Metal NPs *vs* metal oxide NPs

As well-known, the measured parameters of supercapacitors electrode materials are quite comparable between the metal NPs/NC composite and metal oxide NPs/NC composite. Thus, there is a quest for *which type of NPs should be chosen to decorate on NC to obtain high-storage performance*. Let us discuss the pros and cons of both composites. Relatively higher gravimetric capacitance in Cu NPs-carbon aerogel than the Ag counterpart was attributed to the surface area and porosity[164] and also to the additional contribution coming from the native oxide surface of Cu NPs. But Cu-based electrode materials show higher leakage current. It has been shown that a higher amount of native oxide for Ni and Cu NPs decorated vertical graphene and a comparatively lower amount for Au and Ag leads to the highest gravimetric capacitance for Ni and Cu NPs decorated vertical graphene. Actually, those NPs (Cu, Ni, Ti etc.) are highly reactive towards the adsorption of atmospheric oxygen which leads to oxide formation on the surface.[165] Moreover, oxidizing the Cu NPs was recommended to stabilize the electrochemical performances.[141] On the other hand, to avoid the oxide formation on the metal NPs' surface, the NPs were embedded into the NCs rather than the decorating on it.[166] This is because, for example, $Co_3O_4$ undergoes a phase transformation while cycled under an electrolyte. Co NPs embedded carbon nanorods exhibited the higher surface area, micro- and mesopore volume and electrical conductivity compared to the pristine carbon nanorods. As a result, the enhanced gravimetric capacitance with excellent charge transfer kinetics and charge transfer rate has been observed in the carbon nanorod electrode after Co NPs embedment.[166] We want to highlight that the evidence of N-containing functional groups in Co-NPs embedded carbon nanorod from X-ray spectra[166] also contributes pseudocapacitance[8], but this interesting effect was not discussed. It can be said that the final choice of the type of NPs between the metal and metal oxide forms should be made guided by the electrode performance testing results.

### 5.6. Non-oxide NPs/NC

Compared to traditional oxides, the research interest has been directed towards nitrides, sulphides, etc. owing to their higher electrical conductivity, rich redox-active species, and electrochemical stability.[167][168]. Impressive charge-storage performance with a gravimetric capacitance of 500 F/g at 0.5 A/g and 95% capacitance retention after 1000 cycles was obtained from the $SnS_2$/RGO composite in 2M $Na_2SO_4$.[25] The charge-storage performance is much higher than the reports on $SnO_2$-based composites due to the large interlayer spacing of 5.89 Å, which accommodates $Na^+$ ion very comfortably. The choice of 2M $Na_2SO_4$ was due to the higher ionic conductivity compared to the standard use of 1M $Na_2SO_4$.[25] Ion intercalation was decreased and ionic transportation within the electrode materials was boosted with increasing electrolyte concentration.[143] However, unavoidable oxide co-exists with the sulphides and also aggregation of NPs occurs.[169] Although ZnS/RGO showed promising supercapacitive characteristics, the operation window was found to be limited within -0.2 to 0.2 V in KOH vs Ag/AgCl.[170] The decoration of carbide NPs on the NCs surface is also promising.[171] $Mo_2C$/CNT showed a wider CV curve and hence charge-storage capacity, whereas $W_2C$/CNT composite showed better Coulombic efficiency and almost rectangular cyclic voltammogram.[171] With this example, we would like to emphasize that the rate performance and Coulombic efficiency are equally important besides the specific capacitance.

**5.7. NPs on doped NCs**

Later, the research has been extended in decorating the NPs onto the doped/functionalized NC to improve the performance further (Figure 11B).[27,134,172,173] Much better distribution of Ag NPs, reduced average size of NPs, higher electrical conductivity, open pores, and less agglomeration are evidenced when they are coated onto doped RGO compared to their pristine counterpart. As a result, Ag NPs decorated N-doped RGO exhibited higher gravimetric capacitance of 729.2 F/g at 1 A/g compared to that of RGO (293.1 F/g) and Ag-decorated RGO (510.3 F/g) in 3-electrode configuration.[172] Among the dopants (B, N, B/N), $V_2O_5$/N-doped RGO exhibited the highest gravimetric capacitance of 1032.6 F/g in KOH electrolyte *vs* Ag/AgCl.[174] This result also indicates that the NPs decorated on co-doped NCs, where dopants are with opposite charge carriers, may not be effective as well. In spite of the higher capacitance obtained from the $V_2O_5$/doped-RGO, poor rate performance and limited operating potential window (-0.3 to 0.3 V) are the major shortcomings of V-based electrode materials. Apart from the enhancement in charge-storage kinetics, N/P co-doped hierarchically porous carbon framework in-situ armored $Mn_3O_4$ NPs was operated in the potential range of -1.0 to 0.3 V in KOH electrolyte *vs* Hg/HgO.[175] The higher potential window of this composite in the negative region with respect to the reference electrode is basically due to the P-doping. P-doping into NCs shifts the thermodynamic equilibrium of water to the higher side.[8][176] This result indicates that NPs with P-doped NC can be used as negative electrode for asymmetric supercapacitor and doping can enhance the voltage of device, and hence the energy density dramatically. The details of the asymmetric supercapacitor are discussed later.

**5.8. Multi-component composites**

To obtain the high storage performance, introducing multi-metallic NPs into the bare and doped NC matrix is promising (Figure 11A and table 4).[177][178] However, the rate performance of $Co_3O_4$@$RuO_2$/RGO is relatively poor compared to $Co_3O_4$/N-doped RGO. The 35.5% capacitance loss was observed for the $Mn_3O_4$-$Fe_2O_3$/$Fe_3O_4$@RGO hybrid after only 1000 charge/discharge cycles at 0.5 mV/s in 1M KOH. The potential window of this hybrid is also limited within -0.1 to 0.6V *vs*. Ag/AgCl.[178] The poor rate performance of the metal oxide NP decorated NC has been tackled by doping the composite with metals, such as Ag into the $Mn_3O_4$ NP/AC [26]. Doping Ag/N in $TiO_2$ NPs and then decorating onto graphene was also found to stabilize the output voltage of solid-state supercapacitor device in addition to the impressive charge-storage behaviour. However, the lower capacitance retention of Ag/N-doped $TiO_2$ NPs/graphene has been observed and attributed to the amino bond ($C\text{-}NH^+$) formation, which limits the charge extraction.[179] Thus, an obvious question arises: *what is the need of using multicomponent when one can achieve a higher performance from simple metal or metal oxide NPs composites with NCs?* For example, $NiSe_2$/N-doped RGO showed gravimetric capacitance of 99.03 F/g only at 1 mV/s *vs* Ag/AgCl [180], poor rate performance (12* F/g at 100 mV/s) and low operating potential window (-0.25 to 0.25 V).[180] In the multicomponent composite, NCs contributes electric double layer capacitance, serves as a mechanical and conducting platform and protects the NPs in NPs-coated NC from direct interaction with electrolytes; dopant-based functional groups provide pseudocapacitance, and the metal-based component is relying on the intercalation/deintercalation of protons or alkali metal cations and adsorption of anions on the surface. Thus, the total capacitance of multicomponent composite is a combination of EDL capacitance, quantum capacitance, and pseudocapacitance where electric double layer capacitance and quantum capacitance are in series and pseudocapacitance is in parallel with them. Increasing pseudocapacitance will increase the total capacitance, whereas increasing carbon-content improves the rate capacitance. Thus, a significant contribution from each component is very crucial and has to be balanced to obtain high-performance electrode materials.

*Table 4: Physico-chemical changes and supercapacitive performances of the composite after NC incorporation (symbol with an asterisk represents the data estimated either from the plot using WebPlotDigitizer copyright 2010-2020 Ankit Rohatgi or available data from the cited reference)*

| Ref. | Nanocarbons (Synthesis method) | Synthesis of composite or NP decoration technique | Avg. NP size (nm), NP wt% or NP/NC ratio, Surface area ($m^2/g$) | Potential window/voltage, electrolyte, mass loading | specific capacitance (F/g) with capacitance retention | Cap. retention (%), cycles | ESR, $R_{ct}$ (Ω) |
|---|---|---|---|---|---|---|---|
| [61] | RGO | Hummer's method | -, -, - | -0.2 to 0.8 V, 6M KOH, 0.008 g | 50 at 5 mV/s, 50, 0.5 V/s | - | - |
| | Au-RGO | chemical route- stirring + heating | 11-15, -, - | | 100 at 5 mV/s, 102, 0.5 V/s | - | - |
| | Au-rGO | γ-radiation | -, -, - | | 500 at 5 mV/s, 77.46, 10 A/g | ~90%, 50 mA, 600 | - |
| [172] | RGO | Hydrothermal + freeze-drying | -, -, 217.1 | -1.2 to 0V in 6M KOH vs saturated calomel electrode, 5 $mg/cm^2$ | 293.1 at 1 A/g, 47* at 20 A/g | - | 1.21, 0.46 |
| | Ag-RGO | | -, -, 248.5 | | 510.3 at 1 A/g, 51* at 20 A/g | - | 0.73, 0.32 |
| | Ag@N-RGO | | -, -, 269.3 | | 729.2 at 1 A/g, 61* at 20 A/g | - | 0.65, 0.26 |
| | Ag@NS-RGO | | 25, 34.27%, 294.8 | | 923.3 at 1 A/g, 67%* at 20 A/g ( | 91.2%, 10 A/g, 10k | 0.42, 0.18 |
| [166] | Carbon nanorods (CNR) | Electrospinning | -, -, 27.2 | -0.3 to 1V, 0.5 M $H_2SO_4$ in $N_2$ vs Ag/AgCl, 13 $mg/cm^2$ | 8 at 2 mV/s, 25% at 0.1 V/s | - | - |
| | 2.5% Co-CNR | Electrospinning + carbonization | 20-100, -, 30.6 | | 101* at 2 mV/s, 56.6%* at 0.1 V/s | 116%*, 50 mV/s, 5k | - |
| | 5% Co-CNR | | 20-100, -, 31.6 | | 137* at 2 mV/s, 45.61%* at 0.1 V/s | 167*, 50 mV/s, 5k | - |
| | 10% Co-CNR | | 20-100, -, 476.1 | | 146 at 2 mV/s, 73.9 at 0.1 V/s | 139*, 50 mV/s, 5k | - |
| [139] | CNF | Electrospinning + carbonization | 360 (diam.) | 0-1 V, 1 M KOH, - | 102 at 0.5 A/g, 88.2%* at 5 A/g | 68.6%*, 1 A/g, 10k | -, 8.28 |
| | 20CoNi-CNF | Electrospinning + carbonization | 333 (diam.), 20% | | 145 at 0.5 A/g, 89.6%* at 5 A/g | 85.3%, 1 A/g, 10k | -, 6.88 |
| | 40CoNi-CNF | | 280 (diam.), 40% | | 137 at 0.5 A/g, 80.3%* at 5 A/g | 64.6%*, 1 A/g, 10k | -, 16.1 |
| [51] | Vertical graphene (VG) | Plasma- chemical vapor deposition | - | 0-0.9 V vs Ag/AgCl, 1 M $H_2SO_4$, 3.5 μg/$cm^2$ | 6 at 0.5 mV/s | - | - |
| | $RuO_2$ thin film | Sputtering + electrochemical oxidation in $H_2SO_4$ | 20 nm | | 320 at 0.5 mV/s | - | - |
| | $RuO_2$ NP/VG | | 1-2 nm | | 648 at 0.5 mV/s | 70%, 0.5 mV/s, 5k | - |
| [181] | $Fe_3O_4$ nanospheres | hydrothermal | -, -, - | 0-1V, 1 M $Na_2SO_3$ | 153.7 at 5 mV/s, 23%* at 0.2 V/s | 71.2%, 1 A/g, 200 | - |
| | C-dot decorated $Fe_3O_4$ | ultrasonication | -, -, - | | 203.4 at 5 mV/s, 42.5%* at 0.2 V/s | 86%, 1 A/g, 200 | - |
| [182] | $C/Co_3O_4$-600 | Pyrolysis + oxidation. 600, 650, 750 and 800 in the sample name represents the pyrolysis temperature. | 2.3, 43, 332 | 0-0.8 V, 2M KOH, 2E, 19-25 mg/cm$^2$ | 15* at 0.1 A/g, 4% at 10 A/g | - | 0.7-1.4, - |
| | $C/Co_3O_4$-650 | | 2.5, 57, 402 | | 30* at 0.1 A/g, 0.2% at 10 A/g | - | - |
| | $C/Co_3O_4$-750 | | 4.3, 59, 413 | | 56* at 0.1 A/g, 48.7% at 10 A/g | 82% 0.5 A/g, 10k | - |
| | $C/Co_3O_4$-800 | | 7.4, 52, 467 | | 60 at 0.1 A/g, 17.7% at 10 A/g | - | - |
| [183] | Mesoporous carbon nano-sphere (MCN) | Hydrothermal + carbonization + template removal process | -, -, 616 | -1 to 0 V, 6M KOH | 170 at 1 A/g, 79.4%*, 20 A/g | - | 0.32, - |
| | NiO(6.12%)/MCN | | -, 6.12%, 648 | | 236 at 1 A/g, 85%*, 20 A/g | - | 0.35, - |
| | NiO(7.20%)/MCN | | 280 nm, 7.20%, 714 | | 406 at 1 A/g, 77%*, 20 A/g | 91%, 3 A/g, 10k | - |
| | NiO(8.42%)/MCN | | -, 8.42%, 651 | | 284 at 1 A/g, 74%*, 20 A/g | - | - |
| [49] | N-carbon | hydrothermal and post calcination treatment | -, -, 685 | 0 to 1 V vs Ag/AgCl, 1 M $Na_2SO_4$, - | 610* at 5 mV/s, 29*at 0.1 V/s | - | - |
| | $MnO_2$/N-carbon | | 15-20,- , 645 | | 996.21 at 5 mV/s, 39% at 0.1 V/s | - | - |
| [53] | CNTf, f: fiber | CVD spinning method | -, -, 270 | 0 to 0.9* V vs Ag/AgCl, 1 M $Na_2SO_4$, 0.245–0.380 mg/cm$^2$ (of CNTf) | 23* at 5 mV/s, 78%* at 5 mV/s | - | - |
| | $MnO_2$ NPs/CNTf-20 | Electrochemical deposition. 20 and 60 in the sample name represents the deposition time in min. | 150, 0.37, - | | 59* at 5 mV/s, 56%* at 0.2 V/s | - | - |
| | $MnO_2$ NPs/CNTf-60 | | 200, 0.98, 120 | | 81* at 5 mV/s, 52%* at 0.2 V/s | - | - |
| | | | | -1.5 to 1.5 V vs Ag/Ag$^+$, PYR14TFSI | 87 at 5 mV/s, 49.4%* at 0.2 V/s | - | 53 Ω/cm$^2$, - |
| [184] | Ag-MWCNT | Mixing the powder, NPs and surfactant | -, -, - | 0 to 0.8 V vs AgAgCl, 4M LiCl, 0.738 mg/cm$^2$ | 5.2* F/cm$^3$ at 10 mV/s, 25%* at 1 V/s | - | 27.3, 188.5 |
| | $MnO_2$-Ag-MWCNT | | 60-90, -, - | 0 to 0.8 V vs Ag/AgCl, 4 M LiCl, 1.218 mg/cm$^2$ | 30.5 F/cm$^3$ at 10 mV/s, 13.4%* at 1 V/s | - | 8.1, 85.6 |
| [63] | N-rGO | Hummer's method + vacuum-assisted filtration + hydrazination | -, -, - | 0 to 0.8 V *vs* Ag/AgCl, 0.5 M $Na_2SO_4$ | 5.1 at 20 mV/s | - | - |
| | $Fe_3O_4$/N-rGO | Mixing + filtration | 5, 1:8, - | -0.8 to 0.8 V *vs* Ag/AgCl, 0.5 M $Na_2SO_4$ | 166 mF/cm$^2$ at 2 mV/s | - | - |
| | $Fe_3O_4$/N-rGO$_{L1}$ | Sequential filtration of GO and $Fe_3O_4$ | 5, 1:8, - | | 233* mF/cm$^2$ (112 F/g) 2 mV/s, 15.9%* at 100 mV/s | 79 at 20 mV/s, 10 k (2E) | - |
| | $Fe_3O_4$/ N-rGO$_{L2}$ | | 5, 1:16, - | | 162* mF/cm$^2$ (81 F/g) at 2 mV/s, 12.4%* at 0.1 V/s | - | - |
| [185] | AC | - | -, -, 2356.6 | -0.9 – 0.1 V vs Ag/AgCl, 1 M $Na_2SO_4$, - | 99.6 at 2 A/g | - | - |
| | $Fe_2O_3$-oxidized AC | Heating the mixed solution + hydrazination + drying | ~30.6, -, 1965.1 | | 129.6 at 2 A/g | 94% 10 mA/cm$^2$, 2k | - |
| | $Fe_2O_3$ & $Fe_3O_4$-oxidized AC | | ~30.6, -, 1953.5 | | 142.1 at 2 A/g | - | - |
| | $Fe_3O_4$- oxidized AC | | ~30.6, -, 1975.5 | | 168.5 at 2 A/g, 80.2%* at 10 A/g | 93% 10 mA/cm$^2$, 2k | - |
| [186] | $CuCo_2S_4$ NPs | Solvothermal + Freeze drying | 21, -, 28.41 | -0.2 to 0.4 V, KOH vs Hg/HgO, 3.5 mg | - | - | -, - |
| | $CuCo_2S_4$/RGO aerogel | | 21, -, 120.35 | | 668 at 1A/g, 72% at 20 A/g | 84.5% at 20 A/g, 8k | -, - |
| [177] | N-RGO | GO in $NH_4OH$/urea at 95 ℃/12h | -, -, - | 0-0.9 V vs Ag/AgCl, 2M KOH, - | 100-150 at 0.5 A/g | - | -, - |
| | $Co_3O_4$/N-RGO | Hydrothermal + calcination | 70-120, -, - | | 456 at 0.5 A/g, 77.8%* at 3 A/g | 96.8% at 3 A/g, 5k | 8.4, 33.1 |
| | Core-shell $Co_3O_4$@$RuO_2$/N-RGO | | 10-20, -, - | | 472 at 0.5 A/g, 28.2%* at 3 A/g | 97.2% at 3 A/g, 5k | 8.84, 3.03 |
| [187] | $Mn_3O_4$ tetragonal bipyramid (TB) | Stirred at 100 ℃ + refluxed at 100 ℃ | 20, -, 1.85 | 0 to 0.8 V in 1 mol/L $Na_2SO_4$ *vs* SCE, 5 mg/cm$^2$ | 77 at 0.5 A/g, 65% at 10 A/g | - | 0.9, 4.1 |
| | N-hierarchically porous carbon (NHPC) | rapid microwave carbonization | -, -, 1170 | | 117 at 0.5 A/g, 79% at 10 A/g | - | 0.9, 1.2 |
| | $Mn_3O_4$ TB/NHPC | ultrasonic assembly | -, 32%, 334 | | 366 at 0.5 A/g, 76.2% at 10 A/g | - | -, - |
| [188] | Highly N-RGO (HNG) | Reflux the mixed solutions + Hydrothermal | -, -, - | 0 to 0.8 V in 3M KOH vs SCE, 2 mg/cm$^2$ | 136* at 0.1 A/g, 88.3% at 1 A/g | 90.2, 1k | -, - |
| | $Co_3O_4$ NPs | | -, -, - | | 222* at 0.1 A/g, 73% at 1 A/g | 57.7 | -, - |
| | $Co_3O_4$/HNG-4 | | 80-100, 0.25, 312.5 | | 355.9 at 0.1 A/g, 88.7% at 1 A/g | 84.5 | 1.2, 1.0 |
| | $Co_3O_4$/HNG -8 | | -, 0.125, - | | 280* at 0.1 A/g, 78.6% at 1 A/g | - | -, - |
| [174] | $V_2O_5$ | Solvothermal | 60-110, -, - | -0.3 to 0.3 V *vs* Ag/AgCl, 6M KOH, - | 601.5 at 1 mV/s, 0.06%* at 0.1 V/s | - | -, - |
| | $V_2O_5$/RGO | | -, -, - | | 932.7 at 1 mV/s, 0.04%* at 0.1 V/s | - | -, - |
| | $V_2O_5$/B-RGO | | 50-120, -, - | | 818.9 at 1 mV/s, 0.04%* at 0.1 V/s | - | |
| | $V_2O_5$/N-RGO | | 50-120, 90%, 5.472 | | 1032 at 1 mV/s, 0.04%* at 0.1 V/s | 97.78%, 0.1 V/s, 250 | 3.3, - |
| | $V_2O_5$/BN-RGO | | 50-120 nm, -, - | | 634.9 at 1 mV/s, 0.05%* at 0.1 V/s | - | -, - |
| [189] | N-doped carbons | calcination of melamine & glucose | -, -, - | -1.2 to 0 V *vs* Hg/HgO, 2M KOH, 2.5 mg | 23 at 1 A/g, - | - | -, - |
| | 3-Ni/N-Carbons | in-situ: stirring at 80 ℃ + calcination in Nitrogen gas | -, -, - | | 27 at 1 A/g, 37%* at 100 A/g | 89%, 10 A7g, 5k | -, - |
| | VN/ N-Carbons-7 | | -, -, - | | 112 at 1 A/g, 22.3%* at 100 A/g | 77.5%, 10 A/g, 5k | -, - |
| | 3-Ni-VN/ N-Carbons-7 | | -, 2.91%, 58 | | 236 at 1 A/g, 68.6%*at 100 A/g | 85.8%, 10 A/g, 5k | 0.69, - |
| [25] | RGO | modified Hummer's method + chemical reduction | -, -, - | 0 to 1 V, 2M $Na_2SO_4$, - | 34.9 at 0.5 A/g, - | - | 68 Ω/cm$^2$, 0.38 Ω/cm$^2$ |
| | $SnS_2$ nanosheets | Hydrothermal | 120, -,- | | 93.8 at 0.5 A/g, - | 80% at 2 A/g, 100 | 634 Ω/cm$^2$, 3.11 Ω/cm$^2$ |
| | $SnS_2$/RGO | in-situ: modified Hummer's method | 145-155, 50%, - | | 500 at 0.5 A/g, 45%*at 3 A/g | 95% at 2 A/g, 1k | 256 Ω/cm$^2$, 1.81 Ω/cm$^2$ |
| [190] | GONR | longitudinal unzipping of CNT | -, -, - | 0 to 1 V *vs* Ag/AgCl, 0.5 M $Na_2SO_4$, - | 50* at 50 mV/s, - | - | - |
| | $MnO_2$ | Stirring the microemulsion solution followed by drying at oven | -, -, - | | 80* at 50 mV/s, - | - | 6.5, 2.3 |
| | $RuO_2$ | | -, -, - | | 130* at 50 mV/s, - | - | 5.7, 1.3 |
| | $MnO_2$@GNR | | -, -, - | | 120* at 50 mV/s, - | - | 4.7, 1.8 |
| | $RuO_2$@GNR | | -, -, - | | 200* at 50 mV/s, - | - | 3.5, 1 |
| | $MnO_2$-$RuO_2$(8:2)@GNR | | 10, -, - | | 230* at 50 mV/s, 69.6%* at 0.1 V/s | - | 0.4, 2.7 |

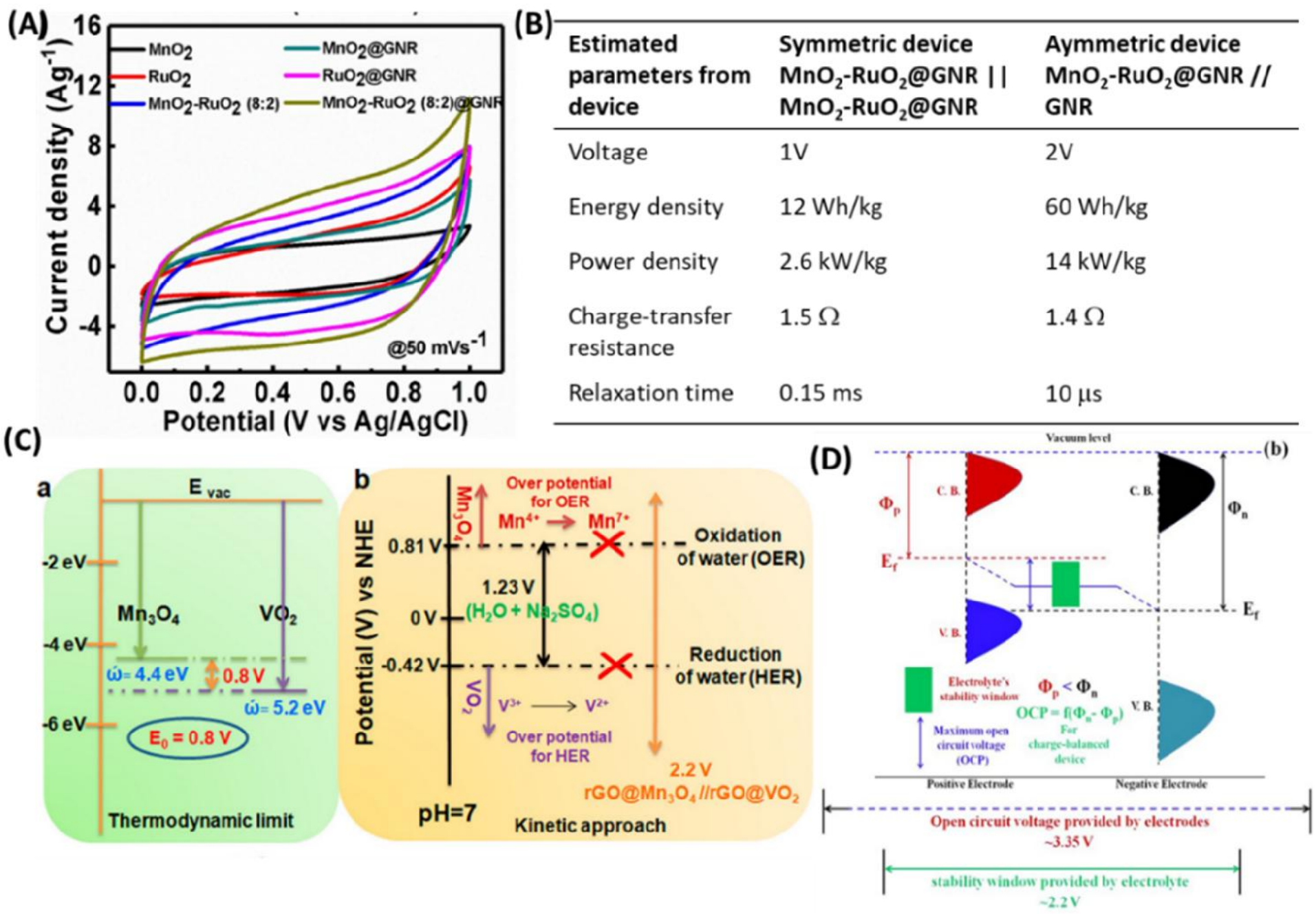

*Figure 11: (A) Cyclic voltammogram of metal oxide NPs and NPs/GO nanoribbons electrode in $Na_2SO_4$ vs Ag/AgCl. Reproduced from Ref. [190] © 2017, with the permission taken from Elsevier. (B) Table of comparison of symmetric and asymmetric device based on $MnO_2$-$RuO_2$@GNR, GNR- graphene oxide nanoribbons.[190] (C) schematic of work function difference and kinetic approach of positive and negative electrode in aqueous medium. Reproduced from Ref. [191] ©2018, with permission from American Chemical Society. (D) Energy band diagram for achieving widened operating voltage window with aqueous electrolyte. Adopted from Ref. [192], © 2017, Springer Nature publishers.*

## 5.9. Anode for asymmetric supercapacitor

Since the energy density is proportional to the square of voltage, widening the potential window is more effective. Incorporation of RGO in the $CeO_2$/RGO composite resulted in a widened operating voltage (-0.8 to 0.6 V) along with the enhancement in gravimetric capacitance, leading to increased energy density.[193] The voltage of device can be increased also by using proper electrolytes. Aqueous electrolytes are limited below the thermodynamic stability of water (1.23 V), while organic and ionic electrolytes are capable of providing voltages higher than 2 V. Aqueous electrolytes are safe in use, cheap, and easy to handle. Thus, the concept of asymmetric supercapacitor device using an aqueous electrolyte became popular, where one can achieve a voltage higher than 2 V by choosing proper electrode pairs (Figure 11B-D). Eventually, asymmetric supercapacitor device made of Ag-$Mn_3O_4$ NPs/activated carbons and activated carbons provides higher gravimetric capacitance (180 F/g at 10 A/g), better cycle life (96% after 1000 CD cycles), higher voltage (1.8 V) and hence higher energy density (81 Wh/kg at 4486 W/kg) in 1M $Na_2SO_4$ compared to the symmetric counterpart (16 F/g at 10 A/g, 0.8 V, 86% after 1000 cycles, 8.1 Wh/kg at 222.6 W/kg).[26] To obtain high-storage performances, the choice of electrode pair is essential. In addition to high specific capacitance, good rate performance, and excellent cycle life, the electrode pair should possess a high work function difference and a larger operating window in electrolyte during the charge/discharge process. In most of the cases, bare NC were used as negative electrodes. However, the problem with pure NC as a negative electrode is the charge balance with the positive electrode, where higher mass loading is desirable. Unfortunately, higher mass loading degrades the electrochemical performance with time. Therefore, this section is dedicated to the materials studied as a negative electrode.

Fe-oxides based composites received the most attention as a negative electrode owing to their large potential window in addition to high theoretical capacity (2606 F/g for $Fe_3O_4$ and 3625 F/g for α-$Fe_2O_3$), availability of many phases, and abundancy.[194][195] A widened potential window of Fe-based composites in aqueous electrolyte is due to the space-charge limited capacitance for an electrode/electrolyte interface.[196] Apart from the pseudocapacitive contribution, Fe-oxide supported CNF enhanced the electric double-layer capacitance (3.6 F/g, whereas electric double-layer capacitance of pristine CNF was 0.74 F/g). The Fe-oxide NPs in this structure consisted of α-$Fe_2O_3$ (hematite), γ-$Fe_2O_3$ (maghemite), and $Fe_3O_4$ (magnetite) phases.[197] The main challenge with the Fe-based oxides is the synthesis of single-phase oxide without the co-existence of another phase. In this regard, the reducing agent hydrazine was optimized in a very controlled way to form those Fe-oxide NPs on the oxygen functionalized activated carbons (O-AC) without coexistence of other phases.[185] The obtained result reveals that co-existed Fe-oxide/O-AC showed the higher capacitance than the $Fe_2O_3$/O-AC, but $Fe_3O_4$/O-AC exhibited the highest specific capacitance in 1M $Na_2SO_3$ electrolyte under the 3-electrode test.[185] One should also note that the NPs were larger and the carbon content was reduced when $Fe_3O_4$/C was transformed into α-$Fe_2O_3$/NCs after calcination, which also had a significant impact on the charge storage properties. Thus, comparing the gravimetric capacitance only between those two composites may not be convincing. Indeed, among Fe-oxides, $Fe_3O_4$ possesses the higher electrical conductivity (~ $10^3$ S/m) than the insulating $Fe_2O_3$ (~$10^{-10}$ S/m) at room temperature, which has a significant influence on the charge-transfer kinetics. The gravimetric capacitance and hence energy density of $Fe_3O_4$ NPs/NCs can be further enhanced by introducing an external magnetic field.[24] The improved energy density in the presence of a magnetic field was attributed to the Lorentz force induced electrolyte convection in the bulk electrolyte, which pushes the electrolyte ions deeper inside the electrode.[24] However, the understanding of the magnetic field induced charge-storage mechanism is the subject of further research aimed at considering the structural changes of the electrode and bulk electrolyte. All of these results ensure that $Fe_3O_4$ NPs/NC composite can be a better choice as a negative electrode. It has been reported that KOH electrolyte suits better for magnetic NPs/RGO in terms of gravimetric capacitance and cycle life than the $Na_2SO_4$ due to the higher ionic mobility of $K^+$ and $OH^-$ ions.[198] $Fe_3O_4$-graphene nanocomposites/few-layered graphene were explored in various electrolyte media namely, $Li_2SO_4$, $Na_2SO_4$, $Cs_2SO_4$, $Rb_2SO_4$ and $MgSO_4$ with 1M concentration (Figure 22A-C).[199] The nanocomposite showed the maximum operation potential of 1.4 V in 1M $Cs_2SO_4$, whereas the maximum power density and the highest gravimetric capacitance were obtained in $RbSO_4$ and excellent cycle life in both electrolyte systems.

Owing to the ability of operating in the negative potential range down to -1.35 V in an aqueous medium along with its high theoretical capacitance of 1380 F/g, vanadium nitride (VN) is another promising negative electrode.[167] It has been seen that VN NPs anchored on N-doped carbon nanosheets can be operated between -1.2 and 0 V in 2M KOH *vs.* SCE (Figure 12).[200] The ratio of carbon content and VN makes an impact on the electrochemical properties of the composite, as reflected from CV.[201] For instance, a near rectangular CV was obtained from the composite when the carbon-content was larger, while CV with a pair of weak redox peaks was evidenced when VN-content was predominant in the structure.[201] The pair of weak redox peaks in the CV of VN/N-doped carbon nanosheets indicates the reversible redox reactions occurring in addition to the adsorption/desorption process and EDLC.[201][189] However, capacitance fading over prolonged cycle life is the major shortcoming for the VN/NC composite due to morphological changes, structural collapse, partial oxidation, and dissolution in the KOH electrolyte.[201] We also stress that there is always some fraction of oxide on the surface, which is also responsible for the poor cycle life of the composite in the alkaline medium. Capacitance fading was found to improve after incorporating Ni NPs into the VN/N-doped NC.[189] Basically, Ni NPs improved the crystallinity of VN and contributed to the electrical contact among the VN NPs.[189]

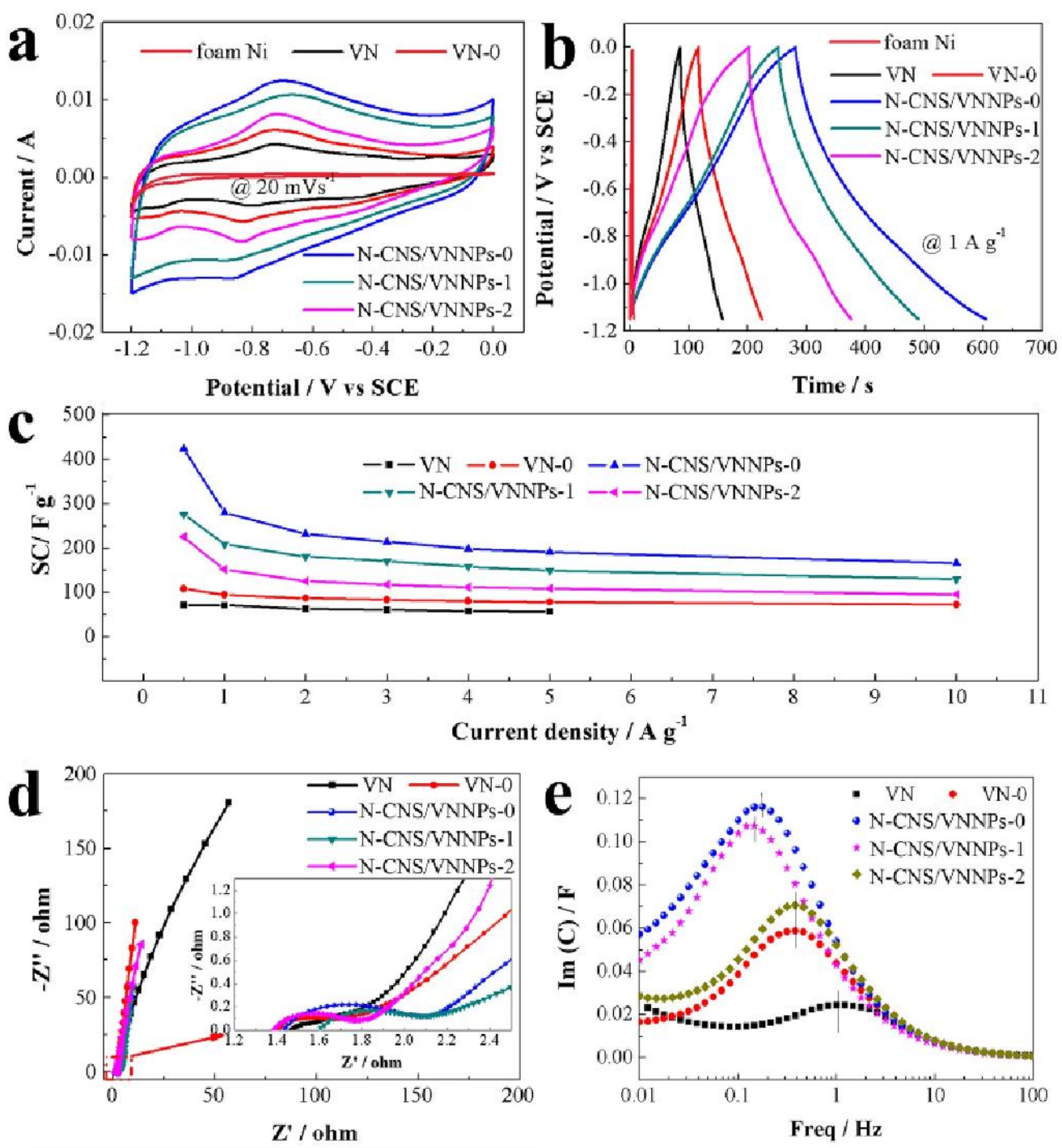

*Figure 12: Electrochemical performances of VN NPs/N-doped carbon nanosheets. In the figure, 0,1, and 2 represent the pH value during the synthesis. VN and VN-0 are prepared by annealing $NH_4VO_3$ and $V_2O_5$ dry xerogel in the same condition, respectively. Adopted from Ref. [200] © The Author(s) 2018, Springer Nature publishers.*

It is also reported that $La_2O_3$/RGO can operate in the negative window of -1.0 to 0 V in 3M KOH *vs.* reference electrode. Other NPs/NC composites studied as a negative electrode are $La_2O_3$/RGO[159], $Mo_xN$/N-doped CNF[202] etc. Although 95% charge-discharge efficiency has been evidenced for $La_2O_3$/RGO after 300 cycles,[159] one has to pay attention to improve the cycle life and specific capacitance of those nanostructures.

### 5.10. Clarification on Ni, Co, Ce -like composite

Materials composed of Ni, Co and Ce-like have been extensively studied in the last decades in spite of their battery-like features as evidenced from both cyclic voltammogram and charge-discharge profile.[133] For example, NiO coated vertical graphene[203], $Co_3O_4$ hollow NPs-CNFs hybrid films[204], $Co_3O_4$ NPs (< 3 nm) decorated phosphorous and nitrogen-doped carbon matrices[205], $CeO_2$/RGO composite[206][207] do not meet the standard supercapacitor features although authors claimed it. Specifically, neither the cyclic voltammogram nor the charge-discharge profile of those NPs/NC composite have a linear profile. Instead, battery-like features have been observed and the capacity should be expressed in this context as C/g instead of F/g.[208][209] This review has not examined this kind of reports of NPs decorated NC for supercapacitor application where the electrochemical features do not follow the basic criteria.

However, there are also reports like NiO NPs/mesoporous carbon nanosphere[183], $NiSe_2$/N-doped RGO,[180] $Co_3O_4$ NPs[188], $CuCo_2S_4$ NPs/RGO aerogel [186], $CeO_2$/RGO[193], $CeO_2/Ce_2O_3$ QD decorated RGO[210] which showed the ideal supercapacitor features. One has to notice here: (i) the range of operation and (ii) the electrolyte used for those studies (Figure 13). It is important to note that NiO NPs/mesoporous carbon nanosphere[183] in 1M KOH was operated within -1.0 to 0 V whereas the operating voltage for NiO coated vertical graphenes[203] was -0.2 to 0.6 V. Thanks to the research work carried out in Ref. [211] for the clarification that the battery-like features of Ni NPs@CNT as positive electrode are obvious in the range of -0.2 to 0.6 V whereas ideal pseudocapacitive behaviour was observed in the same material when explored as a negative electrode in the range of -1.2 to 0 V vs Hg/HgO in 3M KOH (Figure 13). This result suggests that Ni-based electrode has to be operated in negative range of potential window to obtain ideal pseudocapacitive features from it (Figure 13C). Likewise, $CeO_2$/RGO with NPs size of 20-30 nm showed a quite rectangular cyclic voltammogram within -0.4 to 0.4 V in 0.5 M $Na_2SO_4$ *vs* Ag/AgCl[193], whereas $CeO_2$/RGO composite[206] was operated in the range of 0 to 0.6 V in 3M KOH electrolyte. Secondly, Ni NPs decorated RGO showed quasi-rectangular cyclic voltammogram in the range of -0.2 to 0.5 V in 50 mM phosphate buffer saline vs Ag/AgCl.[212] Ni NPs@CNT exhibited rectangular CV in a neutral electrolyte ($K_2SO_4$), indicating electric double layer features of charge-storage (Figure 13D).[211] As it can be seen from many reports, the operating potential window of Ni, Co and Ce-based composite is very low (maximum 0.6 V). Unless one obtains a very high specific capacitance from these materials, implementing them as supercapacitor electrodes may not be useful.

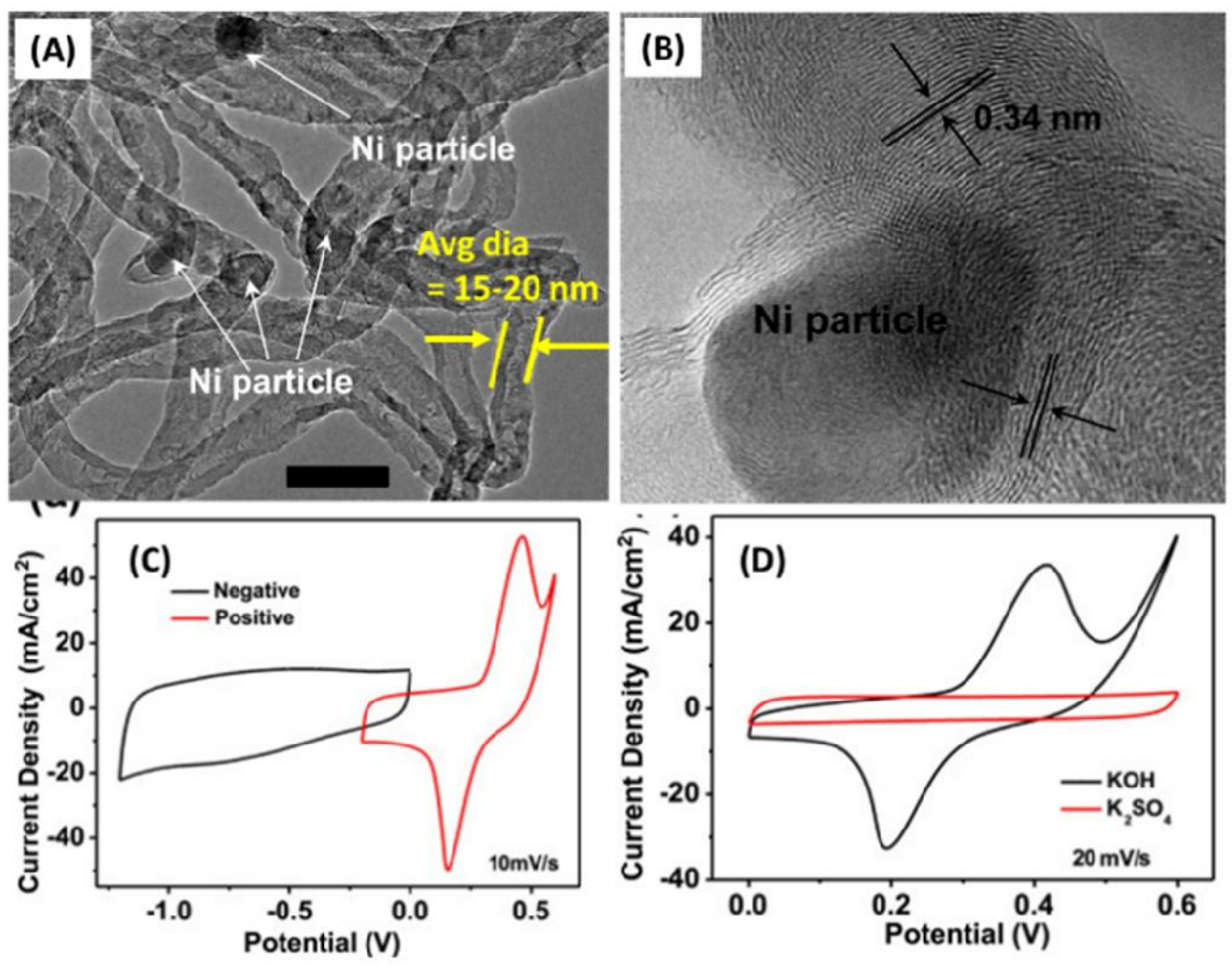

*Figure 13: (A-B) TEM image of Ni@CNT. Since the scale is not labelled in (A), liability has been taken to indicate the average diameter of CNT. CV curves of Ni@CNT (C) in positive and negative potential windows and (D) in KOH and $K_2SO_4$ electrolyte. Reproduced from Ref. [211], © 2017, with permission from Elsevier B.V.*

**5.11. Selection of NCs**

While the research was mainly focused on the synthesis of NPs decorated NC with a variety of NP structures (quantum dots, hollow spheres, yolk-shell, etc.) and oxidation states, a question arises on the suitable choice of NC as a mechanical and conducting backbone. To address that question, a few case studies are discussed here to highlight the role of various NC platforms in the composite with NPs on the electrochemical performances of supercapacitor (Figure 14). Amongst onion-like carbon , multi-wall CNT, RGO, and AC, onion-like carbon stood out as the best NC platform for $Mn_3O_4$ NPs decoration providing excellent electrochemical performances in terms of gravimetric capacitance, rate capability and impedance (Figure 14A-D). Although the onion-like carbon possesses the lower electrical conductivity and surface area compared to other NC, the best charge-performance of $Mn_3O_4$/ onion-like carbon was attributed to the combined intrinsic properties of the composite such as small NPs size, high surface area, broad particle size distribution and higher electrical conductivity.[153] In contrast to the relatively better electrically conducting $Mn_3O_4$/RGO, the higher charge-storage performance of $Mn_3O_4$/GO was attributed to the higher surface area and higher amount of O-functional groups.[213] In another study,[214] $MnO_2$ NPs were decorated on RGO, CNTs and carbon black, and synergistic effects were obtained in $MnO_2$/RGO. Charge-transfer kinetics is shown in Figure 15G, which reveals poor contact and inefficient charge transfer route by 0D and 1D NCs.[214] $SnO_2$ NPs with a size of 5-8 nm on MWCNT were found to be more effective in charge-storage performance than the NPs decorated on Vulcan XC-72 carbon.[215] This implies that MWCNTs is a better platform for the NPs to be decorated on due to more open channels and higher conductivity of MWCNT.[215]

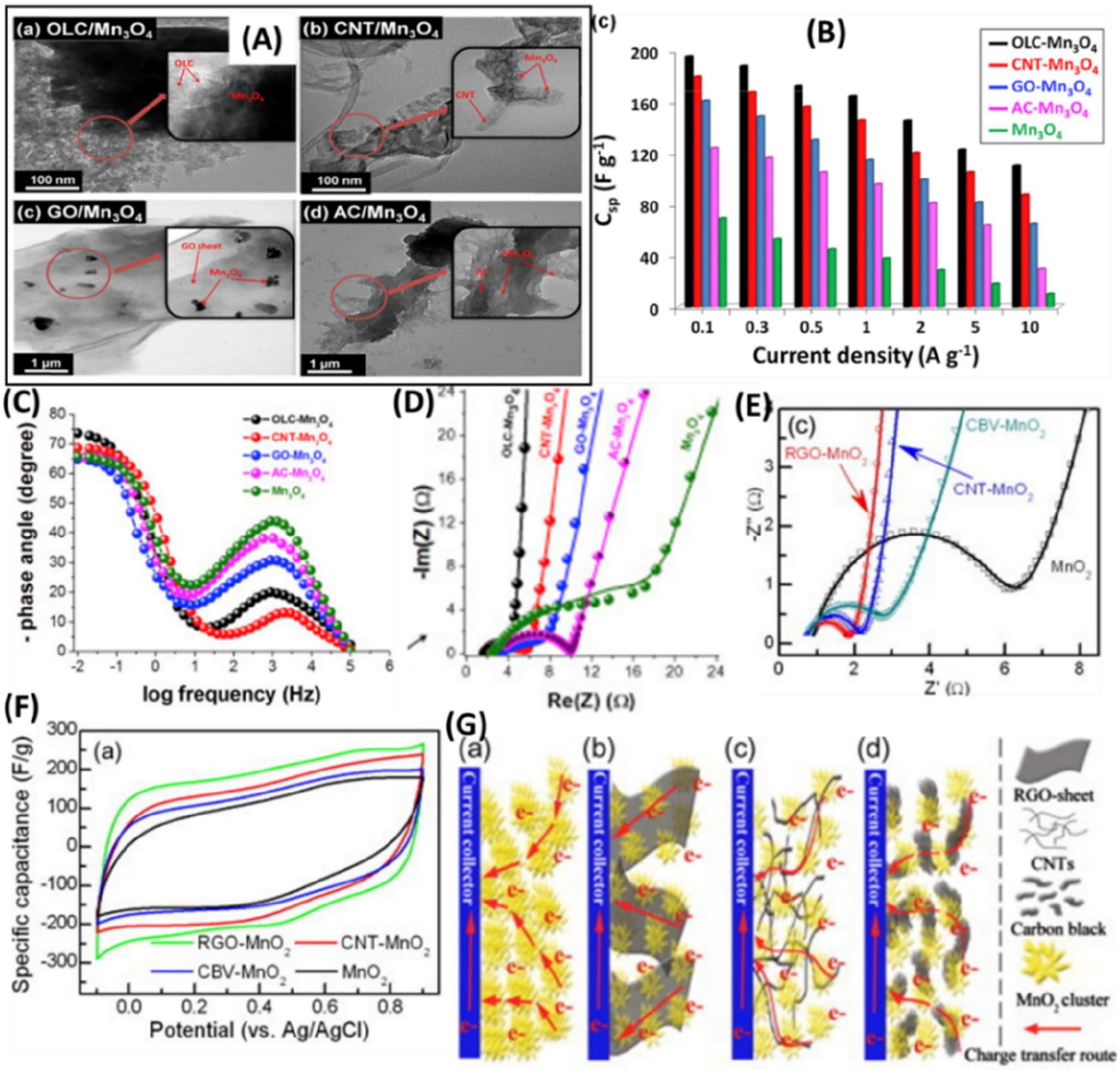

*Figure 14: (A) Scanning electron micrographs of $Mn_3O_4$ NPs decorated on onion-like carbon, CNT, GO and AC. (B) specific capacitance vs current density, and (C-D) Nyquist plot of $Mn_3O_4$/NC symmetric device. Reproduced from Ref. [153], © 2017, with permission from Elsevier Ltd.; (E) Nyquist plot and (F) cyclic voltammogram of $MnO_2$/NCs in 3-electrode configuration. (G) schematic of charge-storage of $MnO_2$/NCs. Reproduced from Ref. [214], ©2012, with permission from Elsevier Ltd.*

Recent results[153][213][214] suggest that a proper choice of NPs and NC is equally significant to obtain the high charge-storage performance. Since the NC are quite different from each other in terms of bonding, hybridizations and intrinsic structural properties, one should expect the possibility of decoration of metal oxide NPs with different number densities and sizes since nucleation and growth of metal oxide NPs depend on the substrate structure. Moreover, EDLC contributions to the composite are also different depending on the used NC. Other factors also impacting on the electrochemical performances of supercapacitor are morphology, $I_D/I_G$ ratio, lattice spacing and metal oxide-content to carbon-content ratio.

### 5.12. Do NPs/NC always work?

There are plenty of reports on the NPs/NC composites as promising electrode materials. After decorating carbon aerogel by Cu and Ag NPs, gravimetric capacitance values of 100 and 76 F/g at 1 A/g in 1M $H_2SO_4$ *vs.* Ag/AgCl were obtained, respectively, whereas pristine carbon aerogel delivered a higher capacitance of 107 F/g at the same testing conditions.[164] As another example, the obtained specific capacitance of a core-shell structure of $Co_3O_4$@$RuO_2$ anchored on RGO[177] was 472 F/g at 0.5 A/g, which is not much higher than that obtained with $Co_3O_4$/NGO (456 F/g at 0.5 A/g) in the same testing 3-electrode configuration.[177] On the contrary, significantly improved electrochemical performances were evidenced when Cu and Ag NPs were decorated on vertical graphenes.[133] The previous results [133][164] indicate that not only the NPs decoration is essential to enhance the charge storage performance, but also the geometry of NC, the way NPs distribute on it, the number density of NPs etc. are the key factors.

## 6. Metal-ion capacitors anode

Various NPs/NC composites explored as a potential anode material for metal-ion capacitors are, mostly $Li^+$-ions and $Na^+$-ions, tabulated in Table 5. The ion-storage performances of the composites in half-cell are same as discussed in section 4. Since the charge-storage mechanisms of the battery and supercapacitor electrodes are completely different, the main challenge here is to balance the kinetics between the anode and the cathode. One can easily manipulate the weight ratio between cathode and anode to get a higher potential window and hence energy density.[20] Besides the storage capacity, one also needs to check the rate performance, the Coulombic efficiency, and the cycle life with respect to the mass ratio. It is also important to mention that one may not always obtain symmetric cyclic voltammogram or triangular charge-discharge for metal-ion capacitors as from a symmetric supercapacitor device since the total charge storage rely on the supercapacitor electrode and battery electrode.[216]

### 6.1. Lithium-ion capacitors

Based on the literature survey, it has been seen that $Nb_2O_5$ is one of the extensively studied materials as anode since (i) it has a relatively higher specific capacity of 200 mAh/g than LTO (175 mAh/g), (ii) rapid capacitive insertion/de-insertion of ions, (iii) it is quite safe to operate in the voltage window (1-3.5 V), (iv) prelithiation is not needed, (v) excellent rate capability originates from its pseudocapacitive intercalation nature, and (vi) low cost.[217] Three phases of $Nb_2O_5$ composite were explored till now, namely amorphous, orthorhombic (T), pseudohexagonal (TT). Compared to amorphous $Nb_2O_5$/rGO paper, T-$Nb_2O_5$/rGO paper exhibited much higher volumetric and gravimetric capacitance owing to its higher crystallinity and expanded interlayers with the mesoporous channel.[218] Due to sufficient storage sites with open structures, single-phase reaction and negligible volume changes, T-$Nb_2O_5$@carbon core-shell nanocrystals showed higher specific capacity and rate capability.[217] The *b*-value of $Nb_2O_5$ QD coated biomass carbon in 0.1 to 1 mV/s, estimated from the relation between peak current (*I*) and scan rate (*v*) ($I \propto v^b$), was 0.92 which ensured a rapid capacitive insertion/deinsertion of $Li^+$-ions.[219] *b* = 1 corresponds to pure capacitive response and 0.5 indicates semi-infinite diffusion-controlled charge-storage. In spite of growing the NPs on CNT scaffold,[220] decorating the NPs on both sides of hollow CNT surfaces could be a promising strategy for maximizing the usage of NC surface.[221] It has been seen that M-$Nb_2O_5$@C/RGO (M stands for metal-organic framework) outperformed compared to M-$Nb_2O_5$@C when they are explored as anodes for LIB due to the smaller particle size and higher surface area. The presence of additional carbon such as graphene lowers the charge-transfer resistance.[87]

To further improve the performance, doping and functionalization are also appealing. AC was functionalized with oxygen and $SnO_2$ NPs with a size of 2-5 nm were decorated on it.[222] $SnO_2$ microparticles in the composite were found to be beneficial for $Li_2O$ activation energy reduction during the decomposition and hence for the effective conversion of Sn-$SnO_2$ during the charging process.[222] Despite the size of NPs, surprisingly RGO decorated with $SnO_2$ NPs (size ~ 250 nm) showed outstanding performance as LIC when assembled with physically activated RGO cathode.[22] An N-doped carbon was coated on hollow $NiNb_2O_6$ NPs, which delivered a very high capacity of 475.4 mAh/g at 0.05 mA/g and the corresponding $Li^+$-ion capacitor provided a very high energy density of 123.9 Wh/kg at 100 W/kg. In the composite, Ni promotes the fast electron transfer and improves the electrical conductivity without contributing any reversible capacity.[223] Although the $MnO_2$/RGO nanoscrolls//AC metal-ion capacitors showed the high (92%) capacitance retention at 5 A/g after 10000 charge-discharge cycles, changes in the structure occurred as observed from the XPS result (Figure 15B). This fact was attributed to the humidity-induced alteration of Mn valence state due to the interaction between $MnO_2$ and water vapour.[216]

Looking at the impressive progress on the metal oxide NPs coated NC as an electrode materials, designing metal carbide NPs became the rapidly evolving approach.[123][224] Eventually, 2D metal carbides, popularly known as MXenes ( also available in nitride/carbonitride forms), are also emerging.[225] However, 3D interconnected TiC NPs chain was reported to deliver the capacity of 450 mAh/g, which is much higher than its theoretical value (30 mAh/g) and also higher than the delaminated $Ti_3C_2T_x$ MXene thin film (410 mAh/g) and 300 mm $Ti_3C_2$ (5.9 mAh/g).[224] The unavoidable presence of TiOC, $TiO_2$ and $Ti_xO_y$ is advantageous for the partial $Li^+$-ion insertion at 1.5 V.[224] In contrast to the TiC NP chain, MXene synthesis by exfoliation from its MAX phase is a time consuming and lengthy process, which leads to a material containing additional functional groups of fluorine and Chlorine.[226] Importantly, the voltage of $Li^+$-ion capacitor made with TiC NP chain as anode with N-doped porous carbon as cathode was found to be 4.5 V, which is one of the highest voltage obtained among the reports.[224]

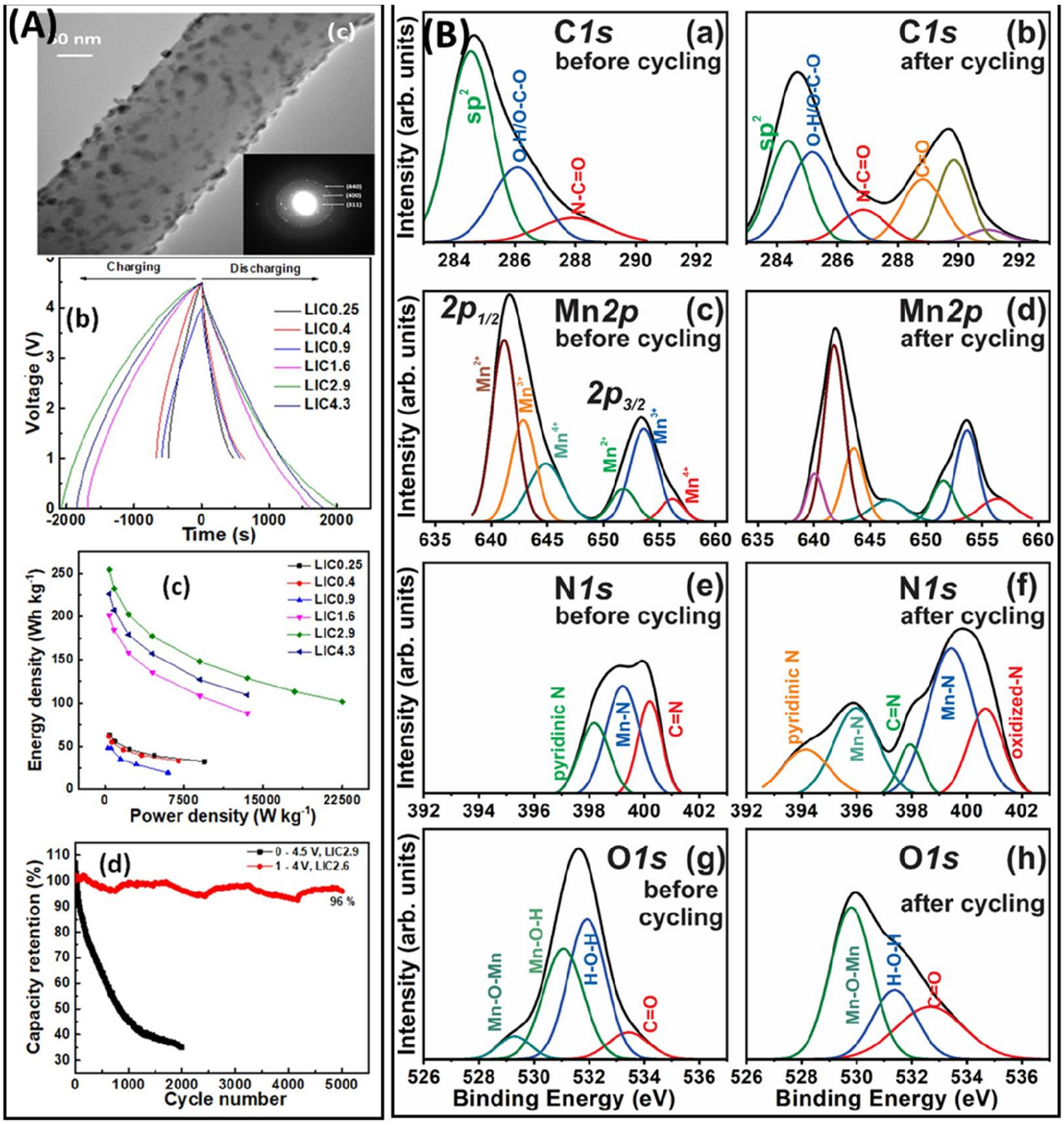

Figure 15: (A) Morphology and $Li^+$-ion capacitor performance of CNFs supported $CoNi_2S_4$ NPs hybrid anode. Adopted from Ref. [20]. Copyright ©2018. Springer-Nature publishers. (B) changes in X-ray spectra of C, Mn, N and O of $MnO_2$/RGO nanoscroll before and after cycling. Adopted from Ref. [216], © 2020 by the authors, License MDPI, Basel, Switzerland.

Compared to sulfides and selenides, transition metal tellurides possess higher density, lower electronegativity, higher conductivity, stability against the atmosphere and thanks to the higher atomic size, they are excellent to accommodate $Li^+$-ions, resulting in improved $Li^+$-ion diffusion kinetics. $CoSe_2$ NPs were embedded in N-doped hard carbon microspheres, in which larger irreversible capacity mainly arose from N-doped hard carbon microspheres. Moreover, when used as anode this composite showed a narrow working potential range during charging/discharging such that the capacitive cathode had a wide working potential and hence the energy density of $Li^+$-ion capacitor increased.[227] Where most of the composite is limited within the volumetric capacity in the range of 100-300 mAh/cm$^3$, tellurides are emerging as an excellent candidate to provide higher volumetric capacity due to their high packing density (e.g. 6.24 g/cm$^3$ for $CoTe_2$).[228] Importantly, although $Co_3O_4$-N doped three-dimensional porous carbon exhibited higher specific capacity, the long-term cycling structural stability of $CoTe_2$-N doped three-dimensional porous carbon was attributed to the higher surface area and reasonable micropores/mesopores ratio.[229] Moreover, the $Li^+$-ion diffusion coefficients of polyhedral $CoTe_2$-C was found to be higher ($1.37 \times 10^{-15}$ cm$^2$ S$^{-1}$) compared to their bulk counterpart ($2.15 \times 10^{-16}$ cm$^2$ S$^{-1}$) revealing the importance of NC inclusions.[228] Telluride NPs/NC composites are also emerging as a promising candidate for $Na^+$-ion storage too.[228] The electrochemical performance of NPs/NCs as anode for Na-ion batteries is discussed in the next section. Despite those advantages, tellurides have a low gravimetric capacity (420 mAh/g) compared to sulfides (1675 mAh/g) and selenides (678 mAh/g).

### 6.2. Sodium-ion capacitors

As an alternative to $Li^+$-ion capacitor, $Na^+$-ion capacitor is also in the pipeline as a viable energy storage device. Graphite is unfavourable for $Na^+$-ion insertion and hard carbon is anticipated as a potential alternative to graphite. However, low reversible capacity of 250 mAh/g, poor initial Coulombic efficiency and safety issues due to the low working potential are the bottleneck challenges of using hard carbon as an anode. Since there are limited anode materials to host $Na^+$-ion, development in this field is still ongoing.

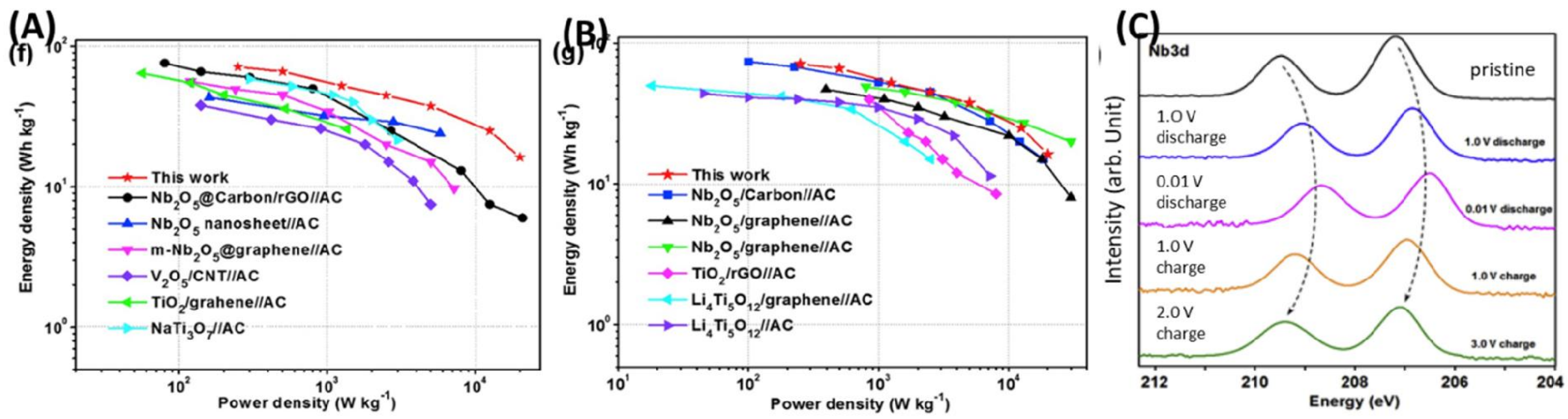

*Figure 16: Ragone plots of the mesoporous-$Nb_2O_5$@C//activated carbon $Na^+$-ion capacitor cell with reported (A) $Na^+$-ion capacitor and (B) $Li^+$-ion capacitor. (C) The changes of Nb 3d XPS peak of mesoporous $Nb_2O_5$/carbon composite during the charge-discharge Reproduced from Ref. [7], © 2018, with permission from Elsevier.*

With the combination of $Nb_2O_5$/carbon composite and activated carbon, one can obtain a very high specific capacity and voltage (Figure 16). In $SnO_2$ nanocrystals anchored on RGO [230], ultrafine nanocrystal of $SnO_2$ (sub-10 nm) with exposed {221} facet maximized the reaction kinetics on the surface and improved the pseudocapacitance. The material with <10 nm is highly capacitive and the {221} facet of $SnO_2$ crystals has the lowest surface adsorption energy for $Na^+$-ion of -10.3 eV compared to {001} (-8.4 eV) and {110} (-6.5 eV). In addition to electric double layer capacitance, graphene keeps the electrode structure stable against volume changes. Importantly, a continuous increase in capacitance for $SnO_2$ nanocrystals anchored on RGO/CNT based $Na^+$-ion capacitor has been observed and attributed to the self-activation of solid-electrolyte-interfacial film formed on the surface.[230] $TiO_2$ based nanocomposite proved to be another promising electrode for $Na^+$-ion capacitor. The capacity contribution of anatase $TiO_2$ mesocage/graphene nanocomposite at below 1 V *vs.* $Na/Na^+$ in contrast to the typical voltage plateau of 1.75-2.1 V *vs.* $Li/Li^+$ ensures the suitability of anatase $TiO_2$ as an anode material for $Na^+$-ion storage.[231]

The several current challenges in the metal-ion capacitors technology are: (i) designing anode materials with high capacity, (ii) anode materials with low lithiation (or sodiation) potential since the low potential allows the cell to have higher voltage (ii) minimizing the kinetic imbalance between the electrodes and (iii) designing suitable cathode materials with high capacity. Since the total capacity is based on the equation: $1/_{C} = 1/_{C_{cathode}} + 1/_{C_{anode}}$ and conventional activated carbons cathode has a low capacity of 35 mAh/g (which limits the energy density of the device), a mass ratio between activated carbons and NPs/NCs composite at least > 3 was used to obtain the highest storage capacity (see Table 5). This also imbalance the charge-transfer kinetics since one has to either increase the amount of activated carbon or lower the loading of anode materials. Thus, recent research has been directed towards searching for an alternative strategy to fabricate cathode materials. For example, doping and/or functionalizing activated carbons by heteroatoms[94][222][232] or using boron carbonitride nanotubes[233] with pseudocapacitive properties as a cathode are the promising approaches to enhance the cycle life and the power density simultaneously. The challenge of dissolution of metal oxide NPs in the organic electrolyte was tackled by wrapping the NPs with a carbon coating.[216] Solid electrolytes can also be the probable solution to the safety issues caused by the flammability, leakage, and internal short-circuit of organic liquid electrolytes.

*Table 5: State-of-art NPs/NCs anode-based metal-ion capacitors. Symbols with an asterisk represents the data estimated either from the plot using WebPlotDigitizer copyright 2010-2020 Ankit Rohatgi or available data from the cited reference. (LIC: $Li^+$-ion capacitor, SIC: $Na^+$-ion capacitor, EC: ethylene carbonate, DEC: diethyl carbonate, DMC: dimythyl carbonate, FEC: fluoroethylene carbonate)*

| **Anode//cathode** | **Synthesis method for anode** | **Type** | **Operating voltage, optimized cathode-to-anode mass ratio, electrolyte** | **Specific capacity** | **Cycle life, current density, cycle number** | **Energy density at Power density** |
|---|---|---|---|---|---|---|
| Sn@N-mesoporous carbon//pomelo peel derived carbon [23] | aerosol-assisted spraying process | LIC | 2.0 – 4.5 V, 3:1, 1 M $LiPF_6$ in EC:DEC (1:1) | - | 70%, 2 A/g, 5k | 195.7 Wh/kg at 731.25 W/kg, 84.6 Wh/kg at 24375 W/kg |
| M-$Nb_2O_5$@C/reduced graphene oxide (RGO)//activated carbon (AC) [87] | Mixing + assembling at ambient temperature + calcination | LIC | 0.7 – 3.2 V, 6:1, 1 M $LiPF_6$ in EC:DEC:DMC (1:1:1) | - | 94%, 0.2 A/g, 2.5k | 71.5 Wh/kg at 0.247 kW/kg, 18.3 Wh/kg at 3.9 kW/kg |
| CNT-$Nb_2O_5$//AC [220] | solvothermal | LIC | 3.0 – 0.5 V, 3.5:1, 1 M $LiPF_6$ in EC:DMC (1:1) | - | 80%, 10C, 1k | 47.17 Wh/kg at 86.46 W/kg, 14.77 Wh/kg at 6753.54 W/kg |
| $CoNi_2S_4$@carbon nanofiber//AC [20] | Electrodeposition | LIC | 1 – 4 V, 2.6:1, 1 M $LiPF_6$ in EC/DEC | 137 F/g at 0.05 A/g, 56 F/g at 5 A/g (41%) | 96%, 2 A/g, 5k | 85.4 Wh/kg at 150 W/kg, 35 Wh/kg at 15 kW/kg |
| T-$Nb_2O_5$ NPs@N- carbon hollow tubes//AC [221] | Hydrothermal + calcination | LIC | 0.5 – 3.0 V, 3:1, 1 M $LiPF_6$ in EC:DMC (1:1) | - | 91.1%, 0.5 mV/s, 1k | 49.7 Wh/kg at 8750 W/kg |
| RGO-$SnO_2$ NPs//physically activated RGO [22] | Mixing+ heating + freeze-drying + thermal reduction | LIC | 1.5 – 4.2 V, 2:1, 1 M $LiPF_6$ in EC:DMC (1:1) | - | 70%, 3 A/g, 5k | 186 Wh/kg at 142 W/kg |
| $SnO_2$-OAC//N-AC [222] | Vacuum melting | LIC | 2.0 – 4.5 V, 2:1, 1 M $LiPF_6$ in DMC:FEC (4:1) | 75* mAh/g, 0.4 A/g, 25* mAh/g, 6.4 A/g | 93%, 1.6 A/g, 9k | 151 Wh/kg at 787 W/kg, 13 Wh/kg at 7753 W/kg |
| $Fe_3O_4$-RGO//AC [234] | Solvothermal | LIC | 0 – 4 V*, 5:1, 1 M $LiPF_6$ in EC:DEC:DMC (1:1:1) | 65 mAh/g at 0.2 A/g, 27 mAh/g at 2 A/g | 78.9%, 0.4 A/g, 1k | 98.8 Wh/kg at 343.8 W/kg, 38.3 Wh/kg at 3.4 kW/kg |
| 3D interconnected $Fe_3O_4$@carbon core@shell//AC [235] | Hydrothermal + calcination | LIC | 0 – 4.0 V, 3:1, 1 M $LiPF_6$ in EC:DEC (1:1) | 39.6 mAh/g at 1 A/g, 19.5 mAh/g at 5 A/g | 95.7% at 1 A/g, 1k | 110.1 Wh/kg at 250 W/kg, 36.8 Wh/kg at 2.5 W/kg |
| mesoporous Si@N-carbon//glucose-derived carbon nanospheres [68] | carbonization + Magnesiothermic reduction + residue removal | LIC | 2 – 4.5 V, 7:1, 1 M $LiPF_6$ in EC:DMC (1:1) | - | 89% at 8 A/g, 20k | 210 Wh/kg at 0.32 kW/kg, 70.1 Wh/kg at 36.1 kW/kg |
| N-carbon coated $NiNb_2O_6$ hollow NP//AC [223] | Dopamine polymerization + carbonization | LIC | 0 – 4 V, 4:1, 1 M $LiPF_6$ in EC:DEC:DMC (1:1:1) | 47.4 F/g at 0.2 A/g, 27.5 F/g at 5 A/g | 86.1%, 1 A/g, 5k | 123 Wh/kg at 100 W/kg, 61.1 Wh/kg at 10k W/kg |
| MnO @ 3D interconnected graphene scroll framework//AC [236] | Sonication + lyophilisation + annealing | LIC | 1 – 4 V, 4:1, 1 M $LiPF_6$ in EC:DEC:DMC (1:1:1) | 144.5 F/g at 0.1 A/g, 68.3 F/g at 10 A/g | 80.8%, 5 A/g, 5k | 179.3 Wh/kg at 139.2 W/kg, 48.2 Wh/kg at 11.7 kW/kg |
| 3D interconnected TiC NPs chain//N-doped porous carbon [224] | Hydrothermal + freeze-dry + vacuum dry + annealing | LIC | 0.0 – 4.5 V, 4:1, 1 M $LiPF_6$ in EC/DMC | 160* F/g at 0.1 A/g, 50* F/g at 30 A/g | 82%, 2 A/g, 5k | 101.5 Wh/kg at 450 W/kg, 23.4 Wh/kg at 67.5 kW/kg |
| SiNPs@RGOA//Boroncarbonitride [233] | Hydrothermal + heating + residue removal + filtration + vacuum drying | LIC | 0 – 4.5 V, 4:1, 1 M $LiPF_6$ in EC:DEC:DMC 1:1:1) | 70.2 F/g at 0.1 A/g, 25.1 F/g at 5 A/g | 82.4%, 10 A/g, 10k | 197.3 Wh/kg at 225 W/kg |
| S-$TiO_2$ embedded N-carbon nanosheets//N-porous carbon [94] | Thermal treatment | LIC | 1.0 – 4.2 V, 2:1, 1 M $LiPF_6$ in EC:DMC (1:1) | 40 F/g at 0.1 A/g, 25 F/g at 1 A/g | 85.8%, 5 A/g, 10k | 92.7 Wh/kg at 260 W/kg, 33.2 Wh/kg at 26000 W/k |
| T-$Nb_2O_5$@Carbon Core-Shell Nanocrystals// AC (MSP-20) [217] | Microemulsion + heat treatment | LIC | – 3.5 V, 3.5:1, 1 M $LiPF_6$ in EC:DMC (1:1) | - | - | 63 Wh/kg at 70 W/kg, 5 Wh/kg at 16528 W/kg |
| G–$MoO_2$//G–$MoO_2$ [237] | Hydrothermal method | LIC | 0 – 3 V*, -, 1 M $LiPF_6$ in EC:DMC (1:1) | 624 F/g at 0.05 A/g, 173.2 F/g at 1 A/g | 91.2%, 0.5 A/g, 0.5k | 142.6 Wh/kg at 150 W/kg, 33.2 Wh/kg at 3000 W/kg |
| $MnO_2$/RGO nanoscrolls//AC [216] | Sonication+ centrifugation+ filtration +annealing | LIC | 0 – 3 V, -, 1 M $LiPF_6$ in EC:DMC (1:1) | 223.2 F/g | 92%, 5 A/g, 10k | 105.3 Wh/kg at 308.1 W/kg, 42.77 Wh/kg at 30800 W/kg |
| $CoSe_2$ NPs embedded N-doped hard carbon microspheres//AC [227] | Hydrothermal + carbonization + selenization | LIC | 2.2 – 4.5 V, -, 1 M $LiPF_6$ in EC:DEC:DMC (1:1:1) | - | 94.38%, 2k | 144 Wh/kg at 335 W/kg |
| Ultra-small $Mn_3O_4$ NPs-porous carbon microrods// porous carbon rods [238] | Vigorous stirring of precursors | HLIC | 0 – 4 V*, 1:1, 1 M $LiPF_6$ in EC:DEC (1:1) | 78.5 F/g at 0.1 A/g, 33.5 F/g at 5 A/g | 93%, 2 A/g, 2k | 174 Wh/kg at 0.2 kW/kg, 74.5 Wh/kg at 10 kW/kg |
| $Co_3ZnC$@N-carbon nanopolyhedra// N/O- microporous carbon [232] | Co-precipitation + heating under Ar/$H_2$ and Ar | HLIC | 1.0 – 4.5 V, -, 1 M $LiPF_6$ in EC:DMC (1:1) | 52.8 F/g at 0.1 A/g, 12* F/g at 5 A/g | 80%, 1 A/g, 1k | 141.4 Wh/kg at 0.275 kW/kg, 15.2 Wh/kg at 10.3 kW/kg |
| FeS QDs embedded in 3D inverse opal-structured N-doped carbon//AC [239] | Annealing the precursors mixture under Ar gas | SIHC | 0.5 – 3.4 V, 3:1, 1 M $NaClO_4$ in EC:PC (1:1) | 129 F/g at 0.1 A/g, 62 F/g at 6.4 A/g | 91%, 1 A/g, 5k | 151.8 Wh/kg at 9280 W/kg |
| Mesoporous orthorhombic $Nb_2O_5$@ carbon composite//AC [7] | Mixing the solutions + calcination at 800 °C in $N_2$ | SIC | 0 – 4 V*, 4:1, 1 M $NaClO_4$ in EC:PC (1:1) + 0.5 wt% FEC | - | - | 73 Wh/kg at 250 W/kg, 16.8 Wh/kg at 20 kW/kg |
| Sub-10 nm $SnO_2$ nanocrystals anchored on RGO//CNT [230] | Hydrothermal method | SIC | 0 – 3.8 V, 3.6-3.8:1, $Na^+$ conducting gel polymer electrolyte | 41 F/g at 0.5 A/g, 27 F/g at 1.2 A/g | 100%, 0.5 A/g, 900 | 86 Wh/kg at 955 W/kg, Max 4100 kW/kg |
| Anatase $TiO_2$ mesocage@graphene nanocomposite//AC [231] | microwave- assisted solvothermal method | SIC | 1 – 3.8 V, 5:1, 1 M $NaClO_4$ in EC:PC (1:1) + 0.5 wt% FEC | - | 90%, 10C, 10k | 64.2 Wh/kg at 56.3 W/kg, 25.8 Wh/kg at 1357 W/kg |

## 7. Summary, challenges and outlook

This review summarized and scrutinized the latest progress of metal-based nanoparticles(NPs)-nanocarbons (NCs) composites as anode materials for batteries, and metal-ion capacitors and electrode materials for supercapacitor applications. The metal-based NPs decorated on the zero-dimensional to three-dimensional nanocarbons (NC) are metal, metal oxide, nitride, sulfide, phosphide, selenide, carbide, chalcogenides etc., and the shape of NPs includes solid, hollow, yolk-shell, core-shell etc. This review also highlighted the impact of doping the NC as well as the NPs to improve the charge-storage performances. It is important to note that an as-prepared composite with specific features is highly desirable and anticipated as an electrode material for a specific energy storage device. For example, $Mn_3O_4$/GO with high surface area and oxygen functional groups represents a better supercapacitor electrode, whereas $Mn_3O_4$/RGO with higher electrical conductivity, high crystalline NPs enwrapped with RGO, and accommodable void spaces in curved RGO was found to be a better anode for $Li^+$-ion batteries (Figure 17B).[213] However, it is difficult to control single property of the materials keeping other properties remain constant experimentally. There is no straightforward relationship can be established unless light given onto it theoretically.

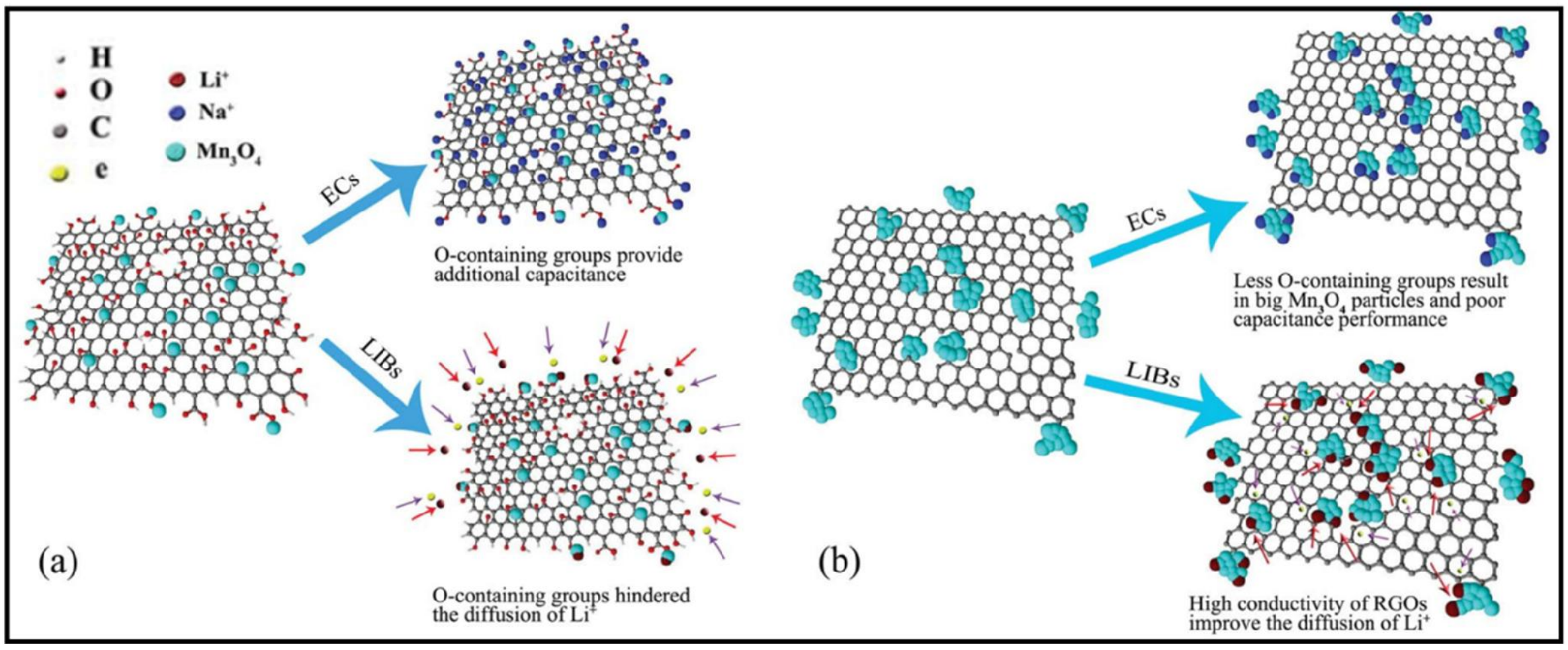

*Figure 17: The charge storage mechanism of $Mn_3O_4$/GO and $Mn_3O_4$/RGO for $Li^+$-ion battery and supercapacitor. Reproduced from Ref. [213], © 2013, with permission from The Royal Society of Chemistry.*

The key challenges of NPs/NC composite for the energy storage technology are outlined below.

(i) There are diverse nanostructures designed in various innovative manners and implemented for energy storage applications. While the advantageous features of NPs/NCs are highlighted in this review, other nanostructures, for example, N-doped hierarchical porous carbon (HPC)/SnS nanorod retained a higher discharge capacity of 717.06 mAh/g even after 100 charge/discharge cycles compared to the N-HPC/$SnS_2$ NPs composite.[240] Even NCs like the as-prepared carbon NPs from candle soot showed high storage capacity of $Li^+$-ion (1240 mAh/g at 0.15 A/g), $Na^+$-ion (300 mAh/g at 0.15 A/g) and $K^+$-ion (140 mAh/g at 1 A/g) [241] compared to the many literature reports. Thus, the development of the preparation methodology of NPs/NC and the control of their intrinsic properties are still open subjects of research to obtain high-performance electrode materials.

(ii) While the maximum effort has been dedicated towards the size, shape and mass loading of NPs, attention has to be paid to the thickness of NC coating to ensure the excellent permeability of electrolyte ions, since diffusion is essential for the charge storage.

(iii) Although NCs were used as a backbone in the composite and provided excellent electrical conductivity, current collectors, conducting agents and binders are still being used, and they increase the dead volume and the dead weight in the electrode, increasing the polarization resistance. Although self-standing nanostructures are shown to be promising, they are far away from industrial implementation. It is also noteworthy that the weight of current collectors, binders and conductive agents is not taken into consideration in many reports when the specific capacity or capacitance are calculated. As a result, there is a huge gap between the value obtained and reported in the literature and the value obtained in real devices after packaging.

(iv) Another important factor for an efficient energy storage device is the power delivery at > 60C rate without sacrificing the energy density.[242]

(v) As can be seen from the Table 4, in spite of significant advances on the composite as post-$Li^{+}$-ion storage, there is still *plenty of rooms* to mature other metal-ion battery technology. Moreover, other metal-ion capacitors based on $Al^{3+}$-ion, $K^{+}$-ion, $Mg^{2+}$-ion, $Zn^{+}$-ion etc. are rarely explored and need attentions.

(vi) An attention should be given on the 'dark side' of clean energy technology such as use of hazardous materials, their waste-management and environmental impacts. Briefly, for example, availability of lithium is like 'salt on salad' and the materials like Ni, Co etc. comes at health and environmental costs.

Therefore, there are plentiful remaining challenges, which need to be addressed by conducting more research on *in-depth* understanding and developing the corresponding technology. Despite all the challenges and pitfalls discussed in this review, the innovative strategies of NPs/NC composite preparation and novel design of electrochemical energy storage device fabrication will be very significant for the future energy technology.

**Acknowledgement**

S.G., A.M., and C.S.C. acknowledge funding from the European Research Council (ERC) under the European Union's Horizon 2020 research and innovation program ERC—Consolidator Grant (ERC CoG 2016 EspLORE grant agreement No. 724610, website: www.esplore.polimi.it). S. G thanks to European commission for Seal of Excellence award under the Horizon 2020's Marie Skłodowska-Curie actions. A. M. acknowledge partial support from Ministry of Education, University and Research (MIUR), Italy, within the "Department of Excellence 2018-2022" award program. K.O. thanks to the Australian Research Council and QUT Centre for Materials Science for partial support. The authors are very thankful to all researchers who have contributed to the relevant research areas and do apologize for not including every relevant publication because of the obvious limitations in their knowledge, time, and available space. We also acknowledge anonymous reviewers for their excellent and constructive comments.